
\input harvmac
\input amssym.def
\input amssym
\baselineskip 14pt
\magnification\magstep1
\font\bigcmsy=cmsy10 scaled 1500
\parskip 6pt
\newdimen\itemindent \itemindent=32pt
\def\textindent#1{\parindent=\itemindent\let\par=\resetpar%
\indent\llap{#1\enspace}\ignorespaces}

\let\oldpar=\par
\def\resetpar{\oldpar\parindent=20pt\let\par=\oldpar}

\font\ninerm=cmr9 \font\ninesy=cmsy9
\font\eightrm=cmr8 \font\sixrm=cmr6
\font\eighti=cmmi8 \font\sixi=cmmi6
\font\eightsy=cmsy8 \font\sixsy=cmsy6
\font\eightbf=cmbx8 \font\sixbf=cmbx6
\font\eightit=cmti8
\def\eightpoint{\def\rm{\fam0\eightrm}
  \textfont0=\eightrm \scriptfont0=\sixrm \scriptscriptfont0=\fiverm
  \textfont1=\eighti  \scriptfont1=\sixi  \scriptscriptfont1=\fivei
  \textfont2=\eightsy \scriptfont2=\sixsy \scriptscriptfont2=\fivesy
  \textfont3=\tenex   \scriptfont3=\tenex \scriptscriptfont3=\tenex
  \textfont\itfam=\eightit  \def\it{\fam\itfam\eightit}%
  \textfont\bffam=\eightbf  \scriptfont\bffam=\sixbf
  \scriptscriptfont\bffam=\fivebf  \def\bf{\fam\bffam\eightbf}%
  \normalbaselineskip=9pt
  \setbox\strutbox=\hbox{\vrule height7pt depth2pt width0pt}%
  \let\big=\eightbig  \normalbaselines\rm}
\catcode`@=11 %
\def\eightbig#1{{\hbox{$\textfont0=\ninerm\textfont2=\ninesy
  \left#1\vbox to6.5pt{}\right.\n@@space$}}}
\def\vfootnote#1{\insert\footins\bgroup\eightpoint
  \interlinepenalty=\interfootnotelinepenalty
  \splittopskip=\ht\strutbox %
  \splitmaxdepth=\dp\strutbox %
  \leftskip=0pt \rightskip=0pt \spaceskip=0pt \xspaceskip=0pt
  \textindent{#1}\footstrut\futurelet\next\fo@t}
\catcode`@=12 %
\def \de{\delta}

\def \si{\sigma}

\def \pr{\partial}
\def \tr{{\rm tr }}

\def \ha{{\hat a}}
\def \hb{{\hat b}}
\def \hk{{\hat k}}
\def \hp{{\hat p}}
\def \hq{{\hat q}}

\def \bs{{\bar s}}

\def \bQ{{\bar Q}}
\def \bS{{\bar S}}
\def \bJ{{\bar J}}

\def \bO{{\bar {\cal O}}}
\def \bm{{\bar m}}

\def \l{\langle}
\def \r{\rangle}

\def \vep{\varepsilon}
\def \half{{\textstyle {1 \over 2}}}

\def \quar{{\textstyle {1 \over 4}}}
\def \ts{\textstyle}

\def \d{{\rm d}}

\def \bs{{\bar s}}
\def \bt{{\bar t}}
\def \P{{\rm P}}
\def \K{{\rm K}}
\def \J{{\rm J}}

\def \A{{\cal A}}
\def \B{{\cal B}}
\def \C{{\cal C}}
\def \D{{\cal D}}
\def \E{{\cal E}}

\def \I{{\cal I}}
\def \J{{\cal J}}

\def \N{{\cal N}}
\def \O{{\cal O}}
\def \Q{{\cal Q}}
\def \S{{\cal S}}
\def \W{{\cal W}}
\def \V{{\cal V}}

\def \1{{\bar 1}}
\def \2{{\bar 2}}
\def \3{{\bar 3}}

\def \vphi{{\varphi}}
\def \Bsw{\!\mathrel{\hbox{\bigcmsy\char'056}}\!}
\def \Bse{\!\mathrel{\hbox{\bigcmsy\char'046}}\!}

\def \bcJ{{\bar \J}}
\def \bj{{\bar \jmath}}

\def \dal{{\dot \alpha}}
\def \dbe{{\dot \beta}}
\def \dga{{\dot \gamma}}
\def \dde{{\dot \delta}}

\def \bsi{\bar \sigma}

\def \mapright#1{\smash{\mathop{\longrightarrow}\limits^{#1}}}
\def \ss{\scriptstyle}
\def \sss{\scriptscriptstyle}
\def \uLambda{{\underline \Lambda}}
\font \bigbf=cmbx10 scaled \magstep1

\lref\hughtwo{J. Erdmenger and H. Osborn, {\it Conserved Currents and the 
Energy Momentum Tensor in Conformally Invariant Theories for General 
Dimensions}, Nucl. Phys. B483 (1997) 431, hep-th/9605009.}
\lref\hughone{H. Osborn and A. Petkou, {\it Implications of Conformal 
Invariance for Quantum Field Theories in $d>2$}
Ann. Phys. (N.Y.) {231} (1994) 311, hep-th/9307010.}
\lref\WCon{B.P. Conlong and P.C. West, {\it Anomalous dimensions of 
fields in a supersymmetric quantum field theory at a renormalization 
group fixed point}, J. Phys. A 26 (1993) 3325.}
\lref\HO{H. Osborn, {\it $\N=1$ Superconformal Symmetry in Four-Dimensional
Quantum Field Theory}, Ann. Phys. (N.Y.) 272 (1999) 243, hep-th/9808041.}
\lref\Sei{S. Lee, S. Minwalla, M. Rangamani and N. Seiberg, {\it Three-Point
Functions of Chiral Operators in $D=4$, $\N=4$ SYM at Large $N$},
Adv. Theor. Math. Phys.  2 (1998) 697, hep-th/9806074.}

\lref\LW{K. Lang and W. R\"uhl, Nucl. Phys. {B402} (1993) 573.}
\lref\Pet{A.C. Petkou, Ann. Phys. (N.Y.) 249 (1996) 180, hep-th/9410093.}

\lref\West{E. D'Hoker, D.Z. Freedman, S.D. Mathur, A. Matusis,
L. Rastelli, {\it in} The Many Faces of the Superworld, ed. M.A. Shifman,
hep-th/9908160.\semi
M. Bianchi and S. Kovacs, Phys. Lett. B468 (1999) 102, hep-th/9910016.}

\lref\Nextreme{B. Eden, P.S. Howe, C. Schubert, E. Sokatchev and P.C. West,
{\it Extremal correlators in four-dimensional SCFT},
Phys. Lett. B472 (2000) 323, hep-th/9910150\semi
J. Erdmenger and M. P\'erez-Victoria, {\it Non-renormalization of
next-to-extremal correlators in $\N=4$ SYM and the AdS/CFT correspondence}.
Phys. Rev. D62 (2000) 045008, hep-th/9912250\semi
E. D'Hoker, J. Erdmenger, D.Z. Freedman and M. P\'erez-Victoria,
{\it Near-extremal correlators and vanishing supergravity couplings in AdS/CFT},
Nucl. Phys. B589 (2000) 3, hep-th/0003218\semi
B. Eden, P.S. Howe, E. Sokatchev and P.C. West, {\it Extremal and
next-to-extremal $n$-point correlators in four dimensional SCFT},
Phys. Lett. B494 (2000) 141, hep-th/0004102.}

\lref\Free{D.Z. Freedman, S.D. Mathur, A. Matsusis and L. Rastelli, 
Nucl. Phys. B546 (1999) 96, hep-th/9804058\semi
D.Z. Freedman, S.D. Mathur, A. Matsusis and L. Rastelli, 
Phys. Lett. B452 (1999) 61, hep-th/9808006\semi
E. D'Hoker and D.Z. Freedman, Nucl. Phys. B550 (1999) 261, hep-th/9811257\semi
E. D'Hoker and D.Z. Freedman, Nucl. Phys. B544 (1999) 612, hep-th/9809179.}
\lref\FreeI{E. D'Hoker, D.Z. Freedman and L. Rastelli, 
Nucl. Phys. B562 (1999) 395, hep-th/9905049.}
\lref\FreeD{E. D'Hoker, D.Z. Freedman, S.D. Mathur, A. Matsusis and 
L. Rastelli, Nucl. Phys. B562 (1999) 353, hep-th/9903196.}

\lref\Bia{M. Bianchi, S. Kovacs, G. Rossi and Y.S. Stanev, {\it On the
logarithmic behaviour in $\N=4$ SYM theory}, JHEP 9908 (1999) 020, 
hep-th/9906188.} 
\lref\Pen{S. Penati and A. Santambrogio, {\it Superspace approach to anomalous
dimensions in ${\N}=4$ SYM}, Nucl. Phys. B614 (2001) 367, hep-th/0107071.}

\lref\Arut{G. Arutyunov and S. Frolov, {\it Four-point Functions of Lowest
Weight CPOs in $\N=4$ SYM${}_4$ in Supergravity Approximation},
Phys. Rev. D62 (2000) 064016, hep-th/0002170.}
\lref\Three{G. Arutyunov and S. Frolov, {\it Three-point function of the 
stress-tensor in the AdS/CFT correspondence}, Phys. Rev. D60 (1999) 026004,
hep-th/9901121.}
\lref\OPEN{G. Arutyunov, S. Frolov and A.C. Petkou, {\it Operator Product
Expansion of the Lowest Weight CPOs in $\N=4$ SYM${}_4$ at Strong Coupling},
Nucl. Phys. B586 (2000) 547, hep-th/0005182; (E) Nucl. Phys. B609 (2001) 539.}
\lref\OPEW{G. Arutyunov, S. Frolov and A.C. Petkou, {\it Perturbative and
instanton corrections to the OPE of CPOs in $\N=4$ SYM${}_4$},
Nucl. Phys. B602 (2001) 238, hep-th/0010137; (E) Nucl. Phys. B609 (2001) 540.}
\lref\one{F.A. Dolan and H. Osborn, {\it Implications of $\N=1$ Superconformal
Symmetry for Chiral Fields}, Nucl. Phys. B593 (2001) 599, hep-th/0006098.}
\lref\Dos{F.A. Dolan and H. Osborn, {\it Conformal four point functions
and the operator product expansion}, Nucl. Phys. B599 (2001) 459, 
hep-th/0011040.}
\lref\Hok{E. D'Hoker, S.D. Mathur, A. Matsusis and L. Rastelli, {\it The 
Operator Product Expansion of $N=4$ SYM and the $4$-point Functions of 
Supergravity}, Nucl. Phys. B589 (2000) 38, hep-th/9911222.}
\lref\Hoff{L.C. Hoffmann, A.C. Petkou and W. R\"uhl, 
Phys. Lett. B478 (2000) 320, hep-th/0002025\semi
L.C. Hoffmann, A.C. Petkou and W. R\"uhl, hep-th/0002154.}
\lref\Witt{E. Witten, Adv. Theor. Math. Phys. 2 (1998) 253, hep-th/9802150.}
\lref\HoffR{L. Hoffmann, L. Mesref and W. R\"uhl, Nucl. Phys. B589 (2000) 337.}
\lref\Howe{P.S. Howe, E. Sokatchev and P.C. West, {\it 3-Point Functions in 
$N=4$ Yang-Mills}, Phys. Lett. B444 (1998) 341, hep-th/9808162.}
\lref\Eden{B. Eden, P.S. Howe, A. Pickering, E. Sokatchev and P.C. West,
{\it Four-point functions in $N=2$ superconformal field theories},
Nucl. Phys. B581 (2000) 523, hep-th/0001138.}
\lref\Edent{B. Eden, A.C. Petkou, C. Schubert and E. Sokatchev, {\it Partial
non-renormalisation of the stress-tensor four-point function in $N=4$
SYM and AdS/CFT}, Nucl. Phys. B607 (2001) 191, hep-th/0009106.}

\lref\HMR{L. Hoffmann, L. Mesref and W. R\"uhl, {\it Conformal partial wave
analysis of AdS amplitudes for dilaton-axion four-point functions},
Nucl. Phys. B608 (2001) 177,
hep-th/0012153.}

\lref\Intr{K. Intriligator, {\it Bonus Symmetries of ${\cal N} =4$ 
Super-Yang-Mills Correlation Functions via AdS Duality}, 
Nucl. Phys. B551 (1999) 575, hep-th/9811047.}
\lref\Bon{K. Intriligator and W. Skiba, {\it Bonus Symmetry and the Operator
Product Expansion of ${\cal N} =4$ Super-Yang-Mills},
Nucl. Phys. B559 (1999) 165, hep-th/9905020.}
\lref\Non{G. Arutyunov, B. Eden and E. Sokatchev, {\it On non-renormalization
and OPE in Superconformal Field Theories}, Nucl. Phys. B619 (2001) 359, 
hep-th/0105254.}
\lref\bpsN{B. Eden and E. Sokatchev, {\it On the OPE of $1/2$ BPS Short
Operators in $N=4$ SCFT${}_4$}, Nucl. Phys. B618 (2001) 259, hep-th/0106249.}
\lref\Hes{P.J. Heslop and P.S. Howe, {\it OPEs and 3-point correlators of
protected operators in $N=4$ SYM}, hep-th/0107212.}
\lref\Except{G. Arutyunov, B. Eden, A.C. Petkou and E. Sokatchev, 
{\it Exceptional non-renorm-alization properties and OPE analysis of
chiral four-point functions in $\N=4$ SYM${}_4$}, hep-th/0103230.}

\lref\BKRS{M. Bianchi, S. Kovacs, G. Rossi and Y.S. Stanev, {\it Properties
of the Konishi multiplet in  $\N=4$ SYM theory}, JHEP 0105 (2001) 042,
hep-th/0104016.}
\lref\BERS{M. Bianchi, B. Eden, G. Rossi and Y.S. Stanev, {\it On operator
mixing in $\N=4$ SYM}, hep-th/0205321.}

\lref\Hoksemi{A.V. Ryzhov, {\it Quarter BPS operators in $\N=4$ SYM}, 
JHEP 0111 (2001) 046, hep-th/0109064\semi
E. D'Hoker and A.V. Ryzhov, {\it Three point functions of quarter BPS 
operators in $\N=4$ SYM}, JHEP 0202 (2002) 047, hep-th/0109065.}
\lref\HH{P.J. Heslop and P.S. Howe, {\it A note on composite operators in
$N=4$ SYM}, Phys. Lett. 516B (2001) 367, hep-th/0106238.}

\lref\NonP{G. Arutyunov, S. Penati, A.C. Petkou, A. Santambrogio  and 
E. Sokatchev, {\it Non protected operators in $\N=4$ SYM and multiparticle
states of ADS${}_5$ SUGRA}, hep-th/0206020.}

\lref\Soh{M.F. Sohnius, {\it The multiplet of currents for $N=2$ extended
supersymmetry}, Phys. Lett. 81B (1979) 8.}
\lref\Class{M. G\"unaydin and N. Marcus, {\it The spectrum of the $S^5$
compactification of the chiral $N=2$, $D=10$ supergravity and the unitary
supermultiplets of $U(2,2/4)$}, Class. and Quantum Gravity, 2 (1985) L11.}
\lref\Gun{M. G\"unaydin, D. Minic and M. Zagermann, 
{\it $4$D doubleton conformal theories, CPT and IIB strings on
AdS${}_5 \times$ S${}_5$},  Nucl. Phys. B534 (1998) 96, (E) B538 (1999), 531,
hep-th/9806042\semi
M. G\"unaydin, D. Minic and M. Zagermann,
{\it Novel supermultiplets of $SU(2,2|4)$ and the AdS${}_4$/CFT${}_4$ duality}, 
Nucl. Phys. B544 (1999) 737, hep-th/9810226.} 

\lref\harm{S. Ferrara and E. Sokatchev, {\it Short representations of
$SU(2,2|N)$ and harmonic superspace analyticity},  
Lett. Math. Phys. 52 (2000) 247, hep-th/9912168\semi
L. Andrianopoli, S. Ferrara, E. Sokatchev and B. Zupnik, {\it Shortening
of primary operators in $N$-extended SCFT${}_4$ and harmonic superspace
analyticity}, Adv. Theor. Math. Phys. 3 (1999) 1149, hep-th/9912007.}
\lref\ESok{S. Ferrara and E. Sokatchev, {\it Superconformal interpretation of 
BPS states in AdS geometries}, Int. J. Theor. Phys. 40 (2001) 935, 
hep-th/0005151.}
\lref\Heslop{P.J. Heslop and P.S. Howe, {\it On harmonic superspaces and
superconformal fields in four-dimensions}, Class. Quant. Grav. 17 (2000) 
3743, hep-th/0005135.}

\lref\OPENP{F.A. Dolan and H. Osborn, {\it Superconformal symmetry, 
correlation functions and the operator product expansion},
Nucl. Phys. B629 (2002) 3, hep-th/0112251.}

\lref\Nirschl{M. Nirschl and H. Osborn, in preparation, H. Osborn, talk
at the 35th International Symposium Ahrenshoop, August 2002.}

\lref\Das{K. Dasgupta, M.M. Sheikh-Jabbari and M. Van Raamsdonk,
{\it Protected multiplets of $M$-theory on a plane wave}, hep-th/0207050.}
\lref\Kim{N. Kim aand J Plefka, {\it On the spectrum of PP-wave matrix
theory}, hep-th/0207034\semi
N. Kim and J-H. Park, {\it Superalgebras for M-theory on a
pp-wave},  hep-th/0207061.}
\lref\Dob{V.K. Dobrev and V.B. Petkova, {\it All positive energy unitary
irreducible representations of extended conformal symmetry}, Phys. Lett.
162B (1985) 127\semi
V.K. Dobrev and V.B. Petkova, {\it Group-theoretical approach to extended
conformal supersymmetry: function space realizations and invariant differential
operators}, Fortschr. der Phys. 35 (1987) 537.}
\lref\Min{S. Minwalla, {\it Restrictions imposed by superconformal invariance
on quantum field theories}, Adv. Theor. Math. Phys. 2 (1998) 781, hep-th/9712074.}
\lref\AF{L. Andrianopoli and S. Ferrara, {\it On short and long $SU(2,2/4)$
multiplets in the AdS/CFT correspondence}, Lett. Math. Phys. 48 (1999) 145, 
hep-th/9812067.}
\lref\Short{S. Ferrara and A. Zaffaroni, {\it Superconformal Field Theories,
Multiplet Shortening, and the AdS${}_5$/SCFT${}_4$ Correspondence},
Proceedings of the Conf\'erence Mosh\'e Flato 1999, vol. 1, ed. G. Dito and 
D. Sternheimer, Kluwer Academic Publishers (2000), hep-th/9908163.}

\lref\RSpeis{J. Fuchs and C Schweigert, {\it Symmetries, Lie Algebras
and Representations}, Cambridge University Press, Cambridge, 1997.}
\lref\Swart{J.J. de Swart, {\it The Octet Model and its Clebsch-Gordan
Coefficients}, Rev. Mod. Phys. 35 (1964) 916.}
\lref\Hump{J.E. Humphreys, {\it Introduction to Lie Algebras and Representation
Theory}, Springer-Verlag, New York, third revised printing, 1980.}

\lref\Ans{D. Anselmi, {\it Theory of higher spin tensor currents and central 
charges}, Nucl. Phys. B541 (1999) 323, hep-th/9808004.}
\lref\Hig{S.E. Konstein, M.A. Vasiliev and V.N. Zaikin, {\it Conformal higher 
spin currents in any dimension and AdS/CFT correspondence}, JHEP 0012 (2000) 018,
hep-th/0010239.}
{\nopagenumbers
\rightline{DAMTP/02-114}
\rightline{hep-th/0209056}
\vskip 1.5truecm
\centerline {\bigbf On Short and Semi-Short Representations for}
\vskip  6pt
\centerline {\bigbf Four Dimensional Superconformal Symmetry}
\vskip 1.5 true cm
\centerline {F.A. Dolan and H. Osborn${}^\dagger$}

\vskip 12pt
\centerline {\ Department of Applied Mathematics and Theoretical Physics,}
\centerline {Silver Street, Cambridge, CB3 9EW, England}
\vskip 1.5 true cm

{\eightpoint
\parindent 1.5cm{

{\narrower\smallskip\parindent 0pt

Possible short and semi-short representations for $\N=2$ and $\N=4$ superconformal
symmetry in four dimensions are discussed. For $\N=4$ the well known short
supermultiplets whose lowest dimension conformal primary operators correspond to 
$\half$-BPS or ${1\over 4}$-BPS states and are scalar fields belonging to the
$SU(4)_r$ symmetry representations $[0,p,0]$ and $[q,p,q]$ and having 
scale dimension $\Delta =p$ and $\Delta = 2q+p$ respectively are recovered.
The representation content of semi-short multiplets, which arise at the
unitarity threshold for long multiplets, is discussed. It is shown how, at the
unitarity threshold, a long multiplet can be decomposed into four semi-short
multiplets. If the conformal primary state is spinless one of these becomes
a short multiplet. For $\N=4$ a ${1\over 4}$-BPS multiplet need not have a 
protected dimension unless the primary state belongs to a $[1,p,1]$ representation.



\narrower}}

\vfill
\line{${}^\dagger$ 
address for correspondence: Trinity College, Cambridge, CB2 1TQ, England\hfill}
\line{\hskip0.2cm emails:
{{\tt fad20@damtp.cam.ac.uk} and \tt ho@damtp.cam.ac.uk}\hfill}
}

\eject}
\pageno=1

\newsec{Introduction}

Since the discovery of the AdS/CFT correspondence there has been a resurgence
of interest in superconformal quantum field theories in four dimensions. 
The maximal $\N=4$ theories for gauge group $SU(N)$ with coupling $g$ have 
been of particular interest and, as is now well known, the supergravity 
approximation to type IIB string theories on $AdS_5 \times S^5$
gives direct information in the limit $N\to \infty$ for large $\lambda$ where
$\lambda = g^2 N$. The $\N=4$ theory is superconformal for any $g$ and it is
possible to also obtain perturbative results valid for small $\lambda$. Of
particular interest are correlation functions of chiral primary operators
which may be written as 
$\C^I{}_{\! r_1 \dots r_p} \tr ( X_{r_1} \dots X_{r_p})$
where $\C^I{}_{\! r_1 \dots r_p}$ are a set of symmetric traceless $SO(6)$ 
tensors and $X_r$ are the lowest dimension scalar fields in the $\N=4$ 
theory, belonging to the adjoint representation of the gauge group and 
the 6-dimensional representation of the $R$-symmetry group 
$SO(6)_R \simeq SU(4)_R$.

Recently there has been extensive analysis of such four point correlation
functions, primarily for the simplest $p=2$ case, in relation to the operator
product expansion \refs{\Hok,\OPEN,\OPEW,\HMR,\Except,\bpsN,\OPENP}. 
The operators appearing belong to various supermultiplets.
For a long supermultiplet, for which the number of states is proportional to
$2^{16}$, the scale dimension $\Delta$ for the lowest dimension state
is unconstrained so long as it satisfies an inequality $\Delta\ge \Delta_0$,
where $\Delta_0$ depends on the spin and $SU(4)_R$ symmetry representation,
necessary for unitarity \refs{\Dob,\Min}, for a useful review see \Short.
In such supermultiplets the scale dimension ranges from $\Delta$ to $\Delta+8$.  
The cardinal example \BKRS\ is the supermultiplet for
which the lowest dimension state is formed by the Konishi scalar, which is
a $SU(4)_R$ singlet and for which $\Delta_0 =2$. In the interacting theory
there is an anomalous dimension so that $\Delta= \Delta_0 + {\rm O}(\lambda)$.
In addition there are various short supermultiplets in which the number of
states involve factors $2^p$, $p<16$.  These result from the lowest
dimension state satisfying BPS like conditions where various supercharges
annihilate the lowest dimension state. The simplest are the supermultiplets
formed from the chiral primary operators described above which are scalars
belonging to the $[0,p,0]$ $SU(4)_R$ representation and with 
$\Delta_{\rm min} =p$. 
These are $\half$-BPS representations and the supermultiplet then has 
operators with a maximum dimension $\Delta_{\rm min}+4$, for $p\ge 4$.
As well as this example the known short supermultiplets also
include $\quar$-BPS representations with the lowest dimension states 
belonging to the $[q,p,q]$ $SU(4)_R$ representation
$\Delta_{\rm min} = 2q+p$ and $\Delta_{\rm max}= \Delta_{\rm min}+6$. For
these examples the BPS shortening conditions apply to both the $Q$ and 
$\bQ$ supercharges
which requires the lowest dimension state to be spinless. The scale
dimensions of such short multiplets are expected to be protected against 
perturbative corrections.

Besides the contributions of these operators and their higher dimension 
descendents other operators with apparently protected scale dimensions have
been identified in the operator product expansion analysis. The first to
be exhibited \refs{\OPEN,\OPEW} was a scalar operator with $\Delta=4$ in the 
20-dimensional $[0,2,0]$ representation. This exactly satisfies the 
unitarity bound so there is no apparent reason to prohibit an anomalous 
dimension. This operator has been identified with the double trace
operator formed from the product of two $[0,2,0]$ chiral primary operators
in the large $N$ limit and its vanishing anomalous dimension confirmed by
field theory calculations \refs{\BKRS,\Pen,\BERS}. 
Multiplets with protected dimensions, and with the possibility of non zero 
spins, were also demonstrated in the general tensor product decomposition of 
two chiral primary $\half$-BPS operators by an analysis of the corresponding 
three point functions for $\N=2$ \Non\ and $\N=4$ \bpsN. 
Recently \OPENP\ we also showed the necessary existence in the operator 
expansion for two $[0,2,0]$ chiral primary operators of families of such 
protected supermultiplets  whose lowest dimension operators satisfied 
$\Delta_0 = 4 + \ell$ where the spin representation was 
$(\half \ell, \half \ell)$ and which belonged to the $[0,2,0], \ [1,0,1]$ 
representations for $\ell$ even, odd. The conformal partial wave analysis 
in  \OPENP\ demonstrates that the associated supermultiplets do not have 
the full range of scale dimensions, and also spins, expected for a long
supermultiplet.

In this paper we reanalyse the possible shortening conditions consistent with
superconformal symmetry directly from the superconformal algebra and
attempt to provide a unified treatment of different possible cases. We
reproduce well known results for finding short supermultiplets 
when the lowest dimension operator satisfies BPS like conditions. If the BPS
conditions are applied for both the $Q$ and $\bQ$ supercharges then the
lowest dimension operator must be spinless, $j=\bj=0$.
There are also various so called semi-short supermultiplets, depending on
the $SU(4)_R$ representation of the lowest dimension state, which may
occur at the threshold of the unitarity bound, $\Delta=\Delta_0$. 
In these examples the semi-shortening conditions may be applied for both
the $Q$ and $\bQ$ supercharges for arbitrary spin representations
$(j,\bj)$ and then the maximum scale dimension in the supermultiplets
is $\Delta_0 + n$, $n=4,5,6,7$ with the total number of states 
involving a factor $2^{2n}$. We also consider multiplets in which the
semi-shortening condition is applied to the $Q$ supercharges while for $\bQ$
there is the full BPS shortening condition, and vice versa. The lowest
dimension state then has spin $(j,0)$ or $(0,\bj)$. 

The structure of superconformal multiplets and possible shortening conditions
has been extensively investigated previously by a variety of methods, in 
particular by using harmonic oscillator methods \refs{\Class,\Gun} and 
harmonic superspace \refs{\harm,\ESok,\Heslop}. Here we construct the 
supermultiplets directly from the superconformal algebra and also determine 
the spins and $R$-symmetry representations for the various fields present in
supermultiplet  which may contribute in a conformal operator product expansion.
We attempt to  give a unified description of possible short and semi-short
multiplets including  special cases where there are conservation equations or
equations of motion.  The particular cases depend on the fraction of the total
number of supercharges  for which the shortening or semi-shortening conditions
are applied. We also  show how a long multiplet at its unitarity threshold may be
decomposed in general into four semi-short multiplets although for a spinless 
lowest dimension state one is a $\quar$-BPS multiplet.

In  $\N=4$ supersymmetric gauge theories, except for $\half$-BPS multiplets
and one particular case for  $\quar$-BPS multiplets, then even if fields are
part of short or semi-short multiplets in the free theory there is no guarantee
that there could not be a non zero anomalous dimension in the interacting theory
for non zero $g$ since they may all be components of a long  supermultiplet. It
is a non trivial exercise, not attempted here, to see how the spectrum of
free field operators may be combined into different supermultiplets which may
then be used to construct possible long multiplets. For the case of the
Konishi scalar then the associated $\N=1$ superfield satisfies a differential
constraint in the free theory, corresponding to a semi-shortening condition
on the supermultiplet, which is related to the presence of a conserved current.
However this superspace equation has an anomaly in the interacting theory,
which implies the usual axial anomaly for the conserved current, and the 
addition of the anomaly ensures that there are then sufficient degrees of
freedom to form a long multiplet which may gain an anomalous dimension.

In our discussion an essential role is played by the Racah Speiser
algorithm \RSpeis\ for determining the different representations that may
be formed when the supercharges act on a superconformal primary state. 
In general the corresponding Dynkin labels are obtained by adding the weights
for the various supercharges to the Dynkin label for the superconformal
primary state, although if one or more of the resulting Dynkin indices obtained 
in this way is negative then either the representation is replaced,
up to a sign, by one with a Dynkin label with all indices positive or
zero as usual or, as when one of the indices is $-1$, the associated representation 
is set to zero and is absent. This prescription allows a straightforward
unification of the results for the different semi-shortening conditions. It also allows
an easy understanding of when the fields satisfy conservation equations or
equations of motion. A simple introduction to the  Racah Speiser algorithm is
given in an appendix. In special cases with this prescription a semi-short
multiplet reduces to a short multiplet.

In detail in this paper in the next section we review the conformal algebra
and describe briefly how unitary positive energy representations are 
constructed from conformal primary states.
In particular we relate the states formed by the action of field operators on
the vacuum to the finite norm positive energy states, as defined by an
operator $H$, in the usual mathematical treatment. The eigenvalues $\Delta$ of 
$H$ correspond to the scale dimensions of the fields as given by the dilation 
operator $D$ (although $D$ is hermitian it has eigenvalues $i\Delta$).
In section 3 we review the superconformal algebra in four dimensions and
relate the generators of the $U(2)_R$ and $SU(4)_R$ $R$-symmetry groups for
$\N=2$ and $\N=4$ to those in the standard basis associated with simple roots.
In section 4 the $\N=2$ supermultiplets and the corresponding BPS and
semi-shortening conditions analysed. It is shown how constraints such
as conservation equations arise from the supersymmetry algebra. The $SU(2)_R$
and spin representations which arise in various supermultiplets are described
pictorially. The case when a long multiplet can be decomposed as a semi-direct
sum into short multiplets  is described. The same analysis is repeated in 
section 5 for the $\N=4$ case although there is a greater range of possibilities 
due to there being BPS conditions for a fraction $s=\quar,\half,{3\over 4},1$
of the supercharges. There is a similar set of cases for the
semi-shortening conditions and these can be applied independently to the action
of the $Q$ and $\bQ$ supercharges. A general long multiplet at the unitarity 
threshold is decomposed into semi-short multiplets, if the initial state is 
spinless this includes a $\quar$-BPS multiplet. In section
6 we discuss some special cases which arise for small representations when
the Racah Speiser algorithm gives rise to negative contributions in the
multiplet. The presence of such terms is identified with the requirement for
imposing equations of motion or conservation equations on particular fields.
We relate the results obtained here to specific supermultiplets which have
been found by different methods previously. Finally in section 7 we discuss
the implications of our results for the spectrum of operators in $\N=4$
supersymmetric gauge theories, in particular when operators which belong to
a short or semi-short multiplet may be expected to have protected scale
dimensions. Some details are deferred to three appendices. In appendix A we show
how the usual two point function may be obtained algebraically using essentially
just group theory. In appendix B we describe the Racah Speiser algorithm for 
decomposing tensor products while appendix C contains some tables of representations
for short and semi-short multiplets.

\newsec{Conformal Algebra}

In $d$ dimensions the conformal algebra is well known, for 
$\eta_{ab} = {\rm diag.} (-1,1\dots,1)$ where $a,b=0,1,\dots d-1$,
\eqn\ConA{\eqalign{
[M_{ab}, P_c ] = {}& i ( \eta_{ac} P_b - \eta_{bc} P_a) \, , \qquad
[M_{ab}, K_c ] = i ( \eta_{ac} K_b - \eta_{bc} K_a) \, , \cr
[M_{ab}, M_{cd}] = {}& i ( \eta_{ac} M_{bd} - \eta_{bc} M_{ad} - 
\eta_{ad} M_{bc} + \eta_{bd} M_{ac}) \, , \cr
[D,P_a] = {}& i P_a \, , \quad [D,K_a] = - i K_a \, , \qquad
[K_a , P_b] = - 2i M_{ab} - 2i\eta_{ab} D \, , \cr}
}
where $P_a, K_a$ are the generators of translations, special conformal transformations,
$M_{ab}=-M_{ba}$ are the generators for $SO(d-1,1)$ and $D$ is the generator of
scale transformations and other commutators not shown are zero. The algebra
corresponds to that for $SO(d,2)$ since defining $M_{AB}$, 
$A,B=0,1,\dots,d+1$  by
\eqn\MAB{
M_{AB} = \pmatrix{M_{ab} & -\half(P_a-K_a) & -\half(P_a+K_a)\cr
\half(P_b-K_b) & 0 & D \cr \half(P_b+K_b) &- D & 0 \cr} \, ,
}
then \ConA\ is summarised by
\eqn\SOD{
[M_{AB}, M_{CD}] = i ( \eta_{AC} M_{BD} - \eta_{BC} M_{AD} -
\eta_{AD} M_{BC} + \eta_{BD} M_{AC}) \, ,
}
with $\eta_{AB} = {\rm diag.} (-1,1\dots,1,-1)$. For physical applications
we require unitary, so that $M_{AB}{}^\dagger = M_{AB}$, positive
energy, where the energy is determined by
\eqn\defH{
H= M_{0\, d{+1}} = - \half (P_0+K_0)\, ,
}  
representations. 

Acting on a multi-component quasi-primary field $\O_I(x)$ we have
\eqn\tO{\eqalign{
[P_a , \O_I(x) ] = {}&i \pr_a \O_I(x) \, , \qquad [ D , \O_I(x) ] = i  
(  x{\cdot \pr} + \Delta) \O_I(x) \, , \cr
[M_{ab} , \O_I(x) ] = {}&i ( x_a \pr_b - x_b \pr_a ) \O_I(x) 
+  \O_J(x) (s_{ab} )^J{}_{\! I}  \, , \cr
[K_a , \O_I(x) ] = {}&i ( x^2 \pr_a - 2 x_a \, x{\cdot \pr}  - 
2 \Delta x_a ) \O_I (x) -  2 \O_J(x) (s_{ab} )^J{}_{\! I} x^b \, , \cr}
}
where $(s_{ab})^I{}_{\! J}$ are the appropriate finite dimensional spin 
matrices, obeying the algebra of $M_{ab}$, and 
$\Delta$ is the scale dimension. For such a quasi-primary field we define
\eqn\Os{
|\O \r_I = \O_I(0) |0\r \, 
}
as a set of conformal primary states satisfying
\eqn\Ost{
K_a |\O \r_I = 0 \, , \qquad D |\O \r_I = i \Delta |\O \r_I \, , \qquad
M_{ab}  |\O \r_I =  |\O \r_J  (s_{ab})^J{}_{\! I}  \, .
}
The space of states, on which the  conformal representation is defined, 
is then spanned by vectors of the form
\eqn\Span{
\prod_n P_{a_n} |\O \r_I \, .
}
For the conjugate field given by $\bO_{\bar I}(x) = \O_I(x)^\dagger$
we may also define
\eqn\Ob{
{}_{\bar I} \l \bO | = \l 0 | \bO_{\bar I}(0) \, .
}
which satisfies
\eqn\Osb{
{}_{\bar I} \l \bO | K_a = 0 \, , \qquad {}_{\bar I} \l \bO | D
= -  i \Delta \, {}_{\bar I} \l \bO | \, , \qquad 
{}_{\bar I} \l \bO | M_{ab} = ({\bar s}_{ab})_{\bar I}{}^{\! {\bar J}}\,
{}_{\bar J} \l \bO | \, , \quad ({\bar s}_{ab})_{\bar I}{}^{\! {\bar J}}
= (s_{ab})^J{}_{\! I}{}^* \, .
}

The relation to the standard treatment of unitary positive energy
representations is obtained by a similarity transformation. With 
$\ha =1, \dots d-1$,
\eqn\uni{\eqalign{
- e^{-{\pi\over 2} M_{0d}} i D e^{{\pi\over 2} M_{0d}} = {}& H \, , \qquad
e^{-{\pi\over 2} M_{0d}} (M_{\smash{\ha \hb}}, - i M_{0\ha}) 
e^{{\pi\over 2} M_{0d}} = (M_{\smash{\ha \hb}},  M_{d\ha} ) \, , \cr
e^{-{\pi\over 2} M_{0d}} (P_{\ha}, - i P_{0}) e^{{\pi\over 2} M_{0d}} 
= {}& (\E^+{}_{\!\!\ha},  \E^+{}_{\!\!d}) \, , \qquad 
e^{-{\pi\over 2} M_{0d}} (K_{\ha}, - i K_{0}) e^{{\pi\over 2} M_{0d}}        
= (\E^-{}_{\!\!\ha},  \E^-{}_{\!\!d}) \, .  \cr}
}
for $H$ given by \defH\ and where
\eqn\defE{
\E^\pm{}_{\!\!r} = M_{d{+1}\, r} \pm i M_{0\, r} \, , \ \ r=1,\dots d \, ,
\qquad \E^-{}_{\!\!r} = \E^+{}_{\!\!r}{}^\dagger \, .
}
These have the commutators
\eqn\com{
[ \E^-{}_{\!\!r} , \E^+{}_{\!\!s} ] = 2\delta_{rs} H - 2i M_{rs} \, , \quad
[ H, \E^\pm{}_{\!\!r} ] = \pm \E^\pm{}_{\!\!s} \, , \quad
[ \E^+{}_{\!\!r} , \E^+{}_{\!\!s} ] = 0 \, ,
}
with $M_{rs}$ generators for the compact $SO(d)$ subgroup given by
\SOD\ with $\eta_{rs}=\de_{rs}$.

If we now define
\eqn\defst{
| \Delta \r_I = e^{-{\pi\over 2} M_{0d}} |\O \r_I \, , \qquad
{}_{\bar I} \l \Delta | = {}_{\bar I} \l \bO | e^{-{\pi\over 2} M_{0d}}
}
then from \Ost,\Osb\ and \uni\ it is easy to see that
\eqn\HE{\eqalign{
H | \Delta \r_I = {}& \Delta  | \Delta \r_I \, , \quad 
\E^-{}_{\!\!i} | \Delta \r_I =0 \, , \quad 
M_{rs} | \Delta \r_I =  | \Delta \r_J ({\hat s}_{rs})^J{}_{\! I}\, , \cr
{}_{\bar I}\l \Delta | H = {}& \Delta {}_{\bar I} \l \Delta | \, , \quad
{}_{\bar I}\l \Delta | \E^+{}_{\!\!i} =0 \, , \quad
{}_{\bar I}\l \Delta | M_{rs} = 
({\hat{\bar s}}_{rs})_{\bar I}{}^{\!{\bar J}}\,{}_{\bar J} \l \Delta | \, , 
\cr}
}
where ${\hat s}_{rs}$ and ${\hat{\bar s}}_{rs}$ are  matrices representing 
$M_{rs}$ given by
\eqn\defss{
{\hat s}_{\smash{\ha\hb}} = s_{\smash{\ha\hb}} \, , \quad 
{\hat s}_{d\ha} = -i s_{0\ha} \, , \qquad
{\hat{\bar s}}_{\smash{\ha\hb}} = {\bar s}_{\smash{\ha\hb}} \, , \quad
{\hat {\bar s}}_{d\ha} = i {\bar s}_{0\ha} \, .
}
Corresponding to \Span\ we have a basis for a positive energy representation 
space given by $\prod_n \E^+{}_{\!\!r_n} | \Delta \r_I  $. These states
are normalisable,
\eqn\norm{
{}_{\bar I} \l \Delta |  \Delta \r_I = N_{{\bar I}I} \, ,
}
where $N=[N_{{\bar I}I}]$ is a positive definite matrix satisfying
\eqn\Ns{
N^{-1} {\hat{\bar s}}_{rs} N = {\hat s}_{rs} \, .
}
The corresponding states defined in \Os\ and \Ob\ are not normaliseable,
unless $\O$ is the identity and $\Delta=0$. The lack of a finite
norm for $|\O \r_I$ is of course necessary for the hermitian operator $D$ to 
have imaginary eigenvalues in  \Ost. For subsequent discussion it is more
convenient to use the non-normalisable states formed by the action
of quantum fields at $x=0$ although this is entirely equivalent to
a  discussion in terms of states with finite norm. In appendix A we show how
the standard two point function for the field operators can be recovered
algebraically from \norm\ using \HE.

\newsec{Superconformal Algebra, Four Dimensions}

As is well known the conformal group in four dimensions may be
extended by including supercharges $Q^i{}_{\! \alpha}, \, \bQ_{i\dal}$
and also $S_i{}^{\!\alpha}, \, \bS{}^{i\dal}$, $i=1,\dots \N$, 
so as to realise the Lie superalgebra for $SU(2,2|\N)$. The usual 
supercharges satisfy
\eqn\algQ{
\{ Q^i{}_{\! \alpha} ,  \bQ_{j\dal} \} =  2 \de^i{}_{\! j}
\P_{\alpha\dal} \, , \qquad  \{ Q^i{}_{\! \alpha} , Q^j{}_{\! \beta} \} =
\{ \bQ_{i\dal} , \bQ_{\smash{j\dbe}} \} = 0 \, ,
}
and their superconformal extensions
\eqn\algS{
\{ \bS{}^{i\dal} , S_j{}^{\!\alpha} \} = 2 \de^i{}_{\! j} 
{\tilde \K}{}^{\dal\alpha} \, , \qquad \{ \bS{}^{i\dal} , \bS{}^{j\dbe} \} =
\{ S_i{}^{\!\alpha} , S_j{}^{\!\beta} \} = 0 \, ,
}
while the anti-commutators of the $Q$'s and the $S$'s are
\eqn\algQS{
\{ Q^i{}_{\! \alpha}  ,   \bS{}^{j\dal} \} =  0 \, , \qquad\qquad 
\{ S_i{}^{\!\alpha} , \bQ_{j\dal} \} = 0 \, , 
}
and also
\eqna\QS
$$\eqalignno{
\{ Q^i{}_{\! \alpha} ,  S_j{}^{\!\beta} \} = {}&  4 \big (
\de^i{}_{\! j} ( M_\alpha{}^{\! \beta} -\half i \, 
\de_\alpha{}^{\! \beta} D )
- \de_\alpha{}^{\! \beta} R^i{}_{\! j} \big ) \, , & \QS{a} \cr
\{ \bS{}^{i\dal} , \bQ_{\smash{j\dbe}} \}  = {}&  4 \big ( \de^i{}_{\! j} 
( {\bar M}{}^\dal{}_{\!\smash{\dbe}} + \half i \,  
\de^\dal{}_{\!\smash{\dbe}} D )
- \de^\dal{}_{\!\smash{\dbe}} R^i{}_{\! j} \big ) \, ,  & \QS{b} \cr}
$$
where
\eqn\defPKM{\eqalign{
\P_{\alpha\dal} = {}& (\si^a)_{\alpha\dal} P_a \, , \qquad\qquad \ \ \
{\tilde \K}{}^{\dal\alpha} = (\bsi^a)^{\dal\alpha} K_a \, , \cr
M_\alpha{}^{\! \beta} = {}& - \quar i ( \si^a \bsi^b)_\alpha{}^{\! \beta}
M_{ab} \, , \quad {\bar M}{}^\dal{}_{\!\smash{\dbe}} = - \quar i
( \bsi^a \si^b)^\dal{}_{\!\smash{\dbe}} M_{ab} \, . \cr}
}
In this basis the commutators for $M_{ab}$ reduce to
\eqn\Mcom{
[M_\alpha{}^{\! \beta} , M_\gamma{}^{\! \delta} ] =
\de_\gamma{}^{\! \beta} M_\alpha {}^{\! \delta} - \de _\alpha {}^{\! \delta}
M_\gamma{}^{\! \beta} \, , \qquad
[{\bar M}{}^\dal{}_{\!\smash{\dbe}} , {\bar M}{}^\dga{}_{\!\smash{\dde}}] =
- \de{}^\dal{}_{\!\smash{\dde}} {\bar M}{}^\dga{}_{\!\smash{\dbe}} +
\de{}^\dga{}_{\!\smash{\dbe}} {\bar M}{}^\dal{}_{\!\smash{\dde}} \, ,
}
and acting on the supercharges we have
\eqn\MQS{\eqalign{
[M_\alpha{}^{\! \beta} , Q^i{}_{\! \gamma} ] = {}& \de_\gamma{}^{\! \beta}
Q^i{}_{\! \alpha} - \half \de_\alpha{}^{\! \beta} Q^i{}_{\! \gamma} \, , 
\qquad \ \ \ \
[M_\alpha{}^{\! \beta} ,  S_i{}^{\!\gamma} ] = - \de_\alpha{}^{\! \gamma}
S_i{}^{\!\beta} + \half  \de_\alpha{}^{\! \beta}  S_i{}^{\!\gamma} \, , \cr
[{\bar M}{}^\dal{}_{\!\smash{\dbe}} ,  \bQ_{i\dga} ] = {}& -
\de{}^\dal{}_{\!\smash{\dga}}  \bQ_{\smash{i\dbe}}  + 
\half \de{}^\dal{}_{\!\smash{\dbe}} \bQ_{i\dga} \, , \qquad
[{\bar M}{}^\dal{}_{\!\smash{\dbe}} , \bS{}^{i\dga} ] = 
\de{}^\dga{}_{\!\smash{\dbe}}  \bS{}^{i\dal} - \half 
\de{}^\dal{}_{\!\smash{\dbe}}  \bS{}^{i\dga} \, . \cr}
}
Under the action of the generator of scale transformations we have
\eqn\DQS{
[D, Q^i{}_{\! \alpha} ] = \half i Q^i{}_{\! \alpha} \, , \quad
[D, \bQ_{i\dal}] = \half i \bQ_{i\dal} \, , \qquad
[D, S_i{}^{\!\alpha} ] = - \half i S_i{}^{\!\alpha} \, , \quad
[D, \bS{}^{i\dal} ] = - \half i \bS{}^{i\dal} \, ,
}
and also we have
\eqna\PKQ
$$\eqalignno{
[K_a , Q^i{}_{\! \alpha} ] ={}& - (\si_a)_{\alpha\dal} \bS{}^{i\dal} \, , 
\qquad [K_a , \bQ_{i\dal}] =  S_i{}^{\! \alpha} (\si_a)_{\alpha\dal}  
\, , & \PKQ{a} \cr
[P_a , \bS{}^{i\dal} ] = {}& - (\bsi_a)^{\dal\alpha}Q^i{}_{\! \alpha} \, , 
\qquad
[P_a , S_i{}^{\!\alpha} ] = \bQ_{i\dal} (\bsi_a)^{\dal\alpha} \, . 
& \PKQ{b} \cr}
$$
The remaining generators involve the $U(\N)$ $R$-symmetry for which
the Lie algebra is
\eqn\LR{
[ R^i{}_{\! j} , R^k{}_{\! l} ] = \de^k{}_{\! j}R^i{}_{\! l} -
\de^i{}_{\! l}R^k{}_{\! j} \, .
}
These act on the supercharges according to
\eqn\RQS{\eqalign{
[ R^i{}_{\! j} ,  Q^k{}_{\! \alpha} ] = {}&\de^k{}_{\! j}  Q^i{}_{\! \alpha}
- \quar  \de^i{}_{\! j}  Q^k{}_{\! \alpha} \, , \qquad \ \
[ R^i{}_{\! j} ,  \bQ_{k\dal}] = - \de^i{}_{\! k} \bQ_{j\dal} +
\quar  \de^i{}_{\! j}  \bQ_{k\dal} \, , \cr
[ R^i{}_{\! j} ,  S_k{}^{\!\alpha} ] = {}& - \de^i{}_{\! k} S_j{}^{\!\alpha} +
\quar  \de^i{}_{\! j}   S_k{}^{\!\alpha} \, , \qquad
[ R^i{}_{\! j} ,  \bS{}^{k\dal} ] = \de^k{}_{\! j}  \bS{}^{i\dal} -
\quar  \de^i{}_{\! j} \bS{}^{k\dal} \, . \cr}
}
For $\N=4$ it is evident that we may impose $R^i{}_{\! i}=0$ so that
the $R$-symmetry is then $SU(4)$ and the supergroup reduces to $PSU(2,2|4)$. 

As operators we also require the hermeticity conditions
\eqn\herm{
Q^i{}_{\! \alpha}{}^\dagger = \bQ_{i\dal} \, , \quad
S_i{}^{\!\alpha}{}^\dagger = \bS{}^{i\dal} \, , \quad
M_\alpha{}^{\! \beta}{}^\dagger = {\bar M}{}^\dbe{}_{\!\smash{\dal}} \, ,
\quad R^i{}_{\! j}{}^\dagger = R^j{}_{\! i} \, .
}
 
Applying the transformation \uni\ to the supercharges, using \PKQ{a,b}, we 
define
\eqn\uniQ{\eqalign{
\Q^{+i}{}_{\! \alpha} = {}& e^{-{\pi\over 2} M_{0d}} Q^i{}_{\! \alpha} 
e^{{\pi\over 2} M_{0d}} = {\ts {1\over \sqrt 2}} \big ( Q^i{}_{\! \alpha}
+ \si_{0\,\alpha\dal} \bS{}^{i\dal}\big )  \, , \cr
{\Q}{}^{-\alpha}{}_{\!\!\!\!\!\! i} \ = {}& e^{-{\pi\over 2} M_{0d}} S_i{}^{\!\alpha}
e^{{\pi\over 2} M_{0d}} = {\ts {1\over \sqrt 2}} \big ( S_i{}^{\!\alpha} +
\bQ_{i\dal} \, \bsi_{0}{}^{\!\dal\alpha}\big ) \, , \cr
- \S^{-i\dal} = {}& e^{-{\pi\over 2} M_{0d}} \bS{}^{i\dal}
e^{{\pi\over 2} M_{0d}} = {\ts {1\over \sqrt 2}} \big ( \bS{}^{i\dal} -
\bsi_{0}{}^{\!\dal\alpha} Q^i{}_{\! \alpha} \big ) \, , \cr
{\S}^+{}_{\!\! i\dal}  = {}& e^{-{\pi\over 2} M_{0d}} \bQ_{i\dal}
e^{{\pi\over 2} M_{0d}} = {\ts {1\over \sqrt 2}} \big ( \bQ_{i\dal} -
S_i{}^{\!\alpha} \si_{0\,\alpha\dal} \big ) \, . \cr}
}
and the algebra \QS{a,b}\ becomes
\eqn\QQSS{\eqalign{
\{\Q^{+i}{}_{\! \alpha} , {\Q}{}^{-\beta}{}_{\!\!\!\!\!\! j}\ \} = {}&
2\,\de^i{}_{\! j} \de_\alpha{}^{\! \beta} H + 4\, \de^i{}_{\! j} 
{\tilde M}_\alpha{}^{\! \beta} - 4\, \de_\alpha{}^{\! \beta}  R^i{}_{\! j} \, , \cr
\{ \S^{-i\dal} , {\S}^+{}_{\smash{\!\!\! j\dbe}}  \} ={}&
2\, \de^i{}_{\! j} \de^\dal{}_{\!\smash{\dbe}} H - 4\, \de^i{}_{\! j} 
{\tilde {\bar M}}{}^\dal{}_{\!\smash{\dbe}} +
4\, \de^\dal{}_{\!\smash{\dbe}} R^i{}_{\! j} \, , \cr}
}
for ${\tilde M}_\alpha{}^{\! \beta} =  e^{-{\pi\over 2} M_{0d}} 
{M}_\alpha{}^{\! \beta} e^{{\pi\over 2} M_{0d}}, \
{\tilde {\bar M}}{}^\dal{}_{\!\smash{\dbe}}=  e^{-{\pi\over 2} M_{0d}} 
{{\bar M}}{}^\dal{}_{\!\smash{\dbe}}  e^{{\pi\over 2} M_{0d}}$.
In addition we have
\eqn\QSQS{\eqalign{
\{\Q^{+i}{}_{\! \alpha} , {\S}^+{}_{\smash{\!\!\! j\dal}}\} = {}& 2 \,\de^i{}_{\! j}
(e_r)_{\alpha\dal} \E^+{}_{\!\!r} \, , \qquad e_r = (\si_{\hat a}, -i \si_0 )\, ,\cr
\{ \S^{-i\dal},  {\Q}{}^{-\alpha}{}_{\!\!\!\!\!\! j}\ \} = {}& 2 \,\de^i{}_{\! j}
({\bar e}_r)^{\dal\alpha} \E^-{}_{\!\!r} \, , \qquad
{\bar e}_r = ( - \bsi_{\hat a}, i \bsi_0 ) \, , \cr}
}
and $[H,\Q^\pm]=\pm \half \Q^\pm , \ [H,\S^\pm]]=\pm \half \S^\pm$. From the
hermeticity condition \herm\ 
\eqn\hermQS{
( \Q^{+i}{}_{\! \alpha} )^\dagger
= {\Q}{}^{-\beta}{}_{\!\!\!\!\!\! i}\ \, \si_{0 \beta\dal} \, , \qquad
( {\S}^+{}_{\smash{\!\! i\dal}} )^\dagger  =  \si_{\smash{0 \alpha\dbe}} 
\S^{-i\dbe} \, ,
}
and also $({\tilde M}_\alpha{}^{\! \beta})^\dagger = \bsi_0{}^{\dbe\gamma}
{\tilde M}_\gamma{}^{\delta}\si_{0\delta \dal}$.
With the conventional choice $\si_0 = \bsi_0=1$ it is clear that
the chiral subalgebra given by $\{\Q^{+i}{}_{\! \alpha}, 
{\Q}{}^{-\alpha}{}_{\!\!\!\!\!\! i}\ ,  {\tilde M}_\alpha{}^{\! \beta}, 
R^i{}_{\! j}, H\}$, or $\{\S^{-i\dal}, {\S}^+{}_{\!\! i\dal} , 
{\tilde {\bar M}}{}^\dal{}_{\!\smash{\dbe}}, R^i{}_{\! j}, H\}$,
are generators for $SU(2|4)$ and $H$ has a positive spectrum. Finding the unitary
finite dimensional representations of these subalgebras, with lowest energy states 
$|\Delta \rangle$ satisfying as well as \HE\ 
${\Q}{}^{-\alpha}{}_{\!\!\!\!\!\! i} \ |\Delta \rangle =
\S^{-i\dal}  |\Delta \rangle = 0$, is a necessary preliminary
subsequently for determining the unitary positive energy representations of the 
full superconformal group.

In this paper we are interested  in discussing just the physical cases of 
$\N=2$ and $\N=4$. For $\N=2$ we may simply write
\eqn\RtwoA{
[ R^i{}_{\! j} ] = \pmatrix{R_3& R_+\cr R_-& -R_3\cr} + \half {\hat R}
\pmatrix{1&0\cr0&1} \, , 
}
where $R_\pm, R_3$ form a standard $SU(2)$ algebra and ${\hat R}$
corresponds to the generator of $U(1)_R$. From \RQS\ it is easy
to see that $ Q^i{}_{\! \alpha} , \, \bQ_{i\dal}$ belong to $R=\half$
representations since we have
\eqn\RQtwo{\eqalign{
[R_+ , Q^1{}_{\! \alpha}] = {}& 0 \, , \qquad [R_3 , Q^1{}_{\! \alpha}]
= \half Q^1{}_{\! \alpha} \, , \qquad [R_- , Q^1{}_{\! \alpha}] =
Q^2{}_{\! \alpha} \, , \cr
[R_+ , \bQ_{2\dal}] = {}& 0 \, ,  \qquad
[R_3 , \bQ_{2\dal}] = \half \bQ_{2\dal} \, , \qquad
[R_- , \bQ_{2\dal} ] = - \bQ_{1\dal} \, . \cr}
}

For $\N=4$ we express
$R^i{}_{\! j}$ in terms of the generators of $SU(4)$ in a Chevalley basis.
For each simple root there is an associated $SU(2)$ algebra given
by $E_i{}^{\!\pm}, \, H_i$, $i=1,\dots r$ so that $r$ is the rank,  
which have the commutation relations
\eqn\comm{
[H_i,H_j]=0 \, , \quad [ E_i{}^{\! +}, E_j{}^{\! -} ] = \de_{ij} H_j \, ,
\quad [H_i ,  E_j{}^{\! \pm}] = \pm K_{ji}  E_j{}^{\! \pm} \, , \
\hbox{no sum on $j$} \, , 
}
where $K_{ji}$ are the elements of the Cartan matrix, $K_{ii}=2$. The
generators for the non simple positive roots are obtained by appropriate
commutators of the $E_i{}^{\! +}$, and for the corresponding negative roots
from the equivalent commutators of the $E_i{}^{\! -}$,
subject to the Serre relations,
\eqn\Ser{
[\underbrace{E_i{}^{\!\pm},[ \dots [E_i{}^{\pm}}_{1 -  K_{ji}}, 
E_j{}^{\!\pm}] \dots ]] = 0 \, , \qquad i\ne j\, . 
}
The remaining commutators for any pair of generators are then determined 
using \comm\ and the Jacobi identity.
For a compact group we may further impose the hermeticity conditions
\eqn\herm{
H_i{}^\dagger = H_i \, , \qquad E_i{}^{\!+\, \dagger} = E_i{}^{\!-} \, , \qquad
i = 1, \dots , r \, .
}
For any representation space a convenient basis is then given by the
eigenvectors of $H_i$
\eqn\Hvec{
H_i | \lambda_1, \lambda_2,\dots \rangle = \lambda_i  
| \lambda_1, \lambda_2,\dots \rangle \, ,
}
where the $\lambda_i$ take integer values. The representation is then
uniquely characterised by the highest weight vector satisfying
\eqn\high{
E_i{}^{\!+} | \lambda_1, \lambda_2,\dots \rangle^{\rm hw} = 0 \, , \qquad 
\lambda_i \ge 0\, ,
}
and the representation then has Dynkin labels $[\lambda_1,\lambda_2, \dots ]$. The 
remaining basis vectors $| \lambda_1, \lambda_2,\dots \rangle$ for the representation
are then obtained by the successive action of $E_i{}^{\!-}$ on the
highest weight vector. We may also note that
\eqn\low{
E_i{}^{\!-} | \lambda_1, \lambda_2,\dots \rangle^{\rm hw} = 0  
\qquad \hbox{if} \qquad  \lambda_i =0 \, .
}

For $SU(4)$, of rank 3, the Cartan matrix is
\eqn\Car{
[K_{ij}] = \pmatrix{2& -1 & 0\cr -1& 2 &-1\cr 0&-1&2} \, .
}
For this case we may satisfy \LR\ by taking
\eqn\Rfour{\hskip -12pt{
[ R^i{}_{\! j} ] = \pmatrix{\quar(3H_1{+2H_2}{+H_3})& E_1{}^{\!+} &
[E_1{}^{\!+},E_2{}^{\!+}]& [E_1{}^{\!+},[E_2{}^{\!+},E_3{}^{\!+}]]\cr
E_1{}^{\!-} & \!\! \quar({-H_1}{+2H_2}{+H_3})& E_2{}^{\!+} &
[E_2{}^{\!+},E_3{}^{\!+}] \cr
-[E_1{}^{\!-},E_2{}^{\!-}]& E_2{}^{\!-} & \!\! -\quar(H_1{+2H_2}{-H_3}) &
E_3{}^{\!+} \cr
[E_1{}^{\!-},[E_2{}^{\!-},E_3{}^{\!-}]] & -[E_2{}^{\!-},E_3{}^{\!-}]&
E_3{}^{\!-} & \!\! - \quar(H_1{+2H_2}{+3H_3}) \cr}  .}
}
All commutators follow from \comm, \Ser\ and the Jacobi identity.
The supercharges $ Q^i{}_{\! \alpha} , \, \bQ_{i\dal}$ correspond to
the $[1,0,0], \, [0,0,1]$ representations and from \RQS\ we then have
\eqn\EQ{\eqalign{
[H_1 ,  Q^1{}_{\! \alpha} ] = {}& Q^1{}_{\! \alpha} \, , \qquad
[H_i ,  Q^1{}_{\! \alpha} ] =0 \, , \ i=2,3 \, , \qquad 
[E_i{}^{\!+},  Q^1{}_{\! \alpha} ] = 0 \, ,\cr
[H_3 , \bQ_{4\dal} ] ={}& \bQ_{4\dal} \, , \qquad 
[H_i , \bQ_{4\dal} ] = 0 \, , \ i=1,2  \, , \qquad
[E_i{}^{\!+},  \bQ_{4\dal} ] = 0 \, . \cr}
}
Under commutation with $E_i{}^{\!-}$ we have
\eqn\EQm{
Q^1{}_{\! \alpha} \ \mapright{E_1{}^{\!-}} \ Q^2{}_{\! \alpha}
\ \mapright{E_2{}^{\!-}} \ Q^3{}_{\! \alpha} \ \mapright{E_3{}^{\!-}} \
Q^4{}_{\! \alpha} \, , \qquad
\bQ_{4\dal} \ \mapright{E_3{}^{\!-}} \  -\bQ_{3\dal} \ 
\mapright{E_2{}^{\!-}}
\ \bQ_{2\dal} \ \mapright{E_1{}^{\!-}} \ -\bQ_{1\dal} \, .
}
In general $SU(4)$ representations with Dynkin labels 
$[\lambda_1,\lambda_2,\lambda_3]$ have dimensions
\eqn\dimfour{
d(\lambda_1,\lambda_2,\lambda_3)={\ts{1\over 12}}(\lambda_1+\lambda_2+\lambda_3+3)
(\lambda_1+\lambda_2+2)(\lambda_2+\lambda_3+2)
(\lambda_1+1)(\lambda_2+1)(\lambda_3+1) \, .
}

\newsec{Superconformal Representations, $\N=2$}

We consider superconformal representations in which the lowest
dimension states belong to a $SU(2)_R$
representation, with $U(1)_R$ charge $r$, 
which is formed by the action of $R_-$ on a 
highest weight $SU(2)$ state $|R,r\rangle^{\rm hw}$ satisfying
\eqn\stwo{
R_+ |R,r\rangle^{\rm hw} = 0 \, , \qquad R_3 |R,r\rangle^{\rm hw}
= R |R,r\rangle^{\rm hw} \, , \qquad {\hat R} |R,r\rangle^{\rm hw}
= r |R,r\rangle^{\rm hw} \, .
}
For the consideration of spin we here also adopt a spinorial basis for
the $(j,\bj)$ representation given by
\eqn\sjj{
|R,r\rangle_{\alpha_1\dots \alpha_{2j}, \dal_1 \dots \dal_{2\bj}}
= |R,r\rangle_{(\alpha_1\dots \alpha_{2j}), (\dal_1 \dots \dal_{2\bj})} \, .
}
Under the action of the spin generators
$M_\alpha{}^{\! \beta},\, {\bar M}{}^\dbe{}_{\!\smash{\dal}}$,
defined in \defPKM, the states given by \sjj\ transform according to
\eqn\Mjj{\eqalign{
M_\alpha{}^{\! \beta}
|R,r\rangle_{\alpha_1\dots \alpha_{2j}, \dal_1 \dots \dal_{2\bj}}
= {}&j \big ( 2 \de_{(\alpha_1}{}^{\!\! \beta}
|R,r\rangle_{\alpha_2\dots \alpha_{2j})\alpha, \dal_1 \dots \dal_{2\bj}}
- \de_\alpha{}^{\! \beta}
|R,r\rangle_{\alpha_1\dots \alpha_{2j}, \dal_1 \dots \dal_{2\bj}} \big )\, ,\cr
{\bar M}{}^\dbe{}_{\!\smash{\dal}}
|R,r\rangle_{\alpha_1\dots \alpha_{2j}, \dal_1 \dots \dal_{2\bj}}
= {}& - \bj \big (
2 |R,r\rangle_{\alpha_1\dots \alpha_{2j}, \dal (\dal_1 \dots \dal_{2\bj-1}}
\de{}^\dbe{}_{\!\smash{\dal_{2\bj})}} - \de{}^\dbe{}_{\!\smash{\dal}}
|R,r\rangle_{\alpha_1\dots \alpha_{2j}, \dal_1 \dots \dal_{2\bj}}\big ) \, .
\cr}
}
The representation formed from states satisfying \stwo\ and \Mjj\ is
labelled $R_{(j,\bj)}$.

A highest weight conformal primary state is then required to satisfy from \Ost
\eqn\Otwo{
K_a |R,r\rangle^{\rm hw}_{\alpha_1\dots \alpha_{2j}, \dal_1 \dots \dal_{2\bj}}
= 0 \, \qquad 
D |R,r\rangle^{\rm hw}_{\alpha_1\dots \alpha_{2j}, \dal_1 \dots \dal_{2\bj}} 
= i \Delta  
|R,r\rangle^{\rm hw}_{\alpha_1\dots \alpha_{2j}, \dal_1 \dots \dal_{2\bj}} \, .
}
In addition from \DQS\ we must impose
\eqn\Stwo{
S_i{}^{\!\alpha}
|R,r\rangle^{\rm hw}_{\alpha_1\dots \alpha_{2j}, \dal_1 \dots \dal_{2\bj}} = 
\bS{}^{i\dal}
|R,r\rangle^{\rm hw}_{\alpha_1\dots \alpha_{2j}, \dal_1 \dots \dal_{2\bj}} = 0 \, ,
}
for this to the  lowest dimension in the supermultiplet and to be a
superconformal primary state.

A basis for the full
representation space is then provided by  acting on states satisfying 
\Otwo\ and \Stwo, as well as \stwo, with  $R_-, P_a$
and the eight supercharges  $ Q^i{}_{\! \alpha} , \, \bQ_{k\dal}$.
The set of states $\S_{l,{\bar l}}$
\eqn\Qstate{
\prod_{i,k,\alpha,\dal} \big ( Q^i{}_{\! \alpha} \big )^{n_{i\alpha}}
\big (  \bQ_{k\dal}  \big )^{{\bar n}_{k\dal}}
|R,r\rangle^{\rm hw}_{\alpha_1\dots \alpha_{2j}, \dal_1 \dots \dal_{2\bj}} \, ,
\qquad n_{i\alpha},{\bar n}_{k\dal} =0,1 \, ,
}
for $l=\sum_{i,\alpha} n_{i\alpha}$, ${\bar l}=\sum_{k,\dal} {\bar n}_{k\dal}$,
$l,{\bar l} = 0,1,2,3,4$, 
and for some appropriate ordering of the $Q$ and $\bQ$ supercharges,
form, together with those obtained by the action of $R_-$, a basis for a 
vector space $\V_{l,{\bar l}}$.  For all  states in $\V_{l,{\bar l}}$
the scale dimension is $\Delta+\half (l + {\bar l})$
while the $U(1)_R$ charge is $r + \half (l - {\bar l})$.
Acting with $K_a$, using \PKQ{a} and \QS{a,b}, \algQS\ with \Otwo, \Stwo, 
$\V_{l,{\bar l}}\to \V_{l-1, {\bar l} -1}$ and $K_a \V_{l,0} = 0 , \
K_a \V_{0,{\bar l}} = 0$.
For general $\Delta$  any state \Qstate\ in 
$\S_{l,{\bar l}}$ is a conformal primary state, annihilated by $K_a$, if
$l=0$ or ${\bar l}=0$ or becomes a  conformal primary by the addition of
a suitable combination of states in $\V_{l-1, {\bar l} -1}$
acted on by the momentum operator. 
The states belonging to $\V_{l,{\bar l}}$ may be decomposed into
representations for $SU(2)_R$ and spin. According to the Racah Speiser
algorithm the resulting representations $R'{}_{\!(j',\bj')}$ 
are obtained by adding the weights associated with each supercharge, from \RQtwo,
\eqn\weQ{
Q^1{}_{\! \alpha} \sim \half_{(\pm {1\over 2},0)} \, , \quad \
Q^2{}_{\! \alpha} \sim (-\half)_{(\pm {1\over 2},0)} \, , \quad \
\bQ_{2\dal} \sim \half_{(0,\pm {1\over 2})}  \, , \quad \
\bQ_{1\dal} \sim (-\half)_{(0,\pm {1\over 2})}  \, , 
}
to $R_{(j,\bj)}$, so long as all $R',j',\bj' \ge 0$, for all possible
$n_{i\alpha},{\bar n}_{k\dal}$ consistent with fixed $l,{\bar l}$. Thus
the $SU(2)_R$ representations are then given by every
$R'=R + \half\sum_\alpha (n_{1\alpha} - n_{2\alpha})
- \half\sum_\dal ({\bar n}_{1\dal} - {\bar n}_{2\dal})$. Altogether there 
are therefore in general ${4 \choose l}{4 \choose {\bar l}}$ representations 
$R'{}_{\!(j',\bj')}$. If one or more of the resulting $R',j',\bj'$ are negative 
then, according the Racah Speiser algorithm which is discussed more fully later, 
there is a simple recipe for cancelling such representations.
The four $Q^i{}_{\! \alpha}$ supercharges 
by themselves acting on a representation $R_{(j,\bj)}$ give 
\eqn\NQ{
R_{(j,\bj)}\  \mapright Q \ {(R+{1\over 2})_{(j\pm {1\over 2},\bj)} \atop 
(R-{1\over 2})_{(j\pm {1\over 2} ,\bj)}} \ \mapright {Q^2} \
{(R\pm 1)_{(j,\bj)}, R_{(j,\bj)} \atop
R_{(j\pm 1 ,\bj)},R_{(j,\bj)}} \ \mapright {Q^3} \
{(R+{1\over 2})_{(j\pm {1\over 2},\bj)} \atop
(R-{1\over 2})_{(j\pm {1\over 2} ,\bj)}} \ \mapright {Q^4} \ R_{(j,\bj)} \ ,
}
corresponding to $l=0,1,2,3,4, \ {\bar l}=0$, so long as $R\ge 1, \, j\ge 1$.
The action of the $\bQ$ supercharges is given by the conjugate of \NQ.

A general long supermultiplet, denoted by $\A^\Delta_{R,r(j,\bj)}$, 
is obtained by the application of all 8  supercharges  
$Q^i{}_{\! \alpha} , \, \bQ_{i\dal}$ to a highest weight state satisfying \stwo\ 
and with $U(2)_R$ and spin quantum numbers $R,r,j,\bj$. The representations
belonging to $\V_{l,0}$ are given by \NQ\ and those forming $\V_{l,{\bar l}}, \, 
{\bar l}>0$ may then be obtained by applying the conjugate of \NQ\ for all
representations in $\V_{l,0}$. Since there are no constraints
it is straightforward to see that
\eqn\dimA{
\dim \A^\Delta_{R,r(j,\bj)} = 256(2R+1)(2j+1)(2\bj+1) \, .
}
For a unitary representation we must have in this case \Dob
\eqn\ineq{
\Delta \ge 2  + 2j + 2R + r \, , \ 2  + 2\bj + 2R - r  \, .
}

For a truncated representation we may impose the BPS-condition  if $j=0$,
\eqn\ttwo{
Q^1{}_{\! \alpha} |R,r\rangle^{\rm hw}_{\dal_1 \dots \dal_{2\bj}}  = 0 \, .
}
In this case we must impose consistency with the result \QS{a} for
the anti-commutator $\{ Q^1{}_{\! \alpha} ,  S_j{}^{\!\beta} \}$ 
and with \stwo\ and \Otwo, since the state is annihilated by $M_\alpha{}^{\! \beta}$,
it is easy to see that this requires
\eqn\Rtwo{
\Delta = 2R + r \, .
}
For the $\bQ$ charges the corresponding condition, if $\bj =0$, is 
\eqn\ttwob{
\bQ_{2\dal} |R,r\rangle^{\rm hw}_{\alpha_1\dots \alpha_{2j}} = 0 \, ,
}
which gives instead from  \QS{b}
\eqn\Rtwob{
\Delta = 2R - r \, .
}
With \ttwo\ or \ttwob\  the multiplet is generated using only
$Q^2{}_{\! \alpha}$ or $\bQ_{1\dal}$ so that $l=0,1,2$ or ${\bar l}=0,1,2$.
The following representations are then obtained for the action of the $Q$
or $\bQ$ supercharges,
\eqn\NQS{
R_{(0,\bj)}\  \mapright {Q} \ 
(R-{\half})_{({1\over 2} ,\bj)} \ \mapright {Q^2} \
(R-1)_{(0,\bj)} \, , \qquad
R_{(j,0)}\  \mapright {\bQ} \ 
(R-\half)_{(j,{1\over 2})} \ \mapright {\bQ^2} \
(R-1)_{(j,0)} \, ,
}
with dimensions $8R(2\bj+1)$,  $8R(2j+1)$ in each case.

If \ttwo\ holds for both $Q^1{}_{\! \alpha}$ and $Q^2{}_{\! \alpha}$ then
it is necessary that $R=0$ and $\Delta = r$. Conversely for $R=0$ \ttwo,
using \RQtwo, implies the corresponding result for $Q^2{}_{\! \alpha}$.

Reduced supermultiplets may be obtained by applying shortening conditions
to the action of the $Q$ or $\bQ$ supercharges or both.
Imposing just \ttwo, and hence requiring \Rtwo, we may construct an
asymmetric supermultiplet, using the conjugate of \NQ, which is
denoted by ${\B}_{R,r(0,\bj)}$. This can then be represented by the diagram,
corresponding to  $l=0,1,2, \ {\bar l}=0,1,2,3,4$,
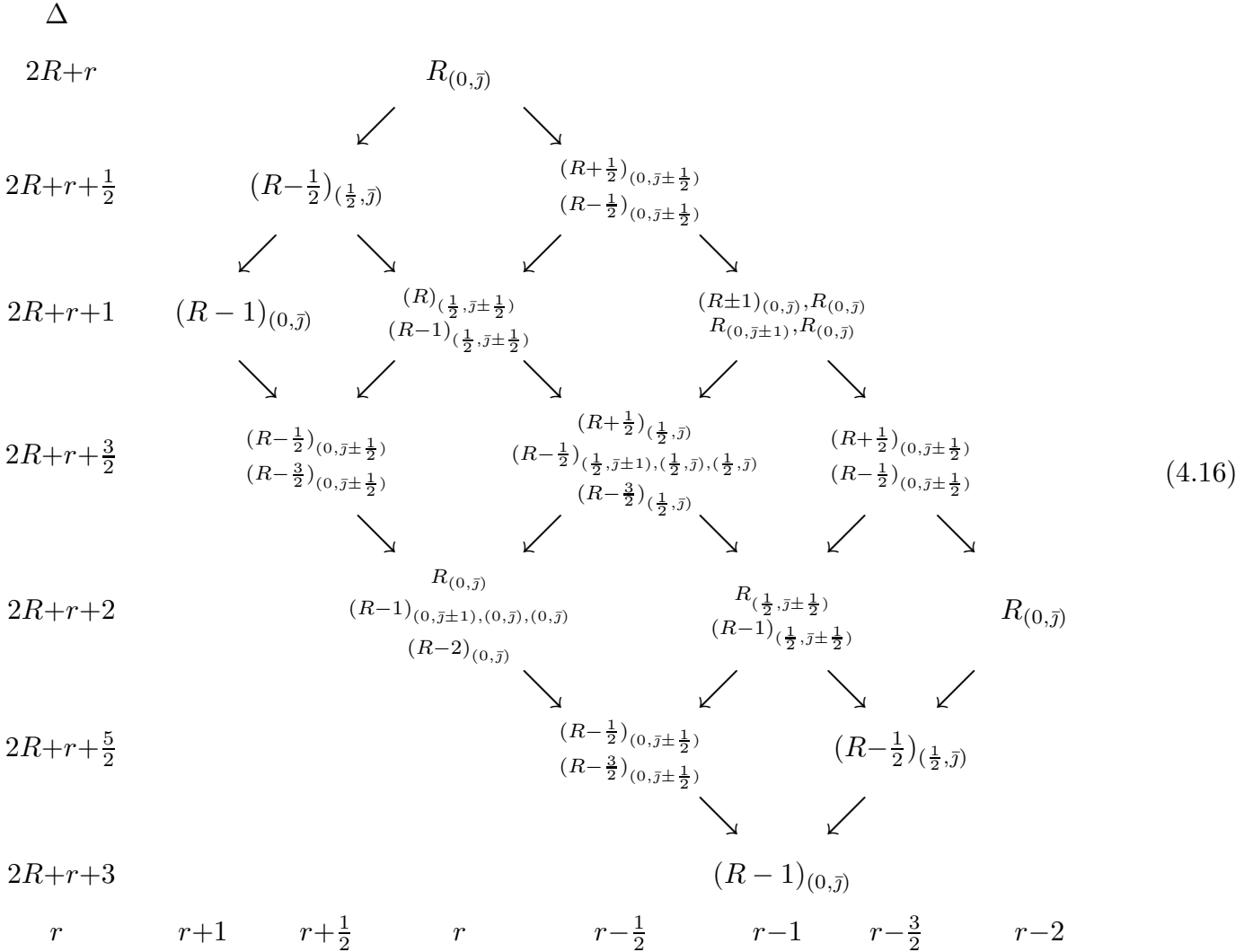
\eqn\RtwoB{\def\normalbaselines{\baselineskip20pt\lineskip3pt
\lineskiplimit3pt}\hskip-1.5cm
\matrix{\Delta~\cr
2R{+r}&&{~~~~}&&{~~~}&&R_{(0,\bj)}&&{~~~}&~~~&{~~~~}&\cr
&&&&&\Bsw&{~~~~~~~~}&\Bse&~~~~~~~~~\cr
2R{+r}{+{\ts{1\over 2}}}&&&&
\hidewidth(R{-{1\over 2}})_{({1\over 2},\bj)}\hidewidth&&
&&\hidewidth{(R{+{1\over
2}})_{(0,\bj{\pm{1\over 2}})}\atop (R{-{1\over 2}})_{(0,\bj{\pm{1\over 2}})}}\hidewidth\cr
&&&\Bsw&&\Bse&&\Bsw&&\Bse&\cr
2R{+r}{+1}&&(R-1)_{(0,\bj)}\hidewidth&&&&\hidewidth{
(R)_{({1\over 2},\bj{\pm{1\over 2}})}\atop
(R{-1})_{({1\over 2},\bj{\pm{1\over 2}})}}\hidewidth &&&&
\hidewidth{(R{\pm 1})_{(0,\bj)},R_{(0,\bj)}\atop 
R_{(0,\bj{\pm 1})},R_{(0,\bj)}}\hidewidth&&\cr
&&&\Bse&&\Bsw&&\Bse&&\Bsw&~~~~~~~\Bse\hidewidth\cr 
2R{+r}{+{\ts{3\over 2}}}&&&&\hidewidth
{(R{-{1\over 2}})_{(0,\bj{\pm{1\over 2}})}\atop
(R{-{3\over 2}})_{(0,\bj{\pm{1\over 2}})}}\hidewidth&&
&&\hidewidth~\matrix{
{\scriptstyle(R{+{1\over 2}})}_{\sss({1\over 2},\bj)}\cr
\noalign{\vskip-9pt}
{\scriptstyle(R{-{1\over 2}})}_{\sss({1\over 2},\bj{\pm1}),({1\over 2},\bj),({1\over
2},\bj)}\cr\noalign{\vskip-6pt} {\scriptstyle(R{-{3\over 2}})}_{\sss({1\over 2},\bj)}\cr}
\hidewidth&&&&\hidewidth{(R{+{1\over 2}})_{(0,\bj{\pm{1\over 2}})}\atop
(R{-{1\over 2}})_{(0,\bj{\pm{1\over 2}})}}\hidewidth\cr
&&&&&\Bse&&\Bsw&&\Bse&&\Bsw\hidewidth&&\Bse\cr
2R{+r}{+2}&&&&&&
\hidewidth\matrix{{\scriptstyle R}_{\sss(0,\bj)}\cr
\noalign{\vskip-9pt}{\scriptstyle(R-1)}_{\sss(0,\bj{\pm1}),
(0,\bj),(0,\bj)}\cr
\noalign{\vskip-6pt}{\scriptstyle(R-2)}_{\sss(0,\bj)}}\hidewidth&&&&
\hidewidth{R_{({1\over 2},\bj{\pm{1\over 2}})}\atop
(R{-1})_{({1\over 2},\bj{\pm{1\over 2}})}}\hidewidth&&&&R_{(0,\bj)}\cr
&&&&&&&\Bse&&\Bsw&&\Bse&&\Bsw\cr
2R{+r}{+{\ts{5\over 2}}}&&&&&&&&\hidewidth
{(R{-{1\over 2}})_{(0,\bj{\pm{1\over 2}})}\atop 
(R{-{3\over 2}})_{(0,\bj{\pm{1\over 2}})}}\hidewidth&&&&
\hidewidth(R{-{1\over 2}})_{({1\over 2},\bj)}\hidewidth&\cr 
&&&&&&&&&\Bse&&\Bsw&\cr
2R{+r}{+3}&&{~~~}&&{~~}&&&&{~~}&
&\hidewidth(R-1)_{(0,\bj)}\hidewidth&&{~~}\cr
r~&~~~~~\hidewidth
&\hidewidth ~~r{+1}\hidewidth&&\hidewidth ~~r{+\half}\hidewidth&&
\hidewidth r\hidewidth&&\hidewidth r{-\half}~~\hidewidth&
&\hidewidth r{-1}~\hidewidth&&
\hidewidth r{-{\ts{3\over 2}}}\hidewidth~&&\hidewidth ~r{-2}\hidewidth \cr}
}
where $\swarrow$ corresponds to the action of the $Q$ supercharges 
and $\searrow$ to $\bQ$. The dimension in this case is given by
\eqn\dimB{
\dim {\B}_{R,r(0,\bj)} = 128R(2\bj+1) \, .
}
Unitarity here requires
\eqn\ins{
r \ge \bj + 1 \, .
}
We may similarly construct the conjugate supermultiplet ${\bar \B}_{R,r(j,0)}$
with, instead of \ins, $-r \ge j+1$ and the lowest scale dimension given by
\Rtwob\ where
\eqn\dimbB{
\dim {\bar \B}_{R,r(j,0)} = 128R(2j+1) \, .
}

For $R=0$, when the lowest state is annihilated by $Q^i{}_{\! \alpha}$ for
both $i=1,2$ and the multiplet is generated solely by the action of the 
$\bQ$ charges, we have a chiral multiplet ${\E}_{r(0,\bj)}$
\eqn\chir{\def\normalbaselines{\baselineskip16pt\lineskip3pt
\lineskiplimit3pt}
\matrix{\Delta&&&
{~~}&&{~~~}
&&{~~~}&&{~~~}&&{~~~}
&&{~~~}\cr
r&&&{~~}0_{(0,\bj)}\cr
&&&&\hidewidth\Bse\hidewidth\cr
r{+{\ts{1\over 2}}}&&&
&&\hidewidth{1\over 2}_{(0,\bj\pm{1\over 2})}\hidewidth\cr
&&&&&&\Bse\cr
r{+1}&&&&&&&&\hidewidth
0_{(0,\bj\pm 1)},0_{(0,\bj)},1_{(0,\bj)}\hidewidth\cr
&&&&&&&&\hidewidth\Bse\hidewidth\cr
r{+{\ts{3\over2}}}~&&
&&&&&&&\hidewidth{1\over 2}_{(0,\bj\pm{1\over2})}\hidewidth\cr
&&&&&&&&&&\hidewidth\Bse\hidewidth\cr
r{+2}&&&&&&&&&&&&\hidewidth 0_{(0,\bj)}\cr
r&&&\hidewidth r~~&\hidewidth
&r{-{\ts{1\over2}}}~\hidewidth&&\hidewidth r{-1}~\hidewidth&
&\hidewidth r{-{\ts{3\over2}}}\hidewidth&&\hidewidth ~r{-2}\hidewidth}
}
For unitarity we must have \ins\ in this case as well and
\eqn\dimE{
\dim {\E}_{r(0,\bj)} = 16 (2\bj+1) \, .
}
Of course there is a corresponding conjugate chiral multiplet 
${\bar\E}_{r(j,0)}$.

Imposing the conditions \ttwo\ and \ttwob\ simultaneously requires $r=0$ and also 
$j=\bj=0$. 
We may describe this $\N=2$ short supermultiplet $\hat \B_R$, generated
by the action of $Q^2,\, \bQ_{1}$ as in \NQS, by the diagram
\eqn\RtwohB{\def\normalbaselines{\baselineskip16pt\lineskip3pt
\lineskiplimit3pt}
\matrix{\Delta&~~~&{~~~}&&{~~~}&&{~~~}&&{~~~}&&{~~~}&&{~~~}&&\cr
2R&&{~~~}&&{~~}&&R_{(0,0)}&&{~~}&&{~~~}&\cr
&&&&&\Bsw&&\Bse&&&&\cr
2R+{\ts{1\over 2}}&&&&\hidewidth~~~~~(R-{\ts{1\over 2}})_{({1\over 2},0)}
\hidewidth&&&&\hidewidth~~(R-{\ts{1\over 2}})_{(0,{1\over 2})}~~\hidewidth&&&\cr&&&
\Bsw&&\Bse&&\Bsw&&\Bse&&\cr
2R+1&&\hidewidth~~~~~~(R-1)_{(0,0)}\hidewidth&&&&
\hidewidth(R-1)_{({1\over 2},{1\over 2})}\hidewidth&&&&
\hidewidth(R-1)_{(0,0)}\hidewidth&\cr
&&&\Bse&&\Bsw&&\Bse&&\Bsw&&\cr
2R+{\ts {3\over 2}}&&&
&\hidewidth~~(R-{\ts{3\over 2}})_{(0,{1\over 2})}\hidewidth&&&
&\hidewidth(R-{\ts{3\over 2}})_{({1\over 2},0)}\hidewidth&&&\cr
&&&&&\Bse&&\Bsw&&&&\cr
2R+2&&&&&
&\hidewidth(R-2)_{(0,0)}\hidewidth&&&&&\cr
r&&1&&{\ts{1\over 2}}&&0&&-{\ts{1\over 2}}&&-1&\cr}
}
It is easy to see that the total
dimension of the representations in \RtwohB\ is
\eqn\dimBs{
\dim \hat \B_R = 16(2R-1) \, .
}

For $j>0$ it is also possible to impose conditions which lead to multiplet
shortening. We consider the condition corresponding to the absence of the
representations with spin ${(j-{1\over2},\bj)}$ at the first level,
\eqn\Ttwoj{
\vep^{\alpha\beta}  Q^i{}_{\! \beta}
|R,r\rangle^{\rm hw}_{\alpha_1\dots \alpha_{2j-1}\alpha, 
\dal_1 \dots \dal_{2\bj}} = 0 \, ,
}      
The necessary consistency conditions then arise from 
$\vep^{\alpha\beta}\{  Q^i{}_{\! \beta} , S_l{}^{\!\gamma} \}
|R,r\rangle^{\rm hw}_{\alpha_1\dots \alpha_{2j-1}\alpha,
\dal_1 \dots \dal_{2\bj}} = 0 $ and we may easily find from \QS{a} and \Mjj
\eqn\TtwoR{
\big ( \de^i{}_{\! l} ( - j - 1 + \half \Delta ) - R^i{}_{\! l} \big )
|R,r\rangle_{\alpha_1\dots \alpha_{2j}, \dal_1 \dots \dal_{2\bj}}^{\rm hw}=0 \,, 
\qquad l=1,2 \, .
}
If we require \Ttwoj\ just for $i=1$ then this gives
\eqn\DRj{
\Delta = 2 +2j + 2R +r \, ,
}
and the action of the $Q$ supercharges on the highest weight state
gives the representations
\eqn\NQj{
R_{(j,\bj)}\  \mapright Q \ {(R + \half)_{(j+{1\over 2},\bj)}\atop
(R - \half)_{(j\pm {1\over 2},\bj)}} \ \mapright {Q^2} \
{R_{(j+1,\bj),(j,\bj)} \atop (R-1)_{(j,\bj)}} \ \mapright {Q^3} \
(R-\half)_{(j+{1\over 2} ,\bj)}  \, ,
}
with a dimension $8(R(4j+3) + j+1)(2\bj+1)$.
If \Ttwoj\ is imposed for both $i=1,2$ then \TtwoR\ requires $R=0$ and
$\Delta = 2+2j+r$. Conversely if $R=0$ the condition \Ttwoj\ for $i=1$ implies
also the $i=2$ case by application of $R_-$.
Assuming \Ttwoj\ for $i=1,2$ the $Q$ supercharges give for general $R$
\eqn\NQjia{
R_{(j,\bj)}\  \mapright Q \ (R \pm \half)_{(j+{1\over 2},\bj)}
 \ \mapright {Q^2} \
R_{(j+1,\bj)} \, . 
}
The dimension of these representations is $8(2R+1)(j+1)(2\bj+1)$.

For $j=0$  the condition \Ttwoj\ is replaced by 
\eqn\Ttwo{
(Q^i)^2 |R,r\rangle^{\rm hw}_{\dal_1\dots \dal_{2\bj}} = 0 \, .
}
The corresponding condition for $j>0$ is of course a consequence of \Ttwoj.
In this case conditions on representations appearing in the supermultiplet
arise only at the second level. The necessary  consistency conditions for
\Ttwo\ then  follow from the requirement
\eqn\Con{
\{ Q^i{}_{\! \alpha} ,  S_l{}^{\!\gamma} \}  Q^i{}_{\! \beta}
 |R,r\rangle^{\rm hw}_{\dal_1\dots \dal_{2\bj}} =  Q^i{}_{\! \alpha}
\{ Q^i{}_{\! \beta} ,  S_l{}^{\!\gamma} \} 
|R,r\rangle^{\rm hw}_{\dal_1\dots \dal_{2\bj}} \, , \qquad l=1,2 \, .
}
Using \QS{a} with \DQS, \MQS\ and \RQS\ then gives
\eqn\Cona{
\big ( \de^i{}_{\! l}(-1 +\half \Delta) - R^i{}_{\! l} \big ) 
|R,r\rangle^{\rm hw}_{\dal_1\dots \dal_{2\bj}} =0 \, , \qquad l=1,2 \, .
}
Taking \Ttwo\ to be valid for just $i=1$ we obtain
\eqn\Conb{
\Delta = 2 + 2R + r \, ,
}
and, instead of \NQj, the action of the $Q$ supercharges on this highest weight 
state gives
\eqn\NQs{
R_{(0,\bj)}\  \mapright Q \ (R\pm \half)_{({1\over 2},\bj)} \ 
\mapright {Q^2} \ {R_{(1,\bj),(0,\bj)} \atop (R-1)_{(0,\bj)}} \ \mapright {Q^3} \
(R-\half)_{({1\over 2} ,\bj)}  \ .
}
For both $i=1,2$ \Ttwo\ requires $R=0$ and $\Delta = 2 + r$. Note also
that $(Q^1)^2 |0,r\rangle = 0$ also entails 
$\vep^{\alpha\beta}Q^1{}_{\! \alpha}Q^2{}_{\! \beta}|0,r\rangle  =
(Q^2)^2 |0,r\rangle  =0$.

If the semi-shortening conditions \Ttwoj\ or \Ttwo\ are applied to
a superconformal primary state, with scale dimensions given by \DRj\ or
\Conb, then together with the action of the $\bQ$ supercharges, given by the
conjugate of \NQ, leads to a supermultiplet ${\C}_{R,r(j,\bj)}$ 
with
\eqn\dimCL{
\dim {\C}_{R,r(j,\bj)} = 128 (R(4j+3) + j+1)(2\bj+1) \, .
}

In a similar fashion we may impose the conjugate conditions to \Ttwoj\ or
\Ttwo,
\eqn\Ttwobj{\eqalign{
\vep^{\dal\dbe} \bQ_{\smash{i\dbe}}
|R,r\rangle^{\rm hw}_{\alpha_1\dots \alpha_{2j},
\dal_1 \dots \dal_{2\bj-1}\dal} = {}& 0 \, , \quad \bj>0 \, , \cr
(\bQ_{i})^2 |R,r\rangle^{\rm hw}_{\alpha_1\dots \alpha_{2j}} 
= {}& 0 \, , \quad \bj=0 \, , \cr}  }
giving for just $i=2$
\eqn\DRbj{
\Delta = 2 +2\bj + 2R -r \, , \qquad \Delta = 2 + 2R -r \, ,
}
in each case. The action of $\bQ$ is then given by the conjugate of \NQj\
or \NQs\  respectively. Requiring
\Ttwobj\ for $i=1,2$ is identical as above after taking $R=0$.
Assuming \Ttwobj\ for $i=2$ leads to a supermultiplet
${\bar \C}_{R,r(j,\bj)}$ which is the conjugate of ${\C}_{R,r(j,\bj)}$. 

If the semi-shortening conditions are applied for both the $Q$ and
$\bQ$ supercharges then a semi-short supermultiplet denoted by 
${\hat \C}_{R(j,\bj)}$ is obtained. Assuming both  \Ttwoj\ and \Ttwobj\
then for compatibility of \DRj\ and \DRbj\ $r=\bj-j$.
Just as in in other cases we may easily construct a diagram displaying
all representations that appear in the supermultiplet in this case by
using \NQj\ and its conjugate to obtain those formed by the action of
both $Q$'s and $\bQ$'s,
\eqn\RtwoE{\def\normalbaselines{\baselineskip20pt\lineskip3pt
\lineskiplimit3pt}\hskip-1.5cm
\matrix{\Delta~\cr
2R{+j}{+\bj}{+2}&&&{~~~}&&{~~~~~}&&{~~~}&&\,~R_{(j,\bj)}~\,&\cr
&&&&&&&&\Bsw&{~~~~~~~~}&\Bse&~~~~~~~~~\cr
2R{+j}{+\bj}{+{\ts{5\over 2}}}&&&&&&&\hidewidth
{(R{+{1\over 2}})_{(j{+{1\over 2}},\bj)}
\atop (R{-{1\over 2}})_{(j{\pm{1\over 2}},\bj)}}
\hidewidth&&&&\hidewidth{(R{+{1\over 2}})_{(j,\bj{+{1\over 2}})}\atop
(R{-{1\over 2}})_{(j,\bj{\pm{1\over 2}})}}\hidewidth\cr
&&&&&&\Bsw&&\Bse&&\Bsw&&\Bse&\cr
2R{+j}{+\bj}{+3}&&&&&\hidewidth {R_{(j{+1},\bj),(j,\bj)}\atop(R-1)_{(j,\bj)}}
\hidewidth&&&&
\hidewidth{(R{+1})_{(j{+{1\over 2}},\bj{+{1\over 2}})},
(R{-1})_{(j{\pm{1\over 2}},\bj{\pm{1\over 2}})}\atop
R_{(j{+{1\over 2}},\bj{\pm{1\over 2}}),(j{\pm{1\over 2}},\bj{+{1\over 2}})}}
\hidewidth&&&&
\hidewidth{R_{(j,\bj{+1}),(j,\bj)}\atop(R-1)_{(j,\bj)}}\hidewidth&&\cr
&&&&\Bsw&&\Bse&&\Bsw&&\Bse&&\Bsw&~~~~~~~\Bse\hidewidth\cr
2R{+j}{+\bj}{+{\ts{7\over 2}}}&&&\hidewidth
(R{-{1\over 2}})_{(j{+{1\over 2}},\bj)}\hidewidth&&&
&\hidewidth
\matrix{{\scriptstyle(R{+{1\over 2}})}_{\sss(j{+1},\bj{+{1\over 2}}),
(j,\bj{+{1\over 2}})}\cr\noalign{\vskip-9pt}
{\scriptstyle(R{-{1\over 2}})}_{\sss(j,\bj{+{1\over 2}}),(j{+1},\bj{\pm{1\over 2}}),
(j,\bj{\pm{1\over 2}})}\cr\noalign{\vskip-6pt}
{\scriptstyle(R{-{3\over 2}})}_{\sss(j,\bj{\pm{1\over 2}})}\cr}
\hidewidth&&&&
\hidewidth~\matrix{
{\scriptstyle(R{+{1\over 2}})}_{\sss(j{+{1\over 2}},\bj{+1}),
(j{+{1\over 2}},\bj)}\cr\noalign{\vskip-9pt}
{\scriptstyle(R{-{1\over 2}})}_{\sss(j{+{1\over 2}},\bj),
(j{\pm{1\over 2}},\bj{+1}), (j{\pm{1\over 2}},\bj)}\cr\noalign{\vskip-6pt}
{\scriptstyle(R{-{3\over 2}})}_{\sss(j{\pm{1\over 2}},\bj)}\cr}
\hidewidth&&&(R{-{1\over 2}})_{(j,\bj{+{1\over 2}})}\cr
&&&&\Bse&&\Bsw&&\Bse&&\Bsw&&\Bse&~~~~~~~\Bsw\hidewidth\cr
2R{+j}{+\bj}{+4}&&&&&\hidewidth{R_{(j{+{1\over 2}},\bj{+{1\over 2}})}\atop
(R{-1})_{(j{+{1\over 2}},\bj{\pm{1\over 2}})}}~~\hidewidth&&&&
\hidewidth\matrix{{\scriptstyle R}_{\sss(j{+1},\bj{+1}),(j{+1},\bj),(j,\bj{+1}),(j,\bj)}\cr
\noalign{\vskip-9pt}{\scriptstyle(R-1)}_{\sss(j{+1},\bj),(j,\bj{+1}),
(j,\bj),(j,\bj)}\cr
\noalign{\vskip-6pt}{\scriptstyle(R-2)}_{\sss(j,\bj)}}\hidewidth&&&&
\hidewidth{R_{(j{+{1\over 2}},\bj{+{1\over 2}})}\atop
(R{-1})_{(j{\pm{1\over 2}},\bj{+{1\over 2}})}}\hidewidth&&\cr
&&&&&&\Bse&&\Bsw&&\Bse&&\Bsw&&&\cr
2R{+j}{+\bj}{+{\ts{9\over 2}}}&&&&&&&\hidewidth
{(R{-{1\over 2}})_{(j{+{1\over 2}},\bj{+1}),(j{+{1\over 2}},\bj)}\atop
(R{-{3\over 2}})_{(j{+{1\over 2}},\bj)}}\hidewidth
&&&&\hidewidth{(R{-{1\over 2}})_{(j{+1},\bj{+{1\over 2}}),(j,\bj{+{1\over 2}})}
\atop (R{-{3\over 2}})_{(j,\bj{+{1\over 2}})}}\hidewidth&&&&\cr
&&&&&&&&\Bse&&\Bsw&&&&&\cr
2R{+j}{+\bj}{+5}&&&{~~}&&{~~~}&&{~~}&&\hidewidth
(R-1)_{(j{+{1\over 2}},\bj{+{1\over 2}})}\hidewidth&&{~~}&
&{~~~}&&{~~}\cr
r~&~~~~\hidewidth&&\hidewidth\bj{-j}{+{\ts{3\over 2}}}\hidewidth&
&\hidewidth\bj{-j}{+1}\hidewidth&&\hidewidth\bj{-j}{+\half}\hidewidth&&
\hidewidth\bj{-j}\hidewidth&&\hidewidth\bj{-j}{-\half}~~\hidewidth&
&\hidewidth\bj{-j}{-1}~\hidewidth&
\hidewidth\bj{-j}{-{\ts{3\over 2}}}\hidewidth~\cr}
}
For this case the dimension formula is
\eqn\dimC{
\dim {\hat \C}_{R(j,\bj)} = 32R(4j+3)(4\bj+3) +32(j+\bj +{\ts{3\over 2}})\, .
}
For $j=\bj=0$ there is no 
restriction on the representations appearing at the first level. 
 
If $R=0$ we may impose both the conditions \Ttwoj\ and \Ttwobj\ simultaneously
for $i=1,2$. In this case compatibility with \algQ\ requires that we have
\eqn\con{
{\tilde {\rm P}}{}^{\dal\alpha} |0,\bj{-j} 
\rangle_{\alpha \alpha_1\dots \alpha_{2j-1}, \dal\dal_1\dots \dal_{2\bj-1}} 
= 0  \, , 
}
where $\Delta = 2 + j +\bj$ for this state. This is of course a generalised
conservation equation. The full multiplet ${\hat \C}_{0(j,\bj)}$ is then 
described by
\eqn\RtwoF{\def\normalbaselines{\baselineskip20pt\lineskip3pt
\lineskiplimit3pt}\hskip-1.5cm
\matrix{\Delta~\cr
j{+\bj}{+2}&&&{~~~}&&{~~~~~}&&{~~~}&&\,~0_{(j,\bj)}~\,&\cr
&&&&&&&&\Bsw&{~~~~~~~~}&\Bse&~~~~~~~~~\cr
j{+\bj}{+{\ts{5\over 2}}}&&&&&&&\hidewidth
{1\over 2}_{(j{+{1\over 2}},\bj)}
\hidewidth&&&&\hidewidth{1\over 2}_{(j,\bj{+{1\over 2}})}\hidewidth\cr
&&&&&&\Bsw&&\Bse&&\Bsw&&\Bse&\cr
j{+\bj}{+3}&&&&&\hidewidth 0_{(j{+1},\bj)}
\hidewidth&&&&
\hidewidth 1_{(j{+{1\over 2}},\bj{+{1\over 2}})},
0_{(j{+{1\over 2}},\bj{+{1\over 2}})}
\hidewidth&&&& \hidewidth 0_{(j,\bj{+1})}\hidewidth&&\cr
&&&&&&\Bse&&\Bsw&&\Bse&&\Bsw&\cr
j{+\bj}{+{\ts{7\over 2}}}&&&&&&
&\hidewidth {1\over 2}_{(j+1,\bj+{1\over 2})} \hidewidth&&&&
\hidewidth {1\over 2}_{(j+{1\over 2},\bj+1)} \hidewidth&&&\cr
&&&&&&&&\Bse&&\Bsw&&&\hidewidth\cr
j{+\bj}{+4}&&&&&&&&& \hidewidth 0_{(j+1,\bj+1)} \hidewidth&&&&&&\cr
r~&~~~~\hidewidth&&&
&\hidewidth\bj{-j}{+1}\hidewidth&&\hidewidth\bj{-j}{+\half}\hidewidth&&
\hidewidth\bj{-j}\hidewidth&&\hidewidth\bj{-j}{-\half}~~\hidewidth&
&\hidewidth\bj{-j}{-1}~\hidewidth& \cr}
}
Every state satisfies a generalised conservation equation akin to \con. Taking
into account these constraints the dimension is still given by \dimC\ for $R=0$.
If $j=0$ then it is appropriate to impose \Ttwo\ and \Ttwobj\ for $i=1,2$.
In this case \algQ\ leads to
\eqn\conj{
{\tilde {\rm P}}{}^{\dal\alpha}  Q^i{}_{\! \alpha}
|0,\bj\rangle_{\dal\dal_1\dots \dal_{2\bj-1}} = 0  \, ,
}
which leads to the conservation equation holding for all representations 
$R_(j,\bj)$ in the multiplet with $j,\bj > 0$. A similar result of course
holds if $\bj=0$. If $j=\bj=0$ then applying \Ttwo\ and  \Ttwobj\ for $i=1,2$
gives
\eqn\conk{
{\tilde {\rm P}}{}^{\dal\alpha} [ Q^i{}_{\! \alpha}, \bQ_{\smash{j\dal}}]
|0,0\rangle = 0  \, ,
}
which implies that the conservation equations apply first at the second level.
This multiplet, in which the lowest dimension state has $r=0$ and $\Delta=2$,
contains the conserved currents for $U(2)_R$ symmetry
as well as the conserved energy momentum tensor and was first constructed 
by Sohnius \Soh. Counting states gives a dimension of 24 in accord with \dimC.

At the unitarity threshold given by \ins\ we may apply \Ttwobj\ to reduce 
further the supermultiplet displayed in \RtwoB\ giving an asymmetric multiplet
which we denote by ${\D}_{R(0,\bj)}$,
\eqn\RtwoD{\def\normalbaselines{\baselineskip20pt\lineskip3pt
\lineskiplimit3pt}\hskip-1.5cm
\matrix{\Delta~\cr
2R{+\bj}{+1}&&{~~~~}&&{~~~}&&R_{(0,\bj)}&&{~~~}&~~~~&{~~~~}&\cr
&&&&&\Bsw&{~~~~~~~~}&\Bse&~~~~~~~~~\cr
2R{+\bj}{+{\ts{3\over 2}}}&&&&
\hidewidth(R{-{1\over 2}})_{({1\over 2},\bj)}\hidewidth&&
&&\hidewidth{(R{+{1\over 2}})_{(0,\bj{+{1\over 2}})}\atop 
(R{-{1\over 2}})_{(0,\bj{\pm{1\over2}})}}\hidewidth\cr 
&&&\Bsw&&\Bse&&\Bsw&&\Bse&\cr
2R{+\bj}{+2}&&(R-1)_{(0,\bj)}\hidewidth&&&&\hidewidth{
(R)_{({1\over 2},\bj{+{1\over 2}})}\atop
(R{-1})_{({1\over 2},\bj{\pm{1\over 2}})}}\hidewidth &&&&
\hidewidth{R_{(0,\bj+1),(0,\bj)}\atop (R-1)_{(0,\bj)}}\hidewidth&&\cr
&&&\Bse&&\Bsw&&\Bse&&\Bsw&~~~~~~~\Bse\hidewidth\cr 
2R{+\bj}{+{\ts{5\over 2}}}&&&&\hidewidth
{(R{-{1\over 2}})_{(0,\bj{+{1\over 2}})}\atop
(R{-{3\over 2}})_{(0,\bj{\pm{1\over 2}})}}\hidewidth&&
&&\hidewidth{(R-{1\over 2})_{({1\over 2},\bj+1),({1\over 2},\bj)}\atop 
(R-{3\over 2})_{({1\over 2},\bj)}}\hidewidth
\hidewidth&&&& 
(R{-{1\over 2}})_{(0,\bj{+{1\over 2}})}\hidewidth\cr
&&&&&\Bse&&\Bsw&&\Bse&&\Bsw\hidewidth&&\cr
2R{+\bj}{+3}&&&&&&
\hidewidth{(R-1)_{(0,\bj+1),(0,\bj)}\atop (R-2)_{(0,\bj)}}\hidewidth
&&&&\hidewidth
(R{-1})_{({1\over 2},\bj{+{1\over 2}})}\hidewidth&&&&\cr
&&&&&&&\Bse&&\Bsw&&&&\cr
2R{+\bj}{+{\ts{7\over 2}}}&&&&&&&&\hidewidth
(R{-{3\over 2}})_{(0,\bj{+{1\over 2}})}\hidewidth&&&&&\cr 
r~&~~~~~\hidewidth
&\hidewidth ~~\bj{+2}\hidewidth&&\hidewidth ~~\bj{+{\ts{3\over2}}}\hidewidth&&
\hidewidth \bj{+1}\hidewidth&&\hidewidth \bj{+\half}~~\hidewidth&
&\hidewidth \bj~\hidewidth&&
~~~\bj{-{\ts{1\over 2}}}\hidewidth~&&\cr}
}
In this case
\eqn\dimD{
\dim {\D}_{R(0,\bj)} = 16(4R-1)(2\bj+1) + 32R \, .
}
We may clearly also define a conjugate multiplet ${\bar \D}_{R(j,0)}$ if
$-r = j+1$ whose dimension is the same as \dimD\ with $\bj\to j$.

For  chiral multiplets, where the lowest dimension states have $R=0$ and 
satisfy $Q^i{}_{\!\alpha} |0,r\rangle_{\dal_1 \dots \dal_{2\bj}}= 0$ or
$\bQ_{i\dal} |0,r\rangle_{\alpha_1\dots \alpha_{2j}} = 0$ for $i=1,2$, we
may also impose the semi-shortening conditions  \Ttwobj, for $\Delta = r = 1+\bj$,
or \Ttwoj, for $\Delta = - r = 1+j$ respectively. For compatibility with
\algQ\ these states must satisfy a generalised Dirac equation, so that
in the latter case we have
\eqn\dirac{
{\tilde {\rm P}}{}^{\dal\alpha}
|0,{-1}{-j}\rangle_{\alpha \alpha_1\dots \alpha_{2j-1}} = 0  \, .
}
Taking $j=\bj$ the reduced chiral multiplet and its conjugate are represented by
\eqn\Nfoura{\def\normalbaselines{\baselineskip16pt\lineskip3pt
 \lineskiplimit3pt}
\matrix{\Delta~~&{~~~~~}&&{~~~}&&{~~~}&&{~~~}&&{~~~}&&{~~~} \cr
j+1~~&&&&&0_{(j,0)}~~~~~~0_{(0,j)}\cr
&&&&\hidewidth\Bsw\hidewidth&&\hidewidth\Bse\hidewidth\cr
j+{\ts{3\over 2}}~~&&&\hidewidth{1\over 2}_{(j+{1\over 2},0)}\hidewidth&&
&&\hidewidth{1\over 2}_{(0,j+{1\over 2})}\hidewidth\cr
&&\hidewidth\Bsw\hidewidth&&&&&&\Bse\cr
j+2~~&\hidewidth 0_{(j+1,0)}\hidewidth&&&&&&&&\hidewidth
0_{(0,j+1)}\hidewidth\cr
r~~&~\hidewidth -j \hidewidth&&\hidewidth-j{-{\ts{1\over2}}}\hidewidth&
&\hidewidth {-j}{-1}~~~~~{j}{+1}~~&
&j{+{\ts{1\over2}}}~\hidewidth&&\hidewidth j~\hidewidth}
}
For a $(j,0)$ or $(0,\bj)$ field obeying the Dirac equation there is
one degree of freedom so that each multiplet in \Nfoura\ has dimension 4.
If $j=0$ the lowest state satisfies $P^2 |0,\pm 1 \rangle =0$ although at
higher levels the states satisfy the generalised Dirac equation.
This describes the standard $\N=2$ vector supermultiplet, whose lowest
dimension state is a massless scalar with scale dimension 1.

The semi-short supermultiplet ${\hat \C}_{R(j,\bj)}$ represented by \RtwoE\ 
occurs at the threshold of the unitarity bound \ineq. It is interesting to
consider how a long supermultiplet $\A^\Delta_{R,r(j,\bj)}$ decomposes at this 
point. For $j,\bj >0$ it can be written as
a semi-direct sum of semi-short representations,
\eqn\ACCC{
\A^{2R{+j}{+\bj}{+2}}_{R,\bj-j(j,\bj)} \simeq {\hat \C}_{R(j,\bj)}  \oplus
{\hat \C}_{R+{1\over 2}(j-{1\over 2},\bj)} \oplus 
{\hat \C}_{R+{1\over 2}(j,\bj-{1\over 2})} 
\oplus {\hat \C}_{R+1(j-{1\over 2},\bj-{1\over 2})} \, .
}
For $\bj=0$ we have
\eqn\ACCB{
\A^{2R{+j}{+2}}_{R,-j(j,0)} \simeq {\hat \C}_{R(j,0)}  \oplus
{\hat \C}_{R+{1\over 2}(j-{1\over 2},0)} \oplus {\bar \D}_{R+1(j,0)} \oplus 
{\bar \D}_{R+{3\over2}(j-{1\over 2},0)}\, ,
}
and for $j=\bj=0$ the decomposition of $\A$  becomes
\eqn\ACBB{
\A^{2R{+2}}_{R,0(0,0)} \simeq {\hat \C}_{R(0,0)}  \oplus
{\D}_{R+1(0,0)} \oplus {\bar \D}_{R+1(0,0)} \oplus {\hat \B}_{R+2} \, ,
}
which involves a short BPS representation.
We also have
\eqn\BDD{
{\B}_{R,\bj+1(0,\bj)} \simeq {\D}_{R(0,\bj)} \oplus 
{\D}_{R+{1\over 2}(0,\bj- {1\over 2})} \, ,
\qquad {\B}_{R,1(0,0)} \simeq {\D}_{R(0,0)} \oplus {\hat \B}_{R+1} \, ,
}
together with their conjugates. In each case the last term represents an 
invariant subspace for the long multiplet.
These decompositions are consistent with \dimA, \dimC, \dimB, \dimbB\ and \dimBs.
Both \ACCC\ and \ACBB\ are valid for $R=0$ when the lowest dimension
multiplet contains conserved currents. We note that \ACCB\ and \ACBB\ are special
cases of \ACCC\ with the use of
\eqn\spec{
{\hat \C}_{R(j,-{1\over 2})} \simeq {\bar \D}_{R+{1\over 2}(j,0)}\, , \qquad
{\hat \C}_{R(-{1\over 2},\bj)} \simeq {\D}_{R+{1\over 2}(0,\bj)} \, , \qquad
{\bar \D}_{R(- {1\over 2},0)} \simeq {\D}_{R(0,- {1\over 2})} \simeq
{\hat \B}_{R+{1\over 2}} \, .
}
If the semi-shortening condition is applied to just the action of the
$Q$ or $\bQ$ supercharges, so that only one of the bounds in \ineq\
is saturated, then instead of \ACCC\ we have
\eqn\ACC{
\A^{2R{+r}{+2j}{+2}}_{R,r(j,\bj)} \simeq {\C}_{R,r(j,\bj)}  \oplus
{\C}_{R+{1\over 2},r+{1\over 2}(j-{1\over 2},\bj)} \, ,
}
together with its conjugate. Note that ${\C}_{R,r(-{1\over 2},\bj)} =
{\B}_{R+{1\over 2},r(0,\bj)}$.

\newsec{Superconformal Representations, $\N=4$}

For the $\N=4$ case the discussion of possible supermultiplets is essentially
similar to $\N=2$. Instead of a spinorial basis as in \sjj\ we now write
\eqn\MJ{
\big [M_\alpha{}^{\! \beta}\big ] = \pmatrix{ J_3 & J_+ \cr J_- & -J_3 } \, , 
\qquad \big [{\bar M}{}^\dbe{}_{\!\smash{\dal}}\big ] =
\pmatrix{ \bJ_3 & \bJ_+ \cr \bJ_- & -\bJ_3 } \, ,
}
which satisfy \Mcom\ subject to the usual commutation relations
for the $J_\pm,J_3$ and $\bJ_\pm,\bJ_3$ generators of $SU(2)_J$ and $SU(2)_\bJ$.
The standard basis states are then given by $|k,p,q;m,\bm\rangle$,
labelled by the eigenvalues $k,p,q$ of $H_1,H_2,H_3$ for $SU(4)_R$ and $m,\bm$ of
$J_3,\bJ_3$. The highest weight states then satisfy
\eqn\EJW{
E_i{}^{\!+} |k,p,q;j,\bj\rangle^{\rm hw} = J_+
|k,p,q;j,\bj\rangle^{\rm hw} = \bJ_+ |k,p,q;j,\bj\rangle^{\rm hw}
= 0 \, ,
}
as well as the analogues of \Otwo\ and \Stwo\ for this to be the lowest dimension 
conformal primary state of a supermultiplet. The representation defined 
by \EJW\ is here denoted by $[k,p,q]_{(j,\bj)}$. 

The supermultiplet is generated by the action of the supercharges
$ Q^i{}_{\! \alpha} , \, \bQ_{i\dal}$ on the state with lowest scale dimension
$\Delta$. The supercharges have weights, corresponding to the change in the 
eigenvalues of $H_i$ and $J_3,\bJ_3$, which are, from \EQ\ and \EQm,
\eqn\QHJ{\hskip -0.5cm \eqalign{
Q^1{}_{\! \alpha} {}& \sim [1,0,0]_{(\pm {1\over 2},0)} \, , \
Q^2{}_{\! \alpha} \sim [-1,1,0]_{(\pm {1\over 2},0)} \, , \
Q^3{}_{\! \alpha} \sim [0,-1,1]_{(\pm {1\over 2},0)} \, , \
Q^4{}_{\! \alpha} \sim [0,0,-1]_{(\pm {1\over 2},0)} \, , \cr
\bQ_{1\dal}  {}& \sim [-1,0,0]_{(0, \pm {1\over 2})} \, , \
\bQ_{2\dal} \sim [1,-1,0]_{(0, \pm {1\over 2})} \, , \
\bQ_{3\dal} \sim [0,1,-1]_{(0,\pm {1\over 2})} \, , \
\bQ_{4\dal} \sim [0,0,1]_{(0,\pm {1\over 2})} \, , \cr}
}
which define the weights of all the $Q$ and $\bQ$ supercharges so that
$Q^1{}_{\! 1}$ and $\bQ_{4\, 1}$ are the highest weight operators
of the $[1,0,0]_{({1\over 2},0)}$ and $[0,0,1]_{(0,{1\over 2})}$ representations
of $SU(4)_R \otimes SU(2)_J \otimes SU(2)_\bJ$. A long supermultiplet
is generated by the unconstrained action of the 16 supercharges on the
lowest dimension state, in a similar fashion to \Qstate. 
For the $Q$ supercharges acting on a general
multiplet $[k,p,q]_{(j,\bj)}$ we have according to the Racah-Speiser 
algorithm generically all possible  $[k',p',q']_{(j',\bj')}$ obtained
by adding the weights in \QHJ\ for each supercharge. 
The action of the supercharges then leads to all  representations with 
$[k',p',q']_{(j',\bj')}$ as the highest weight, so long as $k',p',q',j',\bj'$ 
are all positive or zero. If any of $k',p',q',j',\bj'$ become negative the
appropriate prescription is given in the next section. In general the 
representations which appear are  described  by
\eqnn\NQfour
$$\eqalignno{
[k,p,q]_{(j,\bj)}& \  \mapright Q \ 
{\ss [k{+1},p,q],[k,p{-1},q{+1}]_{(j\pm {1\over 2},\bj)}
\atop \ss [k,p,q{-1}],[k{-1},p{+1},q]_{(j\pm {1\over 2},\bj)}} \cr
\noalign{\vskip 2pt}
& \ \mapright {Q^2} \ \matrix{\ss
[k{+1},p,q{-1}],[k{+1},p{-1},q{+1}],[k,p{+1},q],[k,p{-1},q],[k{-1},p{+1},q{-1}],
[k{-1},p,q{+1}]_{(j{\pm 1},\bj)}\cr
\ss [k{+2},p,q],2[k{+1},p,q{-1}],2[k{+1},p{-1},q{+1}],2[k,p{+1},q],
2[k,p{-1},q]_{(j,\bj)} \cr 
\ss 2[k{-1},p{+1},q{-1}],2[k{-1},p,q{+1}],[k,p,q{-2}],[k,p{-2},q{+2}],
[k{-2},p{+2},q]_{(j,\bj)}}\cr
\noalign{\vskip 2pt}
& \ \mapright {Q^3} \
\matrix{\ss
[k{+1},p{-1},q],[k,p{+1},q{-1}],[k,p,q{+1}],
[k{-1},p,q]_{(j{\pm {3\over 2}},\bj)}\cr
\ss [k{+2},p,q{-1}],[k{+2},p{-1},q{+1}],[k{+1},p{+1},q],[k{+1},p,q{-2}],
[k{+1},p{-2},q{+2}]_{(j{\pm {1\over 2}},\bj)}\cr
\ss 3[k{+1},p{-1},q],
3[k,p{+1},q{-1}],3[k,p,q{+1}],3[k{-1},p,q],[k,p{-1},q{-1}],
[k,p{-2},q{+1}]_{(j{\pm {1\over 2}},\bj)}\cr
\ss [k{-1},p{+2},q],[k{-1},p{+1},q{-2}],[k{-1},p{-1},q{+2}],
[k{-2},p{+2},q{-1}],[k{-2},p{+1},q{+1}]_{(j{\pm {1\over 2}},\bj)}} \cr
\noalign{\vskip 4pt}
& \ \mapright {Q^4} \ \matrix{\ss
[k,p,q]_{(j{\pm 2},\bj)},
[k{+2},p{-1},q{+1}],[k{+1},p{+1},q{-1}],[k{+1},p,q{+1}],[k{+1},p{-1},q{+1}]
_{(j{\pm 1},\bj)} \cr \ss
[k{+1},p{-2},q{+1}],[k,p{+1},q{-2}],4[k,p,q],[k,p{-1},q{+2}],
[k{-1},p{+1},q{+1}]_{(j{\pm 1},\bj)}\cr
\ss [k{-1},p,q{-1}],[k{-1},p{-1},q{+1}],[{k-1},p{+2},q{+1}],
[k{-2},p{+1},q]_{(j{\pm 1},\bj)}\cr
\ss [k{+2},p,q{-2}],2[k{+2},p{-1},q],[k{+2},p{-2},q{+2}]_{(j,\bj)}\cr
\ss 2[k{+1},p{+1},q{-1}],
2[k{+1},p,q{+1}],2[k{+1},p{-1},q{-1}],2[k{+1},p{-2},q{+1}]_{(j,\bj)} \cr
\ss [k,p{+2},q],2[k,p{+1},q{-2}],6[k,p,q],2[k,p{-1},q{+2}],[k,p{-2},q]_{(j,\bj)}\cr
\ss 2[k{-1},p{+2},q{-1}],2[k{-1},p{+1},q{+1}],2[k{-1},p,q{-1}],
2[k{-1},p{-1},q{+1}]_{(j,\bj)}\cr 
\ss [k{-2},p{+2},q{-2}],2[k{-2},p{+1},q],[k{-2},p,q{+2}]_{(j,\bj)}}\cr
\noalign{\vskip 4pt}
& \ \mapright {Q^5} \
\matrix{\ss
[k{+1},p,q],[k,p,q{-1}],[k,p{-1},q{+1}],
[k{-1},p{+1},q]_{(j{\pm {3\over 2}},\bj)}\cr
\ss [k{+2},p{-1},q{-1}],[k{+2},p{-2},q{+1}],[k{+1},p{+1},q{-2}],
[k{+1},p{-1},q{+2}],[k{+1},p{-2},q]_{(j{\pm {1\over 2}},\bj)}\cr
\ss 3[k{+1},p,q],
3[k,p,q{-1}],3[k,p{-1},q{+1}],3[k{-1},p{+1},q],[k,p{+2},q{-1}],
[k,p{+1},q{+1}]_{(j{\pm {1\over 2}},\bj)}\cr
\ss [k{-1},p{+2},q{-2}],[k{-1},p,q{+2}],[k{-1},p{-1},q],
[k{-2},p{+1},q{-1}],[k{-2},p,q{+1}]_{(j{\pm {1\over 2}},\bj)}} \cr
\noalign{\vskip 2pt}
& \ \mapright {Q^6} \ \matrix{\ss
[k{+1},p,q{-1}],[k{+1},p{-1},q{+1}],[k,p{+1},q],[k,p{-1},q],[k{-1},p{+1},q{-1}],
[k{-1},p,q{+1}]_{(j{\pm 1},\bj)}\cr
\ss [k{+2},p{-2},q],2[k{+1},p,q{-1}],2[k{+1},p{-1},q{+1}],2[k,p{+1},q],
2[k,p{-1},q]_{(j,\bj)} \cr
\ss 2[k{-1},p{+1},q{-1}],2[k{-1},p,q{+1}],[k,p{+2},q{-2}],[k,p,q{+2}],
[k{-2},p,q]_{(j,\bj)}}\cr
\noalign{\vskip 2pt}
& \  \mapright {Q^7} \
{\ss [k{+1},p{-1},q],[k,p{+1},q{-1}]_{(j\pm {1\over 2},\bj)}
\atop\ss [k,p,q{+1}],[k{-1},p,q]_{(j\pm {1\over 2},\bj)}} \cr
\noalign{\vskip 1pt}
& \  \mapright {Q^8} \ [k,p,q]_{(j,\bj)} \, . & \NQfour \cr}
$$
At each level $l$ there are $8 \choose l$ representations of 
$SU(4)_R \otimes SU(2)_J \otimes  SU(2)_\bJ$. The total
dimension of this module is $2^8$ times the dimension of the lowest
dimension representation. Defining $\hk=k+1, \, \hp=p+1 , \, \hq=q+1$
and $\J=2j+1, \, \bcJ=2\bj+1$ this becomes,
with the $SU(4)$ dimension formula given in  \dimfour,
\eqn\dimz{
2^8 d(k,p,q) (2j+1)(2\bj+1)=
{\ts 64\over 3}\, \hk \,\hp \,\hq (\hk+\hp)(\hq+\hp)(\hk+\hp+\hq)\J \, \bcJ \, .
}

For multiplet shortening we consider BPS conditions of the form
\eqn\BPSQ{
Q^i{}_{\! \alpha} |k,p,q;j,\bj\rangle^{\rm hw} = 0 \, , \qquad
\alpha=1,2 \, ,
}
which restricts the representation content at level one and subsequent levels.
Since, for the lowest scale dimension state, $ |k,p,q;j,\bj\rangle^{\rm hw}$
is annihilated by $ S_l{}^{\!\beta}$ then applying \QS{a}\ requires that
the state must also be annihilated by $M_\alpha{}^{\! \beta}$ or $j=0$
and then
\eqn\DR{
\big ( \de^i{}_{\! l}\half \Delta - R^i{}_{\! l} \big )
|k,p,q;0,\bj\rangle^{\rm hw} = 0 \, , \qquad l=1,2,3,4 \, .
}
Using \Rfour\ and \Hvec, \high, \low\ it is easy to see that there are the 
following solutions, where $s$ is the fraction of the charges for which \BPSQ\ 
holds,
\eqn\Bps{\eqalign{
i={}& 1 \ \, \quad \qquad\qquad s=\quar \qquad \Delta = \half ( 3k + 2p + q ) \, , \cr
i={}&1,2 \, \ \ \ \ \quad\qquad s = \half \qquad \Delta = \half (  2p + q ) \, , 
\qquad k =0 \, , \cr
i={}&1,2,3 \ \quad\qquad \ s = {\ts{3\over 4}} \qquad \Delta = \half q  \, , \qquad 
\qquad \quad k = q = 0 \, , \cr
i={}&1,2,3,4 \, \ \  \qquad s = 1 \qquad \Delta = 0  \, , \qquad \quad
\qquad k = p = q = 0  \, . \cr }
}
The last case of course corresponds to the trivial singlet representation.
Conversely since from \EQm\ $[E_1{}^{\!-} , Q^1{}_{\! \alpha}] =
Q^2{}_{\! \alpha}$  if $k=0$ the condition \BPSQ\ for $i=1$ implies the
$i=2$ case by virtue of \low. Similarly if $p=0$ or $p=q =0$ as
well as $k=0$ we have the BPS condition for $i=2$ or  $i=2,3$.

For each case  we may obtain the set of representations, analogous to
\NQfour, obtained by the action of the supercharges on a lowest dimension
state $[k,p,q]_{(0,\bj)}$. For the ${1\over 4}$-BPS case  the six supercharges
$Q^i{}_{\! \alpha}$, $i=2,3,4$, give, for general $k,p,q$,
\eqnn\NQfourf
$$\eqalignno{
[k,p,q]_{(0,\bj)}& \  \mapright Q \
{\ss [k,p,q{-1}],[k,p{-1},q{+1}]_{({1\over 2},\bj)}
\atop \ss [k{-1},p{+1},q]_{({1\over 2},\bj)}} \cr
\noalign{\vskip 2pt}
& \ \mapright {Q^2} \ \matrix{\ss
[k,p{-1},q],[k{-1},p{+1},q{-1}],
[k{-1},p,q{+1}]_{(1,\bj)}\cr
\ss [k,p,q{-2}],[k,p{-1},q],[k,p{-2},q{+2}],[k{-1},p{+1},q{-1}],[k{-1},p,q{+1}],
[k{-2},p{+2},q]_{(0,\bj)}}\cr
\noalign{\vskip 2pt}
& \ \mapright {Q^3} \
\matrix{\ss
[k{-1},p,q]_{({3\over 2},\bj)},[k,p{-1},q{-1}],
[k,p{-2},q{+1}]_{({1\over 2},\bj)}\cr
\ss [k{-1},p{+1},q{-2}],2[k{-1},p,q],[k{-1},p{-1},q{+2}],
[k{-2},p{+2},q{-1}],[k{-2},p{+1},q{+1}]_{({1\over 2},\bj)}} \cr
\noalign{\vskip 2pt}
& \ \mapright {Q^4} \ \matrix{\ss \ss 
[k{-1},p,q{-1}],[k{-1},p{-1},q{+1}],[k{-2},p{+1},q]_{(1,\bj)}\cr
\ss [k,p{-2},q],[k{-1},p,q{-1}],[k{-1},p{-1},q{+1}],
[k{-2},p{+2},q{-2}],[k{-2},p{+1},q],[k{-2},p,q{+2}]_{(0,\bj)}}\cr
\noalign{\vskip 2pt}
& \ \mapright {Q^5} \
\matrix{\ss
\ss [k{-1},p{-1},q]_{({1\over 2},\bj)}\cr
\ss [k{-2},p{+1},q{-1}],[k{-2},p,q{+1}]_{({1\over 2},\bj)} \cr} \cr
\noalign{\vskip 2pt}
& \ \mapright {Q^6} \  [k{-2},p,q]_{(0,\bj)} \, . & \NQfourf \cr}
$$
In this case at level $l$ there are $6\choose l$ representations of
$SU(4)_R \otimes SU(2)_\bJ$.
The dimension of this module, which may be obtained by adding the dimensions
of all representations in \NQfourf, is,
\eqn\dims{
2^6 d(k-1,p,q) \bcJ - {\ts {8\over 3}}\, \hp\, \hq(\hp+\hq)(3\hk + 2\hp +\hq
- 3) \bcJ \, .
}

When the $\half$-BPS condition is applied the states are obtained just 
by the action of the $Q^3{}_{\! \alpha},Q^4{}_{\! \alpha}$ supercharges. 
According to \Bps\ we should set $k=0$ in the lowest dimension state 
for this case but we allow for arbitrary $k$ for subsequent appplication. 
The action of the supercharges then gives representations obtained by adding
weights,
\eqnn\NQfours
$$\eqalignno{
[k,p,q]_{(0,\bj)} &  \  \mapright Q \
{\ss [k,p,q{-1}]_{({1\over 2},\bj)}
\atop \ss [k,p{-1},q{+1}]_{({1\over 2},\bj)}} 
\ \mapright {Q^2} \ \matrix{\ss
[k,p{-1},q]_{(1,\bj)}, [k,p,q{-2}]_{(0,\bj)}\cr
\ss [k,p{-1},q],[k,p{-2},q{+2}]_{(0,\bj)}} \cr
& \ \mapright {Q^3} \
{\ss [k,p{-1},q{-1}]_{({1\over 2},\bj)}
\atop \ss [k,p{-2},q{+1}]_{({1\over 2},\bj)}} 
\ \mapright {Q^4} \  [k,p{-2},q]_{(0,\bj)} \, . &  \NQfours \cr}
$$
In \NQfours\ it is easy to see that there are now $4\choose l$ 
representations of $SU(4)_R \otimes SU(2)_\bJ$. The result \NQfours\
is valid for $k\ge 0, \, p,q\ge 2$, restriction to other cases will be given
later.  The dimension for module given by \NQfours\  is, 
\eqn\dimh{
2^4 d(k,p-1,q) \bcJ -  
{\ts{1\over 3}}\hk\, \hq (2\hk^2 - 2\hq^2 - 3) \bcJ \, .
}

For completeness we also note the $3\over 4$-BPS case which is
obtained by the action of $Q^4{}_{\! \alpha}$ alone
\eqn\NQtf{
[k,p,q]_{(0,\bj)} \  \mapright Q \ [k,p,q{-1}]_{({1\over 2},\bj)}
\ \mapright {Q^2} \  [k,p,q{-2}]_{(0,\bj)} \, ,
}
again generalising to arbitrary $k,p,q$, with dimension
\eqn\dime{
2^2 d(k,p,q-1) \bcJ + {\ts {1\over 6}}\, \hk\, \hp(\hk+\hp)(\hk + 2\hp +3\hq
- 3) \bcJ \, .
}

As in the $\N=2$ case there are further shortening conditions that
may be applied for $j\ge 0$. We consider the constraint
\eqn\sshort{
\Big ( Q^i{}_{\! 2} - {\ts {1\over 2j+1}} J_- Q^i{}_{\! 1} \Big )
|k,p,q;j,\bj\rangle^{\rm hw} = 0 \, .
}
Since $J_+$ acting on this state gives zero, using $[J_+ , Q^i{}_{\! 2}] =
Q^i{}_{\! 1}$, this is clearly a state with spin $(j{-1},\bj)$ and hence
\sshort\ is equivalent for $\N=4$ to \Ttwoj\ for $\N=2$. Applying 
$S_l{}^{\! 1}$ to this and using $\{ S_l{}^{\! 1} , Q^i{}_{\! 2} \} =  4
\de^i{}_{\! l} J_-$, $[ S_l{}^{\! 1} , J_- ] =0$, 
$\{ S_l{}^{\! 1} , Q^i{}_{\! 1} \} =  4 \de^i{}_{\! l} ( J_3 - \half i D)
+ 4 R^i{}_{\! l}$, the condition \sshort\ leads to the constraint
\eqn\DRS{
J_- \Big ( \de^i{}_{\! l} - {\ts {1\over 2j+1}} \big ( ( j + \half
\Delta ) \de^i{}_{\! l} -  R^i{}_{\! l} \big ) \Big )
|k,p,q;j,\bj\rangle^{\rm hw} = 0 \, , \quad l=1,2,3,4 \, ,
}
which determines $\Delta$.
Since $S_l{}^{\! 2} = - [J_+ ,S_l{}^{\! 1}]$ it is clear that applying 
$S_l{}^{\! 2}$
to \sshort\ does not give additional conditions. Again there are various
cases according to the various possible choices $i$  for which \sshort\ holds,
\eqn\Bpss{\eqalign{
i={}& 1 \ \, \quad \qquad\qquad t=\quar \qquad \Delta = 2 + 2j + \half 
( 3k + 2p + q ) \, , \cr
i={}&1,2 \, \ \ \ \ \quad\qquad t=\half \qquad \Delta = 2 + 2j + \half (  2p + q ) \, ,
\qquad k =0 \, , \cr
i={}&1,2,3 \ \quad\qquad \ t={\ts {3\over 4}} \qquad
\Delta = 2 + 2j + \half q \, , \qquad \qquad k =p = 0 \, , \cr
i={}&1,2,3,4 \, \ \  \qquad t=1 \qquad \Delta = 2 + 2j  \, , \qquad \quad
\qquad k =p = q = 0  \, , \cr }
}
where $t$ is analogous to $s$ in \Bps.
As discussed earlier $p=0$ requires that \sshort\ for $i=1$ implies $i=2$,
$k=p=0$ implies \sshort\ for $i=2,3$ and $k=p=q=0$ for $i=2,3,4$.

Applying the condition \sshort\ just for $i=1$ leads to the module
\eqnn\NQfourS
$$\eqalignno{
[k,p,q]_{(j,\bj)}& \  \mapright Q \
{\ss [k{+1},p,q]_{(j+{1\over 2},\bj)},[k,p{-1},q{+1}]_{(j\pm{1\over 2},\bj)}
\atop \ss [k,p,q{-1}],[k{-1},p{+1},q]_{(j\pm {1\over 2},\bj)}} \cr
\noalign{\vskip 2pt}
& \ \mapright {Q^2} \ \matrix{\ss
[k{+1},p,q{-1}],[k{+1},p{-1},q{+1}],[k,p{+1},q]_{(j{+ 1},\bj)},
[k,p{-1},q],[k{-1},p{+1},q{-1}],
[k{-1},p,q{+1}]_{(j{\pm 1},\bj)}\cr
\ss [k{+1},p,q{-1}],[k{+1},p{-1},q{+1}],[k,p{+1},q],
2[k,p{-1},q]_{(j,\bj)} \cr
\ss 2[k{-1},p{+1},q{-1}],2[k{-1},p,q{+1}],[k,p,q{-2}],[k,p{-2},q{+2}],
[k{-2},p{+2},q]_{(j,\bj)}}\cr
\noalign{\vskip 2pt}
& \ \mapright {Q^3} \
\matrix{\ss
[k{+1},p{-1},q],[k,p{+1},q{-1}],[k,p,q{+1}]_{(j{+{3\over 2}},\bj)},
[k{-1},p,q]_{(j{\pm {3\over 2}},\bj)}\cr
\ss [k{+1},p,q{-2}],[k{+1},p{-1},q],[k{+1},p{-2},q{+2}],
[k,p{+1},q{-1}],[k,p,q{+1}],k{-1},p{+2},q]_{(j{+{1\over 2}},\bj)}\cr
\ss [k{+1},p{-1},q],
[k,p{+1},q{-1}],[k,p,q{+1}],[k,p{-1},q{-1}],
[k,p{-2},q{+1}]_{(j{\pm {1\over 2}},\bj)}\cr
\ss [k{-1},p{+1},q{-2}],3[k{-1},p,q],[k{-1},p{-1},q{+2}],
[k{-2},p{+2},q{-1}],[k{-2},p{+1},q{+1}]_{(j{\pm {1\over 2}},\bj)}} \cr
\noalign{\vskip 2pt}
& \ \mapright {Q^4} \ \matrix{\ss
[k,p,q]_{(j{+2},\bj)},
[k{+1},p{-1},q{-1}],[k{+1},p{-2},q{+1}],[k,p{+1},q{-2}],2[k,p,q]_{(j{+1},\bj)} \cr 
\ss
[k,p{-1},q{+2}],[k{-1},p{+2},q{-1}],
[k{-1},p{+1},q{+1}]_{(j{+1},\bj)} \cr \ss
[k,p,q],[k{-1},p,q{-1}],[k{-1},p{-1},q{+1}],[k{-2},p{+1},q]_{(j{\pm 1},\bj)}\cr
\ss 
[k{+1},p{-1},q{-1}],[k{+1},p{-2},q{+1}],
[k,p{+1},q{-2}],3[k,p,q],[k,p{-1},q{+2}],[k,p{-2},q]_{(j,\bj)}\cr
\ss [k{-1},p{+2},q{-1}],[k{-1},p{+1},q{+1}],2[k{-1},p,q{-1}],
2[k{-1},p{-1},q{+1}]_{(j,\bj)}\cr
\ss [k{-2},p{+2},q{-2}],2[k{-2},p{+1},q],[k{-2},p,q{+2}]_{(j,\bj)}}\cr
\noalign{\vskip 2pt}
& \ \mapright {Q^5} \
\matrix{\ss
[k,p,q{-1}],[k,p{-1},q{+1}],[k{-1},p{+1},q]_{(j{+{3\over 2}},\bj)}\cr
\ss 
[k{+1},p{-2},q],[k,p,q{-1}],[k,p{-1},q{+1}],[k{-1},p{+2},q{-2}],[k{-1},p{+1},q],
[k{-1},p,q{+2}]_{(j{+{1\over 2}},\bj)}\cr
\ss 
[k,p,q{-1}],[k,p{-1},q{+1}],[k{-1},p{+1},q],
[k{-1},p{-1},q],[k{-2},p{+1},q{-1}],[k{-2},p,q{+1}]_{(j{\pm {1\over 2}},\bj)}} \cr
\noalign{\vskip 2pt}
& \ \mapright {Q^6} \ \matrix{\ss
[k,p{-1},q],[k{-1},p{+1},q{-1}],[k{-1},p,q{+1}]_{(j{+ 1},\bj)}\cr
\ss 
[k,p{-1},q],[k{-1},p{+1},q{-1}],[k{-1},p,q{+1}],[k{-2},p,q]_{(j,\bj)}}\cr
\noalign{\vskip 2pt}
& \  \mapright {Q^7} \ [k{-1},p,q]_{(j+{1\over 2},\bj)} \, . & \NQfourS \cr}
$$
For this case there are $7 \choose l$ representations of
$SU(4)_R \otimes SU(2)_J \otimes  SU(2)_\bJ$ at level $l$. By adding up the
dimensions of all representations we find a total dimension of
\eqn\dimU{\eqalign{
\big ( 2^7 {}& d(k-\half,p,q)(\J+{\ts{1\over 2}}) - {\ts{4\over 3}}\, \hp\hq(\hp+\hq)
(6\, \hk +4\, \hp +2\, \hq - 3)(\J + {\ts{3\over 2}}) \cr
&{} +{\ts {8\over 3}}\, \hp\hq(\hp+\hq) (4\, \hk \hp + 2\,  \hk\hq + \hp\hq
+ \hp^2 + 3\, \hk^2 - 2) \big ) \bcJ \, . \cr}
}

If  \sshort\ is applied  for $i=1,2$  we obtain
\eqnn\NQfourT
$$\eqalignno{
[k,p,q]_{(j,\bj)}& \  \mapright Q \
{\ss [k{+1},p,q],[k{-1},p{+1},q]_{(j+{1\over 2},\bj)}
\atop \ss [k,p,q{-1}],[k,p{-1},q{+1}]_{(j\pm {1\over 2},\bj)}} \cr
\noalign{\vskip 2pt}
& \ \mapright {Q^2} \ \matrix{\ss
[k{+1},p,q{-1}],[k{+1},p{-1},q{+1}],[k,p{+1},q],
[k{-1},p{+1},q{-1}],[k{-1},p,q{+1}]_{(j{+1},\bj)},[k,p{-1},q]_{(j{\pm 1},\bj)}\cr
\ss [k{+1},p,q{-1}],[k{+1},p{-1},q{+1}],[k,p,q{-2}],2[k,p{-1},q],[k,p{-2},q{+2}],
[k{-1},p{+1},q{-1}],[k{-1},p,q{+1}]_{(j,\bj)}}\cr
\noalign{\vskip 2pt}
& \ \mapright {Q^3} \
\matrix{\ss
[k{+1},p{-1},q],[k,p{+1},q{-1}],[k,p,q{+1}],[k{-1},p,q]_{(j{+{3\over 2}},\bj)}\cr
\ss [k{+1},p,q{-2}],[k{+1},p{-1},q],[k{+1},p{-2},q{+2}],
[k,p{+1},q{-1}],[k,p,q{+1}]_{(j{+{1\over 2}},\bj)}\cr
\ss 
[k{-1},p{+1},q{-2}],[k{-1},p,q],[k{-1},p{-1},q{+2}]_{(j{+{1\over 2}},\bj)}\cr
\ss [k{+1},p{-1},q],[k,p{-1},q{-1}],[k,p{-2},q{+1}],[k{-1},p,q]
_{(j{\pm {1\over 2}},\bj)}} \cr
\noalign{\vskip 2pt}
& \ \mapright {Q^4} \ \matrix{\ss
[k,p,q]_{(j{+2},\bj)},
[k{+1},p{-1},q{-1}],[k{+1},p{-2},q{+1}],[k,p{+1},q{-2}]_{(j{+1},\bj)} \cr
\ss
2[k,p,q],[k,p{-1},q{+2}],[k{-1},p,q{-1}],[k{-1},p{-1},q{+1}] _{(j{+1},\bj)} \cr \ss
[k{+1},p{-1},q{-1}],[k{+1},p{-2},q{+1}],[k,p,q],[k,p{-2},q],
[k{-1},p,q{-1}],[k{-1},p{-1},q{+1}]_{(j,\bj)}}\cr
\noalign{\vskip 2pt}
& \ \mapright {Q^5} \
\matrix{\ss
[k,p,q{-1}],[k,p{-1},q{+1}]_{(j{+{3\over 2}},\bj)}\cr
\ss [k{+1},p{-2},q],[k,p,q{-1}],[k,p{-1},q{+1}],
[k{-1},p{-1},q]_{(j{+{1\over 2}},\bj)}\cr} \cr
\noalign{\vskip 2pt}
& \ \mapright {Q^6} \ [k,p{-1},q]_{(j{+1},\bj)}
\, . & \NQfourT \cr}
$$
Here there are $6 \choose l$ representations of
$SU(4)_R \otimes SU(2)_J \otimes  SU(2)_\bJ$ at level $l$. The total
dimension of all representations is
\eqn\dimT{\eqalign{
\big ( 2^6 {}& d(k,p-\half,q)(\J+1) - {\ts{1\over 3}}\, \hk\hq
(4\, \hk^2 - 4\, \hq^2  - 3)(\J + 1) \cr
&{} +{\ts {4\over 3}}\, \hk\hq (\hk+2\, \hp+\hq-1)(2\, \hk\hp+2\, \hq\hp
+2\, \hp^2 +\hk\hq-\hk-2\, \hp-\hq +1 ) \big )  \bcJ \, . \cr}
}

With $i=1,2,3$ we have
\eqnn\NQfourU
$$\eqalignno{
[k,p,q]_{(j,\bj)}& \  \mapright Q \
{\ss [k{+1},p,q],[k,p{-1},q{+1}],[k{-1},p{+1},q]_{(j+{1\over 2},\bj)}
\atop \ss [k,p,q{-1}]_{(j\pm {1\over 2},\bj)}} \cr
\noalign{\vskip 2pt}
& \ \mapright {Q^2} \ \matrix{\ss
[k{+1},p,q{-1}],[k{+1},p{-1},q{+1}],[k,p{+1},q],[k,p{-1},q],
[k{-1},p{+1},q{-1}],[k{-1},p,q{+1}]_{(j{+1},\bj)}\cr
\ss [k{+1},p,q{-1}],[k,p,q{-2}],[k,p{-1},q],
[k{-1},p{+1},q{-1}]_{(j,\bj)}}\cr
\noalign{\vskip 2pt}
& \ \mapright {Q^3} \
\matrix{\ss
[k{+1},p{-1},q],[k,p{+1},q{-1}],[k,p,q{+1}],[k{-1},p,q]_{(j{+{3\over 2}},\bj)}\cr
\ss [k{+1},p,q{-2}],[k{+1},p{-1},q],[k,p{+1},q{-1}],
[k,p{-1},q{-1}],[k{-1},p{+1},q{-2}],[k{-1},p,q]_{(j{+{1\over 2}},\bj)}}\cr
\noalign{\vskip 2pt}
& \ \mapright {Q^4} \ \matrix{\ss
[k,p,q]_{(j{+2},\bj)}\cr\ss
[k{+1},p{-1},q{-1}],[k,p{+1},q{-2}],[k,p,q],[k{-1},p,q{-1}]_{(j{+1},\bj)}}\cr
\noalign{\vskip 2pt}
& \ \mapright {Q^5} \ 
[k,p,q{-1}]_{(j{+{3\over 2}},\bj)}
\, . & \NQfourU \cr}
$$
The dimensions here add up to
\eqn\dimS{\eqalign{
\big (2^5 {}& d(k,p,q-\half)(\J+{\ts{3\over 2}}) + {\ts{1\over 3}}\, \hk\hp(\hk+\hp)
(2\, \hk +4\, \hp +6\, \hq - 3)(\J + \half) \cr
&{} +{\ts {1\over 3}}\, \hk\hp(\hk+\hp) (2\, \hk \hp + 4\,  \hk\hq + 8\,  \hp\hq 
+ 2\,\hp^2 + 6\, \hq^2 - 1)\big ) \bcJ \, . \cr}
}

Finally for  $i=1,2,3,4$ in \sshort,
\eqnn\NQfourV
$$\eqalignno{
[k,p,q]_{(j,\bj)}& \  \mapright Q \
{\ss [k{+1},p,q],[k,p,q{-1}]_{(j+{1\over 2},\bj)}
\atop \ss [k,p{-1},q{+1},[k{-1},p{+1},q]_{(j+{1\over 2},\bj)}} 
\ \mapright {Q^2} \ \matrix{\ss
[k{+1},p,q{-1}],[k{+1},p{-1},q{+1}],[k,p{+1},q]_{(j{+1},\bj)}\cr
\ss [k,p{-1},q],
[k{-1},p{+1},q{-1}],[k{-1},p,q{+1}]_{(j{+1},\bj)}}\cr
\noalign{\vskip 2pt}
& \ \mapright {Q^3} \
\matrix{\ss
[k{+1},p{-1},q],[k,p{+1},q{-1}]_{(j{+{3\over 2}},\bj)}\cr
\ss [k,p,q{+1}],[k{-1},p,q]_{(j{+{3\over 2}},\bj)}}
\ \mapright {Q^4} \ [k,p,q]_{(j{+2},\bj)}
\, , & \NQfourV \cr}
$$
In this case the dimension formula is
\eqn\dimV{
2^4 d(k,p,q) (2j+3)(2\bj+1) =
{\ts {4\over 3}}\, \hk \,\hp \,\hq (\hk+\hp)(\hq+\hp)(\hk+\hp+\hq)(\J+2)\, 
\bcJ \, .
}

Equivalent shortening conditions may be obtained for the action of the
$\bQ$ supercharges. Instead of \BPSQ\ we may impose
\eqn\BPSQb{
\bQ_{\! i\dal} |k,p,q;j,\bj\rangle^{\rm hw} = 0 \, , \qquad
\dal=1,2 \, .
}
Consistency requires $\bj=0$ and then
\eqn\DR{
\big ( \de^l{}_{\! i}\half \Delta + R^l{}_{\! i} \big )
|k,p,q ;0,\bj\rangle^{\rm hw} = 0 \, , \qquad l=1,2,3,4 \, ,
}
which leads to, corresponding to \Bps,
\eqn\Bpsb{\eqalign{
i={}& 4 \ \, \quad \qquad\qquad \bs = \quar \qquad 
\Delta = \half ( k + 2p + 3q ) \, , \cr
i={}&3,4 \, \ \ \ \ \quad\qquad \, \bs = \half \qquad \Delta = \half ( k + 2p ) \, ,
\qquad q =0 \, , \cr
i={}&2,3,4 \ \quad\qquad \ \bs ={\ts {3\over 4}} \qquad \Delta = \half k  \, , 
\qquad\qquad \quad p = q = 0 \, , \cr
i={}&1,2,3,4 \, \ \  \qquad \bs = 1 \qquad \, \Delta = 0  \, , \qquad \quad
\qquad k =p = q = 0  \, . \cr }
}
For the semi-short case we may impose, corresponding to \sshort,
\eqn\sshortb{
\Big ( \bQ{}_{i 1} + {\ts {1\over 2\bj+1}} \bJ_- \bQ{}_{i 2} \Big )
|k,p,q;j,\bj\rangle^{\rm hw} = 0 \, ,
}
and this leads to
\eqn\Bpssb{\eqalign{
i={}& 4 \ \, \quad \qquad\qquad \bt = \quar \qquad \Delta = 2 + 2\bj + \half
( k + 2p + 3q ) \, , \cr
i={}&3,4 \, \ \ \ \ \quad\qquad \bt = \half \qquad
\Delta = 2 + 2\bj + \half (k +  2p ) \, , \qquad q =0 \, , \cr
i={}&2,3,4 \ \quad\qquad \ \bt={\ts {3\over 4}} \qquad
\Delta = 2 + 2\bj + \half k  \, , \qquad \qquad p = q = 0 \, , \cr
i={}&1,2,3,4 \, \ \  \qquad \bt=1 \qquad \Delta = 2 + 2\bj  \, , \qquad \quad
\qquad k =p = q = 0  \, . \cr }
}
For both the short and semi-short cases the action of the $\bQ$ charges may
be obtained by the conjugation of the results for $Q$ given above.

The above results,  summarised in \Bps\ and \Bpss, correspond to representations
of the chiral subalgebra formed by $\{ Q^i{}_{\! \alpha} ,  S_j{}^{\!\beta},
M_\alpha{}^{\! \beta}, R^i{}_{\! j}, D\}$.
Neglecting $\bj$ which is irrelevant
the long multiplet given by \NQfour\ may be denoted by
\eqn\long{
a^\Delta_{[k,p,q]j} \, ,
}
while for the BPS short multiplets we have, for $j=0$,
\eqn\shortb{
b^{1\over 4}_{[k,p,q]} \, , \qquad b^{1\over 2}_{[0,p,q]} \, , \qquad
b^{3\over 4}_{[0,0,q]} \, ,
}
which correspond to \NQfourf, \NQfours\ and \NQtf, and the semi-short multiplets
are
\eqn\shortc{
c^{1\over 4}_{[k,p,q]j} \, , \qquad c^{1\over 2}_{[0,p,q]j} \, , \qquad
c^{3\over 4}_{[0,0,q]j} \, , \qquad c^{1\vphantom{1\over 2}}_{[0,0,0]j} \, ,
}
corresponding to  \NQfourS\ for $k\ge 1$, \NQfourT\ for $k=0, \, p\ge 1$, \NQfourU\
for $k=p=0, \, q\ge 1$ and \NQfourV\ $k=p=q=0$ respectively. At the 
unitarity threshold it is easy to then verify that the long multiplet
may be decomposed as semi-direct sum
\eqnn\decomp
$$\eqalignno{
a^{2+2j+{1\over 2}(3k+2p+q)}_{[k,p,q]j} \simeq {}& 
c^{1\over 4}_{[k,p,q]j} \oplus c^{1\over 4}_{[k+1,p,q]j-{1\over 2}} \, , \qquad
a^{2+2j+{1\over 2}(2p+q)}_{[0,p,q]j} \simeq
c^{1\over 2}_{[0,p,q]j} \oplus c^{1\over 4}_{[1,p,q]j-{1\over 2}} \, , \cr
a^{2+2j+{1\over 2}q}_{[0,0,q]j} \simeq {}&
c^{3\over 4}_{[0,0,q]j} \oplus c^{1\over 4}_{[1,0,q]j-{1\over 2}} \, , \qquad\quad
a^{2+2j}_{[0,0,0]j} \simeq
c^{1\vphantom{1\over 2}}_{[0,0,0]j} 
\oplus c^{1\over 4}_{[1,0,0]j-{1\over 2}} \, , & \decomp \cr}
$$
which can be checked with the aid of the various dimension formulae.
These results also hold for $j=0$ if we make the identification, justified
later, that
\eqn\bc{
c^{1\over 4}_{[k,p,q]-{1\over 2}} \simeq b^{1\over 4}_{[k+1,p,q]} \, ,
}
which is consistent with the dimension results \dimU, for $\J=0$, and \dims.
The first case in \decomp\ has also been obtained recently in \Das\ and 
BPS multiplets further discussed in \Kim.

To construct all conformal primary states for complete supermultiplets it is 
necessary to combine the results for the $Q$ and $\bQ$ supercharges. 
The basic long supermultiplet
is labelled in terms of its scale dimension and $SU(4)$, spin representations
for the lowest dimension state,
\eqn\long{
\A^\Delta_{[k,p,q](j,\bj)} \, , \qquad \dim \A^\Delta_{[k,p,q](j,\bj)}
= 2^{16} d(k,p,q) (2j+1)(2\bj+1) \, .
}
All conformal primary representations may be found with the aid of \NQfour\ and 
its conjugate, applying this
to each representation appearing in \NQfour. If $k,p$ or $q$ are $<4$ there
will be a cancellation of representations according to rules specified later.
For unitarity \Dob
\eqn\unit{
\Delta \ge 2 + 2j + \half (3k+2p+q) , \  2 + 2\bj + \half (k+2p+3q) \, .
}

{}From the above it is simple to obtain the  short supermultiplets where the 
shortening conditions apply only to the action of the $Q$ supercharges while
$\bQ$ are unconstrained, and vice versa. The full supermultiplets are
easily obtained from the results above and in consequence  we do not consider 
further such cases here. 

If the conditions \Bps\ and \Bpsb\ are both applied we must 
have $j=\bj=0$ and we denote these short supermultiplets
as $\B^{s,\bs}_{[k,p,q](0,0)}$. It is easy to see, from compatibility of the
formulae for $\Delta$ in \Bps\ and \Bpsb, that it is necessary that 
$s=\bs$ if they are non zero. For $s=\bs=1$, $k=p=q=0$, 
we must also have from \algQ\ $P_a | 0,0,0;0,0\rangle =0$ and of course this
corresponds to the trivial vacuum representation.
Otherwise only the two well known cases are possible,
\eqn\qB{
\B^{{1\over 4},{1\over 4}}_{[q,p,q](0,0)} \, , \qquad \Delta = p +2q \, ,
}
and
\eqn\hB{
\B^{{1\over 2},{1\over 2}}_{[0,p,0](0,0)} \, , \qquad \Delta = p \, .
}
For these multiplets the dimensions are
\eqn\dshort{\eqalign{
\dim \B^{{1\over 2},{1\over 2}}_{[0,p,0](0,0)} = {}& 2^8 \, d(0,p-2,0)
= {\ts {64\over 3}}\,  \hp (\hp-1)^2 (\hp-2) \, , \cr
\dim \B^{{1\over 4},{1\over 4}}_{[q,p,q](0,0)} = {}&
2^{12} d(q-2,p,q) + {\ts{256\over 3}} \, \hp(2\hp+1)(2\hp-1)
(2\hq+\hp-2) \, . \cr}
}
The $\half$-BPS multiplet corresponds to the chiral primary operators in
$\N=4$ theories. The details of the representations for all conformal primary
states in the  multiplet 
are well known, being first given in \Class\ and were described 
diagrammatically in a fashion similar to that for the $\N=2$ case in \OPEN. 
The representations of conformal primary states are obtained, as described above
for a long multiplet, from \NQfourf\ or \NQfour\ and their conjugates.
Beyond the simple $\half$-BPS case the numbers of representations proliferate 
hugely. We list the self conjugate representations, of the form 
$[q,p,q]_\ell \equiv [q,p,q]_{({1\over 2}\ell,{1\over 2}\ell)}$ which are 
generated by equal
numbers of $Q$ and $\bQ$ supercharges, refered to here as diagonal. This
are just the representations that should contribute to the operator
product expansion of two chiral primary operators. For the $\half$-BPS
and $\quar$-BPS cases the results for the scale dimension and associated 
representations are

\vskip 4pt
\vbox{\tabskip=0pt \offinterlineskip
\hrule
\halign{&\vrule# &\strut \ \hfil#\  \cr
height2pt&\omit&&\omit&&\omit&&\omit&&\omit&\cr
& $p$ \hfil && $p+1$\hfil && $p+2$\hfil && $p+3$\hfil
&& $p+4$\hfil &\cr
height2pt&\omit&&\omit&&\omit&&\omit&&\omit&\cr
\noalign{\hrule}
height4pt&\omit&&\omit&&\omit&&\omit&&\omit&\cr
&\ $[0,p,0]_0$ \ && \ $[1,p-2,1]_1$ \ && \ ${[0,p-2,0]_2\atop [2,p-4,2]_0}$
\ && \ $[1,p-4,1]_1$  \ && \ $[0,p-4,0]_0$ \  &\cr
height4pt&\omit&&\omit&&\omit&&\omit&&\omit&\cr}
\hrule}

\noindent
Table 1. Diagonal representations in 
$\B^{{1\over 2},{1\over 2}}_{[0,p,0](0,0)}$.

\vskip 4pt
\vbox{\tabskip=0pt \offinterlineskip
\hrule
\halign{&\vrule# &\strut \ \hfil#\  \cr
height2pt&\omit&&\omit&&\omit&\cr
& $2q+p$\hfil && $2q+p+1$\hfil  && $2q+p+2$\hfil &\cr
height2pt&\omit&&\omit&&\omit&\cr
\noalign{\hrule}
height4pt&\omit&&\omit&&\omit&\cr
& $ [q,p,q]_0$ &&  $\matrix{\ss[q-1,p,q-1]_1,[q-1,p+2,q-1]_1\cr
\ss 2[q,p,q]_1,[q+1,p-2,q+1]_1}$  &
&  $\matrix{\ss [q-2,p+2,q-2]_2,2[q-1,p,q-1]_2,[q,p-2,q]_2,[q,p,q]_2\cr
\ss [q-2,p,q-2]_0,[q-2,p+2,q-2]_0,[q-2,p+4,q-2]_0\cr
\ss 2[q-1,p,q-1]_0, 2[q-1,p+2,q-1]_0, [q,p-2,q]_0, 3[q,p,q]_0\cr
\ss 2[q+1,p-2,q+1]_0, [q+2,p-4,q+2]_0} $ &\cr
height4pt&\omit&&\omit&&\omit&\cr}
\hrule}

\vbox{\tabskip=0pt \offinterlineskip
\hrule
\halign{&\vrule# &\strut \ \hfil#\  \cr
height2pt&\omit&&\omit&\cr
& $2q+p+3$ \hfil
&& $2q+p+4$\hfil &\cr
height2pt&\omit&&\omit&\cr
\noalign{\hrule}
height4pt&\omit&&\omit&\cr
& $\matrix{\ss[q-1,p,q-1]_3,[q-3,p+2,q-3]_1,[q-3,p+4,q-3]_1\cr
\ss 2[q-2,p,q-2]_1,4[q-2,p+2,q-2]_1,[q-1,p-2,q-1]_1\cr
\ss 6[q-1,p,q-1]_1,[q-1,p+2,q-1]_1,4[q,p-2,q]_1\cr
\ss 2[q,p,q]_1,[q+1,p-4,q+1]_1,[q+1,p-2,q+1]_1}$ &
& $\matrix{\ss [q-2,p,q-2]_2,[q-2,p+2,q-2]_2,2[q-1,p,q-1]_2\cr
\ss [q,p-2,q]_2,[q-4,p+4,q-4]_0,2[q-3,p+2,q-3]_0\cr
\ss 3[q-2,p,q-2]_0,[q-2,p+2,q-2]_0,2[q-1,p-2,q-1]_0\cr
\ss 2[q-1,p,q-1]_0,[q,p-4,q]_0,[q,p-2,q]_0,[q,p,q]_0}$&\cr
height4pt&\omit&&\omit&\cr}
\hrule}

\vbox{\tabskip=0pt \offinterlineskip
\hrule
\halign{&\vrule# &\strut \ \hfil#\  \cr
height2pt&\omit&&\omit&\cr
& $2q+p+5$\hfil && $2q+p+6$ \hfil &\cr
\noalign{\hrule}
height4pt&\omit&&\omit&\cr
& $\matrix{\ss[q-3,p+2,q-3]_1,2[q-2,p,q-2]_1,
[q-1,p-2,q-1]_1,[q-1,p,q-1]_1}$  && $ [q-2,p,q-2]_0 $  &\cr
height4pt&\omit&&\omit&\cr}
\hrule}

\noindent
Table 2. Diagonal representations in $\B^{{1\over 4},{1\over 4}}_{[q,p,q](0,0)}$.

\noindent
If $p<4$ or $p,q<4$ there are cancellations which will be considered later.

If both the semi-short conditions \Bpss\ and \Bpssb\ are applied we obtain
a semi-short supermultiplet $\C^{t,\bt}_{[k,p,q](j,\bj)}$ for various $t,\bt$. 
The main possible examples, with scale dimensions for the lowest weight states, 
are given by
\eqn\qhC{\eqalign{
\C^{{1\over 4},{1\over 4}}_{[k,p,q](j,\bj)}&\, , \qquad k-q = 2(\bj-j)\, ,
\qquad \ \Delta = 2 + j+ \bj + k+ p + q \, , \cr
\C^{{1\over 2},{1\over 2}}_{[0,p,0](j,j)}&\, , \qquad\qquad\qquad\qquad\qquad
\qquad \Delta = 2 + 2j + p \, , \cr
\C^{{1\over 4},{1\over 2}}_{[k,p,0](j,\bj)}&\, , \qquad k = 2(\bj-j)\, ,
\qquad\qquad \, \Delta = 2 + j+ \bj + k + p \, , \cr
\C^{{1\over 4},{3\over 4}}_{[k,0,0](j,\bj)}&\, , \qquad k = 2(\bj-j)\, ,
\qquad\qquad \, \Delta = 2 + j+ \bj + k \, , \cr
\C^{1,1}_{[0,0,0](j,j)}&\, , \qquad\qquad\qquad\qquad\qquad
\qquad \Delta = 2 + 2j  \, , \cr}
}
where other cases are obtained under conjugation $\C^{t,\bt}_{[k,p,q](j,\bj)}
\leftrightarrow \C^{\bt,t}_{[q,p,k](\bj,j)}$. For the dimension of 
$\C^{{1\over 4},{1\over 4}}$ we have
\eqnn\dimCf
$$\eqalignno{
\dim \C^{{1\over 4},{1\over 4}}_{[k,p,q](j,\bj)}{}&  =  
a(k,p,q)\J\bcJ + b(k,p,q) \J + b(q,p,k) \bcJ + c(k,p,q) \, , \cr
a(k,p,q) ={}& 2^{14}d(k-\half , p , q- \half ) \cr
{}& - {\ts {256\over 3}}\, \hp(\hk+\hp+\hq-1)\big ( 4\hk^2 + 4\hq^2 + 8 \hk \hp
+ 8 \hp \hq + 8 \hk \hq - 8 \hk - 8 \hp - 8 \hq + 1 \big ) \, , \cr
b(k,p,q) ={}& 2^{13}d(k-\half , p , q ) \cr
{}& - {\ts {256\over 3}}\, \hp \big ( 4\hk^3 + 2 \hq^3 + 12\hk^2\hp + 8\hk\hp^2
+ 12\hk^2 \hq + 6 \hk \hq^2 + 4 \hp^2 \hq + 6 \hp \hq^2 + 18 \hk\hp\hq \cr
& \qquad\qquad{}-6\hk^2 - 4\hp^2 -3\hq^2 - 12 \hk\hp - 12 \hk \hq - 9 \hp \hq
+1 \big ) \, , \cr
c(k,p,q) ={}& 2^{12}d(k , p , q )
- {\ts {256\over 3}}\, \hp(\hk+\hp+\hq)\big ( 2\hk^2 + 2\hq^2 + 4 \hk \hp
+ 4 \hp \hq + 4 \hk \hq - 3 \big ) \,  . & \dimCf \cr}
$$
The dimensions for other examples may be obtained as special cases of \dimCf,
thus $\dim \C^{{1\over 4},{1\over 2}}_{[k,p,0](j,\bj)}$ is obtained by
setting $q=0$. In particular we may obtain
\eqn\dimCh{\hskip -0.1cm
\dim \C^{{1\over 2},{1\over 2}}_{[0,p,0](j,j)} = 
2^{12}d(0,p-1,0) \J(\J+2) +{\ts {256\over 3}}
\, \hp \, (2\hp-1)(2\hp+1) (\hp + 2\J + 2 ) \, . 
}
For $\C^{1,1}_{[0,0,0](j,\bj)}$, relaxing the constraint $j=\bj$, we have
from imposing both \Bpss\ and \Bpssb, with the basis defined by \MJ,
\eqn\conjj{
\big ( 4j\bj \, {\rm P}_{21} + 2j \, {\rm P}_{22}\, J_- - 2 \bj \,{\rm P}_{11}
\, \bJ_- - {\rm P}_{12} \, J_- \bJ_- \big ) |0,0,0;j,\bj \rangle^{\rm hw} = 0 \, ,
}
which is equivalent to the conservation equation \conj. Using \NQfourV\ we may
easily construct the full supermultiplet in this case, 
\eqnn\Nall
$$\eqalignno{\def\normalbaselines{\baselineskip20pt\lineskip3pt
 \lineskiplimit3pt}\hskip-1.5cm
\matrix{
\Delta&\cr
j{+\bj}{+2}&~~~~~&&{~~~}&&{~~~~~}&&{~~~}&&
\hidewidth[0,0,0]_{(j,\bj)}\hidewidth&&{~}&&{~}&&{~}&&\cr
&&&&&&&&\Bsw&{~~~~~~~~}&\Bse&&&&&&&\cr
j{+\bj}{+{\ts{5\over 2}}}&&&&&&
&\hidewidth[1,0,0]_{(j+{1\over 2},\bj)}\hidewidth&&&&
\hidewidth[0,0,1]_{(j,\bj+{1\over 2})}\hidewidth&&&&&&\cr
&&&&&&\Bsw&&\Bse&&\Bsw&&\Bse&&&&&\cr
j{+\bj}{+3}&&&&&\hidewidth [0,1,0]_{(j+1,\bj)} \hidewidth&&&&
\hidewidth{[1,0,1]_{(j+{1\over 2},\bj+{1\over 2})}\atop
[0,0,0]_{(j+{1\over 2},\bj+{1\over 2})}}\hidewidth&&&&
\hidewidth [0,1,0]_{(j,\bj+1)} \hidewidth&&&&\cr
&&&&\Bsw&&\Bse&&\Bsw&&\Bse&&\Bsw&&\Bse&&&\cr
j{+\bj}{+{\ts{7\over 2}}}&&
&\hidewidth[0,0,1]_{(j+{3\over 2},\bj)}\hidewidth&&&
&\hidewidth{[0,1,1]_{(j+1,\bj+{1\over 2})}\atop[1,0,0]_{(j+1,\bj+{1\over 2})}}
\hidewidth&&&&
\hidewidth{[1,1,0]_{(j+{1\over 2},\bj+1)}\atop[0,0,1]_{(j+{1\over 2},\bj+1)}}
\hidewidth&&&&\hidewidth[1,0,0]_{(j,\bj+{3\over 2})}\hidewidth&&\cr
&&\Bsw&&\Bse&&\Bsw&&\Bse&&\Bsw&&\Bse&&\Bsw&&\Bse&\cr
j{+\bj}{+4}&[0,0,0]_{(j+2,\bj)}\hidewidth&&&&
\hidewidth{[0,1,0]_{(j+{3\over 2},\bj+{1\over 2})}\atop
[0,0,2]_{(j+{3\over 2},\bj+{1\over 2})}}\hidewidth&&&&
\hidewidth {[0,2,0],[1,0,1]_{(j+1,\bj+1)}\atop
[0,0,0]_{(j+1,\bj+1)}} \hidewidth
&&&&\hidewidth{[0,1,0]_{(j+{1\over 2},\bj+{3\over 2})}\atop
[2,0,0]_{(j+{1\over 2},\bj+{3\over 2})}}\hidewidth
&&&&\hidewidth[0,0,0]_{(j,\bj+2)}\hidewidth\cr
&&\Bse&&\Bsw&&\Bse&&\Bsw&&\Bse&&\Bsw&&\Bse&&\Bsw&\cr
j{+\bj}{+{\ts{9\over 2}}}&&
&\hidewidth[0,0,1]_{(j+2,\bj+{1\over 2})}\hidewidth&&&&\hidewidth
{[0,1,1]_{(j+{3\over 2},\bj+1)}\atop[1,0,0]_{(j+{3\over 2},\bj+1)}}\hidewidth&&&&
\hidewidth{[1,1,0]_{(j+1,\bj+{3\over 2})}\atop[0,0,1]_{(j+1,\bj+{3\over 2})}}
\hidewidth&&&&\hidewidth[1,0,0]_{(j+{1\over 2},\bj+2)}\hidewidth&&\cr
&&&&\Bse&&\Bsw&&\Bse&&\Bsw&&\Bse&&\Bsw&&&\cr
j{+\bj}{+5}&&&&&\hidewidth [0,1,0]_{(j+2,\bj+1)}\hidewidth&&&&
\hidewidth{[1,0,1]_{(j+{3\over 2},\bj+{3\over 2})}\atop
[0,0,0]_{(j+{3\over 2},\bj+{3\over 2})}}\hidewidth&&&&
\hidewidth[0,1,0]_{(j+1,\bj+1)}\hidewidth&&&&\cr
&&&&&&\Bse&&\Bsw&&\Bse&&\Bsw&&&&&\cr
j{+\bj}{+{\ts{11\over 2}}}&&&&&&
&\hidewidth[1,0,0]_{(j+2,\bj+{3\over 2})}\hidewidth&&&&
\hidewidth[0,0,1]_{(j+{3\over 2},\bj+2)}\hidewidth&&&&&&\cr
&&&&&&&&\Bse&&\Bsw&&&&&&&\cr
j{+\bj}{+6}&&&{~~}&&{~~~}&&{~~}&
&\hidewidth[0,0,0]_{(j+2,\bj+2)}\hidewidth&\cr}\cr 
\noalign{\vskip-12pt} & {\Nall} \cr }$$
For $j,\bj >0$ all representations correspond to conserved currents. It is easy
to see that
\eqn\dimCone{
\dim \C^{1,1}_{[0,0,0](j,\bj)} = 2^8(\J+\bcJ+3) \, .
}
For $j=\bj=0$ this would correspond to the Konishi scalar in free theory,
all states obtained for $l,{\bar l}=1,2,3,4$ satisfy conservation equations,
for $l={\bar l}=1$ there is a conserved vector curent.
Such higher spin conserved currents (for $j=\bj=\half \ell$ they are described by
symmetric traceless rank $\ell$ tensors) are found in free field theories but
should be absent in interacting theories\foot{Higher spin currents in
conformal theories have been considered by Anselmi \Ans\ and also in \Hig.}. 
There is a crucial difference with $\N=2$ for $j=\bj=0$, described by 
\RtwoF, since the maximal spin in that case corresponds to the the conserved 
energy momentum tensor and there are no higher spin currents.

As earlier we list the self-conjugate representations obtained by
the action of equal numbers of $Q$ and $\bQ$ supercharges

\vskip 4pt
\vbox{\tabskip=0pt \offinterlineskip
\hrule
\halign{&\vrule# &\strut \ \hfil#\  \cr
height2pt&\omit&&\omit&&\omit&\cr
& $\ell+p+2$\hfil && $\ell+p+3$\hfil  && $\ell+p+4$\hfil &\cr
height2pt&\omit&&\omit&&\omit&\cr
\noalign{\hrule}
height4pt&\omit&&\omit&&\omit&\cr
& $ [0,p,0]_\ell$ &&  $\matrix{\ss 2[0,p,0]_{\ell+1},[1,p-2,1]_{\ell+1}\cr
\ss [1,p,1]_{\ell+1},[1,p-2,1]_{\ell-1}}$  &
&  $\matrix{\ss [0,p-2,0]_{\ell+2},3[0,p,0]_{\ell+2},[0,p+2,0]_{\ell+2},
2[1,p-2,1]_{\ell+2}\cr
\ss 2[1,p,1]_{\ell+2},[2,p-2,2]_{\ell+2},[0,p-2,0]_\ell\cr
\ss [0,p,0]_\ell, 4[1,p-2,1]_\ell, [2,p-4,2]_\ell, [2,p-2,2]_\ell\cr
\ss [0,p-2,0]_{\ell-2}} $ &\cr
height4pt&\omit&&\omit&&\omit&\cr}
\hrule}

\vbox{\tabskip=0pt \offinterlineskip
\hrule
\halign{&\vrule# &\strut \ \hfil#\  \cr
height2pt&\omit&&\omit&\cr
& $\ell+p+5$ \hfil
&& $\ell+p+6$\hfil &\cr
height2pt&\omit&&\omit&\cr
\noalign{\hrule}
height4pt&\omit&&\omit&\cr
& $\matrix{\ss 2[0,p,0]_{\ell+3},[1,p-2,1]_{\ell+3},[1,p,1]_{\ell+3}\cr
\ss 2[0,p-2,0]_{\ell+1},2[0,p,0]_{\ell+1},[1,p-4,1]_{\ell+1},
6[1,p-2,1]_{\ell+1}\cr
\ss [1,p,1]_{\ell+1},2[2,p-4,2]_{\ell+1}, 2[2,p-2,2]_{\ell+1},
[3,p-4,3]_{\ell+1}\cr
\ss 2[0,p-2,0]_{\ell-1},[1,p-4,1]_{\ell-1},[1,p-2,1]_{\ell-1} }$ &
& $\matrix{\ss [0,p,0]_{\ell+4},[0,p-2,0]_{\ell+2},[0,p,0]_{\ell+2}\cr
\ss 4[1,p-2,1]_{\ell+2},[2,p-4,2]_{\ell+2},[2,p-2,2]_{\ell+2}\cr
\ss [0,p-4,0]_\ell,3[0,p-2,0]_\ell,[0,p,0]_\ell\cr
\ss 2[1,p-4,1]_\ell,2[1,p-2,1]_\ell,[2,p-4,2]_\ell}$&\cr
height4pt&\omit&&\omit&\cr}
\hrule}
\vbox{\tabskip=0pt \offinterlineskip
\hrule
\halign{&\vrule# &\strut \ \hfil#\  \cr
height2pt&\omit&&\omit&\cr
& $\ell+p+7$\hfil && $\ell+p+8$ \hfil &\cr
\noalign{\hrule}
height4pt&\omit&&\omit&\cr
& $\matrix{\ss[1,p-2,1]_{\ell+3},2[0,p-2,0]_{\ell+1}\cr
\ss [1,p-4,1]_{\ell+1},[1,p-2,1]_{\ell+1}}$  && $ [0,p-2,0]_{\ell+2} $  &\cr
height4pt&\omit&&\omit&\cr}
\hrule}

\noindent
Table 3. Diagonal representations in
$\C^{{1\over 2},{1\over 2}}_{[0,p,0]({1\over 2}\ell,{1\over 2}\ell)}$.

\noindent
The rather more lengthy result for 
$\C^{{1\over 4},{1\over 4}}_{[q,p,q]({1\over 2}\ell,{1\over 2}\ell)}$ is given in
appendix C.

If we apply the semi-shortening conditions \BPSQ\ together with \sshortb,
so that $j=0$, we obtain supermultiplets denoted here by
$\D^{s,\bt}_{[k,p,q](0,\bj)}$. The main case is
\eqn\qC{
\D^{{1\over 4},{1\over 4}}_{[k,p,q](0,\bj)}\, , \qquad
k - q = 2+2\bj \, , \qquad \Delta = 1 + \bj + k + p + q \, .
}
We may also allow $\bt=\half,{3\over 4}$ if $q=0,p=q=0$ respectively but
$s =\half, {3\over 4}$ are not possible. Of course there are also
the conjugate supermultiplets ${\bar \D}^{t,\bs}_{[k,p,q](j,0)}$
for $\bs=\quar$, $t=\quar,\half,{3\over 4}$. We have
\eqn\dimDf{\eqalign{
\dim {\D}^{{1\over 4},{1\over 4}}_{[k,p,q](0,\bj)} = {}&
b(q,p,k{-1})\bcJ + c(k{-1},p,q) \, , \cr
\dim {\bar\D}^{{1\over 4},{1\over 4}}_{[k,p,q](j,0)} = {}&
b(k,p,q{-1})\J + c(k,p,q{-1}) \, , \cr}
}
with $b,c$ given in \dimCf, and
\eqnn\dimD
$$\eqalignno{
\dim {\bar \D}^{{1\over 2},{1\over 4}}_{[0,p,q](j,0)} = {}& 
2^{12} d(0,p,q-1)(\J+1)\cr
{} - {\ts {256\over 3}} \hp & \big ( (2\hp+2\hq-1)(4\hp\hq+2\hq^2 - 2\hp-2\hq-1)
\J + (\hp+\hq)(4\hp\hq+2\hq^2-3) \big ) \, , \cr
\dim {\bar \D}^{{3\over 4},{1\over 4}}_{[0,0,q](j,0)} = {}& 2^{11} d(0,0,q-1)
(\J+{\ts {3\over 2}})- 256\big ( (2\hq^2 -1)\J + \hq \big ) \, . & \dimD
\cr}
$$
In \dimCf\ or \dimDf\ and \dimD\ we have not imposed the relations which
determine $\J - \bcJ$ or $\J$ in terms of $k,p,q$ given in \qhC\ or \qC.

Just as for $\N=2$ we may decompose the long supermultiplet
$\A^\Delta_{[k,p,q](j,\bj)}$ at the unitarity threshold given by \unit.
When both conditions in \unit\ hold simultaneously we have, extending \decomp,
the semi-direct sum
\eqnn\decompa
$$\eqalignno{
\A^{2-j+3\bj+p+2q}_{[k,p,q](j,\bj)} & \big |_{k-q=2(\bj-j)} \cr
\simeq {}& \C^{{1\over 4},{1\over 4}}_{[k,p,q](j,\bj)} \oplus
\C^{{1\over 4},{1\over 4}}_{[k+1,p,q](j-{1\over 2},\bj)} \oplus
\C^{{1\over 4},{1\over 4}}_{[k,p,q+1](j,\bj-{1\over 2})} \oplus
\C^{{1\over 4},{1\over 4}}_{[k+1,p,q+1](j-{1\over 2},\bj-{1\over 2})}\, ,\cr
\A^{2+2j+p}_{[0,p,0](j,j)}  \simeq {}&
\C^{{1\over 2},{1\over 2}}_{[0,p,0](j,j)} \oplus
\C^{{1\over 4},{1\over 2}}_{[1,p,0](j-{1\over 2},j)} \oplus
\C^{{1\over 2},{1\over 4}}_{[0,p,1](j,j-{1\over 2})} \oplus
\C^{{1\over 4},{1\over 4}}_{[1,p,1](j-{1\over 2},j-{1\over 2})}\, ,&
\decompa \cr
\A^{2+2j}_{[0,0,0](j,j)}  \simeq {}&
\C^{1,1}_{[0,0,0](j,j)} \oplus
\C^{{1\over 4},{3\over 4}}_{[1,0,0](j-{1\over 2},j)} \oplus
\C^{{3\over 4},{1\over 4}}_{[0,0,1](j,j-{1\over 2})} \oplus
\C^{{1\over 4},{1\over 4}}_{[1,0,1](j-{1\over 2},j-{1\over 2})}\, , \cr}
$$
as well as, for $j=\bj=0$,
\eqn\decompb{\eqalign{
\A^{2+p+2q}_{[q,p,q](0,0)}  \simeq {}&
\C^{{1\over 4},{1\over 4}}_{[q,p,q](0,0)} \oplus
{\D}^{{1\over 4},{1\over 4}}_{[q+2,p,q](0,0)} \oplus
{\bar \D}^{{1\over 4},{1\over 4}}_{[q,p,q+2](0,0)} \oplus
\B^{{1\over 4},{1\over 4}}_{[q+2,p,q+2](0,0)}\, ,\cr
\A^{2+p}_{[0,p,0](0,0)}  \simeq {}&
\C^{{1\over 2},{1\over 2}}_{[0,p,0](0,0)} \oplus
{\D}^{{1\over 4},{1\over 2}}_{[2,p,0](0,0)} \oplus
{\bar \D}^{{1\over 2},{1\over 4}}_{[0,p,2](0,0)} \oplus
\B^{{1\over 4},{1\over 4}}_{[2,p,2](0,0)}\, ,\cr
\A^{2}_{[0,0,0](0,0)}  \simeq {}&
\C^{1,1}_{[0,0,0](0,0)} \oplus
{\D}^{{1\over 4},{3\over 4}}_{[2,0,0](0,0)} \oplus
{\bar \D}^{{3\over 3},{1\over 4}}_{[0,0,2](0,0)} \oplus
\B^{{1\over 4},{1\over 4}}_{[2,0,2](0,0)}\,  .\cr}
}
Note that, corresponding to \bc, we may identify
\eqn\BCD{
\C^{{1\over 4},{1\over 4}}_{[k,p,q](j,-{1\over 2})} \simeq
{\bar \D}^{{1\over 4},{1\over 4}}_{[k,p,q+1](j,0)} \, , \qquad
\C^{{1\over 4},{1\over 4}}_{[k,p,q](-{1\over 2},-{1\over 2})}
\simeq \B^{{1\over 4},{1\over 4}}_{[k+1,p,q+1](0,0)} \, ,
}
which ensures that \decompb\ is a special case of \decompa.

For $\N=4$ it is also of interest to consider the central extension
which is obtained by letting in \QS{a,b}\ 
$ R^i{}_{\! j} \to  R^i{}_{\! j} + \half \de^i{}_{\! j}Z$ when the
superconformal group is $SU(2,2|4)$.
The multiplet structure is unchanged but in \Bps, \Bpss\ $\Delta \to
\Delta - Z$ whereas in \Bpsb, \Bpssb\ $\Delta \to \Delta + Z$. For
short supermultiplets $\B^{s,\bs}_{[k,p,q](0,0)}$ we no longer require
$s=\bs$ or $k=q$ if $s=\bs=\quar$. As well as \qB\ and \hB\ we may
consider
\eqn\qqB{
\B^{{1\over 4},{3\over 4}}_{[k,0,0](0,0)} \, , \quad \Delta = k \, , \ \
Z = - \half k \, , \qquad \B^{{3\over 4},{1\over 4}}_{[0,0,q](0,0)} \, , \quad 
\Delta = q \, , \ \ Z =  \half q \, ,
}
with
\eqn\dimqB{
\dim \B^{{1\over 4},{3\over 4}}_{[k,0,0](0,0)} = 2^8 d(k-2,0,0) \, , \qquad
\dim \B^{{3\over 4},{1\over 4}}_{[0,0,q](0,0)} = 2^8 d(0,0,q-2) \, .
}
For the semi-short multiplets
$\C^{t,\bt}_{[k,p,q](j,\bj)}$ in \qhC\ there is no longer any constraint 
on $\bj-j$ although the formulae for $\Delta$ are unchanged. The
semi-short/short multiplet $\D^{s,\bt}_{[k,p,q](0,\bj)}$, and its
conjugate ${\D}^{\bs,t}_{[k,p,q](j,0)}$, are also no longer
restricted to $s,\bs=\quar$. For $\D^{{1\over 4},{1\over 4}}_{[k,p,q](0,\bj)}$
there is now no constraint on $k-q$ as in \qC\ although $\Delta$ is the same.
Of particular interest are the chiral multiplets $\D^{1,1}_{[0,0,0](0,\bj)}$ and 
${\bar \D}^{1,1}_{[0,0,0](j,0)}$. For the latter case consistency of
\sshort\ and  \BPSQb\ for $i=1,2,3,4$ with \algQ\ leads to the
Dirac equation, similarly to \dirac, which takes the form
\eqn\Dir{
\big ( 2j \, {\rm P}_{2\dal} - {\rm P}_{1\dal}\,  J_- \big )
|0,0,0;j,0\rangle^{\rm hw} = 0 \, ,
}
where the state $|0,0,0;j,0\rangle^{\rm hw}$ has dimension $\Delta=1+j$.

\newsec{Applications in Special Cases}

In the previous two sections we gave results for constructing supermultiplets
by the action of supercharges on the lowest dimension state when this
was assumed to belong to a general representation of $SU(2)_R$ or $SU(4)_R$.
In such a generic case the Dynkin labels for the resulting representations are 
obtained simply by adding the weights of the relevant supercharges to Dynkin 
indices for the representation to which the lowest dimension state belongs.
For particular examples the indices obtained are not necessarily positive or
zero. In such circumstances the  Racah Speiser algorithm, described more fully
in appendix B, provides a straightforward procedure for dealing with
these situations.

We first consider the $\N=2$ case. Identifying representations with their
labels $R_{(j,\bj)}$ as in section 4 the algorithm  requires that in
tensor product expansions we should identify
\eqn\Rjj{
(-R)_{(j,\bj)} = - (R-1)_{(j,\bj)} \, , \qquad
R_{(-j,\bj)} = - R_{(j-1,\bj)} \, , \qquad R_{(j,-\bj)} = - R_{(j,\bj-1)} \, .
}
Necessarily in a representation  $R_{(j,\bj)}$ if any of $R,j,\bj$ is equal to 
$-\half$ the contribution of this representation in a tensor product expansion
must be set to zero. Results following from
\Rjj\ have been used implicitly at various points in section 4. As an
illustration we may consider \NQS\ for $R=\half$ when there is a further
truncation,
\eqn\NQSs{
\half_{(0,\bj)}\  \mapright {Q} \ 0_{({1\over 2} ,\bj)}  \, , \qquad
\half_{(j,0)}\  \mapright {\bQ} \
0_{(j,{1\over 2})} \, .
}
If we apply the results to the short multiplet ${\hat \B}_R$, illustrated
in \RtwohB, for $R=\half,1$ we find
\eqn\RtwoS{\def\normalbaselines{\baselineskip16pt\lineskip3pt
\lineskiplimit3pt}
\matrix{\Delta&~~~~&{~~~}&{~~~}&&{~~~}&&{~~~}&\Delta&{~~~}&&{~~~}&&{~~~}&&{~~~}&&\cr
1&{~~}&&\half_{(0,0)}&&{~~}&&{~~~}&2&{~~}&&&&1_{(0,0)}&\cr
&&\hidewidth\Bsw\hidewidth&&\hidewidth\Bse\hidewidth&&&&&&
&&\hidewidth\Bsw\hidewidth&&\hidewidth\Bse\hidewidth\cr
{\ts{3\over 2}}&\hidewidth~~~~~0_{({1\over 2},0)}
\hidewidth&&&&\hidewidth~~0_{(0,{1\over 2})}~~\hidewidth&&&{\ts {5\over 2}}
&&&\hidewidth{1\over 2}_{({1\over 2},0)}
\hidewidth&&&&\hidewidth~~{1\over 2}_{(0,{1\over 2})}~~\hidewidth&\cr
&&&&&&&&&&\hidewidth\Bsw\hidewidth&&\hidewidth\Bse\hidewidth&
&\hidewidth\Bsw\hidewidth&&\hidewidth\Bse\hidewidth\cr
2&&&\hidewidth&&&&\hidewidth&3
&0_{(0,0)}\hidewidth&&&&0_{({1\over 2},{1\over 2})}&&&&&\hidewidth~~0_{(0,0)}\cr
&&&&&&&&&\cr
{\ts {5\over 2}}&\hidewidth~~-0_{(0,{1\over 2})}\hidewidth&&&
&\hidewidth -0_{({1\over 2},0)}\hidewidth&&&{\ts {7\over 2}}\cr
&&\hidewidth\Bse\hidewidth&&\hidewidth\Bsw\hidewidth&&&&\cr
3&&&\hidewidth -\half_{(0,0)}\hidewidth&&&&&4
&&&&&\hidewidth -0_{(0,0)}\hidewidth&&&&&\cr
r&~~~{\ts{1\over 2}}\hidewidth&&0&&-{\ts{1\over 2}}~~&&&&1&&{\ts{1\over 2}}&
&0&&-{\ts{1\over 2}}&&-1\cr}
}
The negative contributions may be cancelled by imposing the equations of
motion or conservation equations. Thus in the $R=\half$ case representing 
the contribution with $\Delta =1$ by a scalar field $\vphi^i$ and for
$\Delta ={3\over 2}$ by two fermions $\psi_\alpha, \, \chi^\dal$ it is easy
to see that $\pr^2 \vphi^i$ and $\pr^{\dal\alpha}\psi_\alpha, \, 
\pr_{\alpha\dal}\chi^\dal$ exactly correspond in dimensions and spins to the
negative representations with $\Delta =3$ and $\Delta ={5\over 2}$
respectively. Thus the requirement of the equations of motion cancels
these negative terms and this represents the usual $\N=2$ hypermultiplet.
In the case of ${\hat \B}_1$ the multiplet contains a $SU(2)_R$ singlet
current $J_{\alpha\dal}$ with $\Delta=3$, belonging to the representation
$0_{({1\over 2},{1\over 2})}$, for which the divergence
$\pr^{\dal\alpha}J_{\alpha\dal}$ corresponds to the negative contribution in
\RtwoS\ and so the conservation of the current cancels the negative piece.
This case is the simplest gauge invariant short multiplet in $\N=2$ theories.

For the semi-short modules we may note, using \Rjj,  that \NQj\ reduces to
\NQjia\ for $R=0$. The semi-short supermultiplet $\C_{R(j,\bj)}$,
displayed in \RtwoE\ reduces to $\C_{0(j,\bj)}$, as given in \RtwoF, together
with negative contributions which are cancelled by imposing conservation
equations on all fields belonging to the supermultiplet. We may also
note that in \NQj\ setting $j=-\half$ removes the first term and this 
module is identical to the short module exhibited in \NQS\ for $R\to R+\half$.
This provides an explanation of \spec. 
A further special case, following from
\NQSs, is obtained by considering \RtwoD\ for $R=\half$,
\eqn\RtwoDs{\def\normalbaselines{\baselineskip20pt\lineskip3pt
\lineskiplimit3pt}\hskip-1.5cm
\matrix{\Delta~\cr
{\bj}{+2}&&{~~~~}&&{~~~}&&
{1\over 2}_{(0,\bj)}&&{~~~}&~~~~&{~~~~}&\cr
&&&&&\Bsw&{~~~~~~~~}&\Bse&~~~~~~~~~\cr
{\bj}{+{\ts{5\over 2}}}&&&&
\hidewidth 0_{({1\over 2},\bj)}\hidewidth&&
&&\hidewidth{1_{(0,\bj{+{1\over 2}})}\atop 0_{(0,\bj{\pm{1\over2}})}}\hidewidth\cr 
&&&&&\Bse&&\Bsw&&\Bse&\cr
{\bj}{+3}&&&&&&\hidewidth
{1\over 2}_{({1\over 2},\bj{+{1\over 2}})}\hidewidth&&&&
\hidewidth{1\over 2}_{(0,\bj+1),(0,\bj)}\hidewidth&&\cr
&&&&&&&\Bse&&\Bsw&~~~~~~~\Bse\hidewidth\cr
{\bj}{+{\ts{7\over 2}}}&&&&&&
&&\hidewidth 0_{({1\over 2},\bj+1)}\hidewidth&&&&
0_{(0,\bj{+{1\over 2}})}\hidewidth\cr
r~&~~~~~\hidewidth
&&&\hidewidth~~\bj{+{\ts{3\over2}}}\hidewidth&&
\hidewidth \bj{+1}\hidewidth&&\hidewidth \bj{+\half}~~\hidewidth&
&\hidewidth \bj~\hidewidth&&
~~\bj{-{\ts{1\over 2}}}\hidewidth~&&\cr}
}
neglecting negative terms
which are removed by imposing conservation of the currents for the
representations $0_{({1\over 2},\bj)}$, if $\bj>0$,
${1\over 2}_{({1\over 2},\bj+{1\over 2})}$ and $0_{({1\over 2},\bj+1)}$. In 
addition we may obtain from \RtwoE\ and \RtwohB, using \Rjj,
\eqn\ChB{
\C_{R(-1,-1)} = {\hat \B}_{R(0,0)} \, .
}

A similar pattern emerges for $\N=4$. In this case for $SU(4)$ representations
$[k,p,q]$ in tensor product expansions we should identify
\eqn\kpq{\eqalign{
[k,p,q] = {}& - [-k-2,p+k+1,q] \, , \qquad
[k,p,q]= - [k+p+1,-p-2,q+p+1] \, , \cr
& [k,p,q]=- [k,p+q+1,-q-2] \, .\cr}
}
These identities generate a group of order 24 so there are 23 non trivial
relations which allow $[k,p,q] = \pm [k',p',q']$ with $k',p',q'\ge 0$ or
$[k,p,q] = 0$, as when any of $k,p,q$ are equal to $-1$. These rules 
allow the construction of supermultiplets described previously to be extended to
cases when the lowest state representation has low $k,p,q$. Some examples
are given in tables in appendix C.

Using \kpq\ it is easy to see that there is also a truncation of the basic short
modules for special choices of $k,p,q$. 
For $b^{1\over 4}_{[k,p,q]}$, given by \NQfourf, if $k=1$
\eqnn\NQfourfs
$$\eqalignno{
[1,p,q]_{(0,\bj)}& \  \mapright Q \
{\ss [1,p,q{-1}],[1,p{-1},q{+1}]_{({1\over 2},\bj)}
\atop \ss [0,p{+1},q]_{({1\over 2},\bj)}}
\ \mapright {Q^2} \ \matrix{\ss
[1,p{-1},q],[0,p{+1},q{-1}],[0,p,q{+1}]_{(1,\bj)}\cr
\ss [1,p,q{-2}],[1,p{-1},q],[1,p{-2},q{+2}],[0,p{+1},q{-1}],
[0,p,q{+1}]_{(0,\bj)}}\cr
\noalign{\vskip 2pt}
& \ \mapright {Q^3} \
\matrix{\ss
[0,p,q]_{({3\over 2},\bj)},[1,p{-1},q{-1}],
[1,p{-2},q{+1}]_{({1\over 2},\bj)}\cr
\ss [0,p{+1},q{-2}],2[0,p,q],[0,p{-1},q{+2}]_{({1\over 2},\bj)}} 
\ \mapright {Q^4} \ 
\matrix{\ss
[0,p,q{-1}],[0,p{-1},q{+1}]_{(1,\bj)}\cr
\ss [1,p{-2},q],[0,p,q{-1}],[0,p{-1},q{+1}]_{(0,\bj)}}\cr
\noalign{\vskip 2pt}
& \ \mapright {Q^5} \
[0,p{-1},q]_{({1\over 2},\bj)} \, .  & \NQfourfs \cr}
$$
For $b^{1\over 2}_{[0,p,q]}$, given by \NQfours, we have if $p=1$
\eqn\NQfourss{
[0,1,q]_{(0,\bj)} \  \mapright Q \
{\ss [0,1,q{-1}]_{({1\over 2},\bj)}
\atop \ss [0,0,q{+1}]_{({1\over 2},\bj)}}
\ \mapright {Q^2} \ \matrix{\ss
[0,0,q]_{(1,\bj)}\cr
\ss [0,1,q{-2}],[0,0,q]_{(0,\bj)}} \ \mapright {Q^3} \
[0,0,q{-1}]_{({1\over 2},\bj)} \, , 
}
and for $b^{3\over 4}_{[0,0,q]}$, given by \NQtf, for $q=1$ we have just,
\eqn\NQtf{
[0,0,1]_{(0,\bj)} \  \mapright Q \ [0,0,0]_{({1\over 2},\bj)} \, .
}
In each example there is one less level than for the general case. In 
\NQfourfs\ there is one further stage of shortening if $p=0$ and in
\NQfourss\ if $q=0$. In general for these modules we have
\eqn\mod{
b^{1\over 4}_{[0,p,q]} \simeq  b^{1\over 2}_{[0,p,q]} \ominus
b^{1\over 2}_{[0,p+1,q]} \, , \qquad
b^{1\over 2}_{[0,0,q]} \simeq  b^{3\over 4}_{[0,0,q]} \, .
}
For the semi-short modules $c^t_{[k,p,q]j}$ there
is no similar truncation but it easy to check, with the aid of results
flowing from \kpq, that these are nested in that 
$c^{1\over 4}_{[0,p,q]j} \simeq c^{1\over 2}_{[0,p,q]j}$ and similarly if
$p$ and then $q$ are set to zero for $t={3\over 4},1$. 
Thus \dimU\ and \dimT\ are identical for $k=0$. 
We may also note that in \NQfourS\ if $j=-\half$ the first representation
is absent and the module is identical with \NQfourf\  with $k\to k+1$, 
justifying \bc. In a similar fashion we may note that
\eqn\cbb{
c^{1\over 2}_{[0,p,q]-1} \simeq \ominus b^{1\over 2}_{[0,p,q]}  \oplus
b^{1\over 2}_{[0,p+1,q]} \, .
}

These results extend to complete supermultiplets. 
The simplest non trivial $\half$-BPS multiplets are
$\B^{{1\over 2},{1\over 2}}_{[0,1,0](0,0)}$, which has maximal shortening,
\eqnn\elem
$$\eqalignno{ 
{\def\normalbaselines{\baselineskip16pt\lineskip3pt\lineskiplimit3pt}
\matrix{\Delta&~~~~~&{~~~}&&{~~~}&&{~~~~~}&&{~~~}&&{~~}&&&&\cr
1&&{~~~}&&{~~}&&\hidewidth[0,1,0]_{(0,0)}\hidewidth&&{~~}&&{~~~}&\cr
&&&&&\Bsw&&\Bse&&&&\cr
{\ts{3\over 2}}&&&&\hidewidth~~[0,0,1]_{({1\over 2},0)}
\hidewidth&&&&\hidewidth~~[1,0,0]_{(0,{1\over 2})}~~\hidewidth&&&\cr
&&&\Bsw&&&&&&\Bse&&\cr
2&&\hidewidth[0,0,0]_{(1,0)}\hidewidth&&&&&&&&
\hidewidth[0,0,0]_{(0,1)}\hidewidth&\cr
&&&\Bse&&&&&&\Bsw&&\cr
{\ts{5\over 2}}&&&&\hidewidth~-[0,0,1]_{(0,{1\over 2})}
\hidewidth&&&&\hidewidth~-[1,0,0]_{({1\over 2},0)}~~\hidewidth&&&\cr
&&&\Bsw&&\Bse&&\Bsw&&\Bse&&\cr
3&-[0,0,0]_{({1\over 2},{1\over 2})}\hidewidth&&&{~~}&
&\hidewidth-[0,1,0]_{(0,0)}\hidewidth&&
&-[0,0,0]_{({1\over 2},{1\over 2})}\hidewidth&\cr
&&&{\vphantom\Bsw}&&&&&&&&\cr
{7\over 2}\cr
&&&{\vphantom\Bsw}&&&&&&&&\cr
4&&\hidewidth[0,0,0]_{(0,0)}\hidewidth&&&&&&&&
\hidewidth[0,0,0]_{(0,0)}\hidewidth&\cr}}
& \elem \cr}
$$
and $\B^{{1\over 2},{1\over 2}}_{[0,2,0](0,0)}$
\eqn\BPSf{\def\normalbaselines{\baselineskip20pt\lineskip3pt
 \lineskiplimit3pt}\hskip-1.0cm
\matrix{
\noalign{\vskip-6pt}
\Delta~\cr
2~&~~~~~&&{~~~}&&{~~~~~}&&{~~~}&&\hidewidth[0,2,0]_{(0,0)}\hidewidth&
&{~}&&{~}&&{~}&&\cr
&&&&&&&&\Bsw&{~~~~~~~~}&\Bse&&&&&&&\cr
{5\over 2}~&&&&&&&\hidewidth[0,1,1]_{({1\over 2},0)}\hidewidth&&&&
\hidewidth[1,1,0]_{(0,{1\over 2})}\hidewidth&&&&&&\cr
&&&&&&\Bsw&&\Bse&&\Bsw&&\Bse&&&&&\cr
3~&&&&&\hidewidth {[0,1,0]_{(1,0)}\atop[0,0,2]_{(0,0)}}\hidewidth&&&&
\hidewidth [1,0,1]_{({1\over 2},{1\over 2})}\hidewidth&&&&
\hidewidth{[0,1,0]_{(0,1)}\atop[2,0,0]_{(0,0)}}\hidewidth&&&&\cr
&&&&\Bsw&&\Bse&&\Bsw&&\Bse&&\Bsw&&\Bse&&&\cr
{\ts{7\over 2}}~&&&\hidewidth[0,0,1]_{({1\over 2},0)}\hidewidth&&&
&\hidewidth[1,0,0]_{(1,{1\over 2})}\hidewidth&&&&
\hidewidth[0,0,1]_{({1\over 2},1)}
\hidewidth&&&&\hidewidth[1,0,0]_{(0,{1\over 2})}\hidewidth&&\cr 
&&\Bsw&&&&&&\Bse&&\Bsw&&&&&&\Bse&\cr
4~&[0,0,0]_{(0,0)}\hidewidth&&&&&&&&
\hidewidth {[0,0,0]_{(1,1)}\atop-[1,0,1]_{(0,0)}}\hidewidth 
&&&&&&&&\hidewidth[0,0,0]_{(0,0)}\hidewidth\cr
&&&&&&&&\Bsw&&\Bse&&&&&&&\cr
{\ts{9\over 2}}~&&&&&&&\hidewidth-[1,0,0]_{({1\over 2},0)}\hidewidth&&&&
\hidewidth-[0,0,1]_{(0,{1\over 2})}\hidewidth&&&&&&\cr
&&&&&&&&\Bse&&\Bsw&&&&&&&\cr
5~&&&&&&&&&
\hidewidth -[0,0,0]_{({1\over 2},{1\over 2})}\hidewidth&&&&&&&&\cr}
}
In \elem,
as in the $\N=2$ case, the negative contributions are cancelled by imposing
equations of motion on all fields (for the representation $[0,0,0]_{(1,0)}$
with $\Delta=2$ corresponding to a field $f_{\alpha\beta}= f_{\beta\alpha}$ 
there is associated vector $V_{\alpha\dal}$, $[0,0,0]_{({1\over 2},{1\over 2})}$,
with $\Delta=3$ which contributes negatively and a scalar $\vphi$, 
$[0,0,0]_{(0,0)}$, with $\Delta=4$, the equation of motion is
$\pr^{\dal\alpha}f_{\alpha\beta}=0$ and we may note that
$f \, \mapright{\pr} \, V  \, \mapright{\pr}\, \vphi$ with $\pr\pr f=0$). 
For each column there are zero total degrees of freedom reflecting the 
vanishing of the dimension formula \dshort\ in this case. This of course 
corresponds to the elementary multiplet in $\N=4$ gauge theories. In \BPSf\ 
it is necessary to impose conservation equations for
the $[1,0,1]_{({1\over 2},{1\over 2})}$ $SU(4)$ current, 
as well as the energy momentum tensor, $[0,0,0]_{(1,1)}$, and spinor 
currents, $[1,0,0]_{(1,{1\over 2})}$ and $[0,0,1]_{({1\over 2},1)}$. For $p>2$ 
there are no such constraints on the corresponding fields since no negative 
representations appear in $\B^{{1\over 2},{1\over 2}}_{[0,p,0](0,0)}$. 
As a consequence of \NQfourfs\ we may also note that
$\B^{{1\over 4},{1\over 4}}_{[1,p,1](0,0)}$ is shortened from the generic
$\quar$-BPS case, the scale dimension ranges just from $p+2$ to $p+7$.
Using \NQfourfs\ we also obtain the singleton multiplet
$\B^{{1\over 4},{3\over 4}}_{[1,0,0](0,0)}$ is described by
\eqn\sing{\def\normalbaselines{\baselineskip20pt\lineskip3pt
\lineskiplimit3pt}\hskip-1.5cm
\matrix{\Delta~\cr
1&&&{~~~}&&{~~~~~}&&{~~~}&&[1,0,0]_{(0,0)}&\cr
&&&&&&&&\Bsw&{~~~~~~~~}&\Bse&~~~~~~~~~\cr
{\ts{3\over 2}}&&&&&&&\hidewidth
[0,1,0]_{({1\over 2},0)}
\hidewidth&&&&\hidewidth [0,0,0]_{(0,{1\over 2})}\hidewidth\cr
&&&&&&\Bsw&&&&&&&\cr
2&&&&&\hidewidth [0,0,1]_{(1,0)}\hidewidth&&&&\cr
&&&&\Bsw&&\cr
{\ts{5\over 2}}&&&\hidewidth [0,0,0]_{({3\over 2},0)}\hidewidth&&&\cr}
}
with all fields required to satisfy equations of motion. The singleton
representations $\B^{{1\over 2},{1\over 2}}_{[0,1,0](0,0)}$,
$\B^{{1\over 4},{3\over 4}}_{[1,0,0](0,0)}$ and
$\B^{{3\over 4},{1\over 4}}_{[0,0,1](0,0)}$ are building blocks for all
supermultiplets. We may also note that
$\B^{{1\over 4},{3\over 4}}_{[2,0,0](0,0)}$ contains conserved currents
for the representations $[0,1,0]_{({1\over 2},{1\over 2})}, \,
[0,0,1]_{(1,{1\over 2})}$ and $[0,0,0]_{({3\over 2},{1\over 2})}$.

For the semi-short multiplets $\C^{t,\bt}_{[k,p,q](j,\bj)}$ we have verified 
that $\C^{{1\over 4},{1\over 4}}_{[0,p,0](j,\bj)}
\simeq \C^{{1\over 2},{1\over 2}}_{[0,p,0](j,\bj)}$ and also
$\C^{{1\over 2},{1\over 2}}_{[0,0,0](j,\bj)}$
reduces to $\C^{1,1}_{[0,0,0](j,\bj)}$, shown in \Nall, together with
negative terms corresponding to the divergences of all currents with
spins $j,\bj >0$. In a similar fashion we may show that
${\bar \D}^{{1\over 2},{1\over 2}}_{[0,1,0](j,0)}$ contains higher spin
conserved currents with the representations $[1,0,0]_{(j+{1\over 2},{1\over 2}),
(j+{1\over 2},1),(j+2,{1\over 2})}$, $[0,1,0]_{(j+{1\over 2},{1\over 2}),
(j+1,{1\over 2})}$, $[2,0,0]_{(j+{1\over 2},{1\over 2})}$,
$[0,0,0]_{(j,1),(j+{3\over 2},{1\over 2}),(j+2,1)}$,
$[0,0,1]_{(j+1,{1\over 2}),(j+{3\over 2},1)}$, 
$[1,0,1]_{(j+{3\over 2},{1\over 2})}$ and $[1,1,0]_{(j+1,{1\over 2})}$, in
each case with scale dimension $j+\bj+2$ as is necessary for consistency with
conformal invariance. The relation \BCD\ demonstrates that semi-short
multiplets are connected to short multiplets in special cases, in addition
from \cbb\ we may note that
\eqn\CBBB{
\C^{{1\over 2},{1\over 2}}_{[0,p,0](-1,-1)} \simeq
\B^{{1\over 2},{1\over 2}}_{[0,p,0](0,0)} \ominus 2
\B^{{1\over 2},{1\over 2}}_{[0,p+1,0](0,0)} \oplus
\B^{{1\over 2},{1\over 2}}_{[0,p+2,0](0,0)} \, .
}

Allowing for a central extension there are several cases leading to 
quite small multiplets. These  have been discussed in 
\Gun\foot{In detail for the first paper in
\Gun\ Table 1 corresponds to $\B^{{1\over 2},{1\over 2}}_{[0,1,0](0,0)}$,
Table 2 to $\B^{{1\over 4},{3\over 4}}_{[1,0,0](0,0)}$,
Table 3 to $\B^{{3\over 4},{1\over 4}}_{[0,0,1](0,0)}$,
Table 4 to ${\bar \D}^{1,1}_{[0,0,0](0,0)}$,
Table 5 to ${\D}^{1,1}_{[0,0,0](0,0)}$,
Table 6 to ${\bar \D}^{1,1}_{[0,0,0](j-1,0)}$,
Table 7 to $\D^{1,1}_{[0,0,0](0,j-1)}$,
Table 8 to $\B^{{1\over 4},{1\over 2}}_{[1,1,0](0,0)}$,
Table 9 to $\B^{{1\over 2},{1\over 4}}_{[0,1,1](0,0)}$,
Table 10 to ${\bar \D}^{{1\over 2},{1\over 2}}_{[0,1,0](j-1,0)}$,
Table 11 to ${\D}^{{1\over 2},{1\over 2}}_{[0,1,0](0,j-1)}$,
Table 12 to $\C^{1,1}_{[0,0,0](j_L-1,j_R-1)}$, while
Table 13 corresponds to $\B^{{1\over 2},{1\over 2}}_{[0,2,0](0,0)}$.
In the second paper among the new multiplets
Table 6 corresponds to $\B^{{1\over 4},{3\over 4}}_{[2,0,0](0,0)}$,
Table 7 to $\B^{{3\over 4},{1\over 4}}_{[0,0,2](0,0)}$,
Table 8 to $\B^{0,1}_{[0,0,0](0,0)}$ and
Table 9 to $\B^{1,0}_{[0,0,0](0,0)}$.}.

\newsec{Discussion}

An initial motivation for this work was to identify supermultiplets with the
sets of operators that arise in the operator product expansion analysis
of four point functions of the $\N=4$ superconformal $SU(N)$ gauge theory.
In the analysis in \OPENP\ of superconformal Ward identities for
the four point function for $[0,2,0]$ chiral primary operators we showed
how the solution of these identities required the presence of sets of
operators in the operator product expansion with related scale dimensions
and spins. The spectra in each case match exactly with what we have termed
diagonal states, formed by equal numbers of $Q$ and $\bQ$ supercharges acting
on a superconformal primary state, for particular short and semi-short
multiplets\foot{In particular in \OPENP\ the states labelled by
$\B_\ell$ correspond to
$\C^{1,1}_{[0,0,0]({1\over 2}\ell-1,{1\over 2}\ell-1)}$, $\D_{\ell}$ to
$\C^{{1\over 2},{1\over 2}}_{[0,2,0]({1\over 2}\ell-1,{1\over 2}\ell-1)}$,
$\E_{\ell}$ to
$\C^{{1\over 4},{1\over 4}}_{[1,0,1]({1\over 2}\ell-1,{1\over 2}\ell-1)}$,
${\hat B}_0$ to $\B^{{1\over 2},{1\over 2}}_{[0,2,0](0,0)}$,
${\hat D}_0$ to $\B^{{1\over 2},{1\over 2}}_{[0,4,0](0,0)}$ and
$\E_0$ to $\B^{{1\over 4},{1\over 4}}_{[2,0,2](0,0)}$.}. This is as
expected according to $U(1)_Y$ bonus symmetry \refs {\Intr,\Bon}. Assuming a 
non renormalisation theorem such that there is only a single undetermined
function of two variables in the solution of the Ward identities then
there must be present in the operator product expansion analysis 
semi-short multiplets $\C^{{1\over 2},{1\over 2}}_{[0,2,0](j,j)}$, $
j=0,1,2\dots$ and $\C^{{1\over 4},{1\over 4}}_{[1,0,1](j,j)}$, 
$j=\half,{3\over 2},\dots$. A similar pattern
emerges in an extension \Nirschl\ of this analysis to next extremal \Nextreme\
four point correlators of chiral primary operators. Assuming non 
renormalisation theorems the operator product expansion requires the
presence of semi-short multiplets $\C^{{1\over 2},{1\over 2}}_{[0,p,0](j,j)}$
for all $p=2,3,\dots$ and $j=0,\half,1,\dots$ (a related discussion of next
extremal correlators is also given in \bpsN).

More generally it remains to be understood under what circumstances short
or semi-short multiplets are protected against renormalisation effects
giving rise to anomalous dimensions. If these are to occur it is necessary 
the multiplet is part of a possible long multiplet. As a consequence
of \decompa\ and \decompb\ the only multiplets satisfying shortening conditions
for the action of both the $Q$ and $\bQ$ supercharges that cannot form part of
a long multiplet and therefore must be protected are
\eqn\prot{
\B^{{1\over 2},{1\over 2}}_{[0,p,0](0,0)} \, , \qquad
\B^{{1\over 4},{1\over 4}}_{[1,p,1](0,0)} \, .
}
For other cases there is the possibility of joining with other multiplets
to form a long multiplet so that such multiplets by themselves cannot be
guaranteed to be protected against renormalisation effects.
The number of states in the free and interacting theories should
be the same but the way in which they are arranged into supermultiplets
may change for $g$ non zero. Given the results \decompa\ and \decompb\ 
and a knowledge of the free theory supermultiplet struture then, although
individual semi-short multiplets can be part of more than one
long multiplet, it may be possible, following \Das, to set up index theorems 
which would ensure, for a non zero index, the existence of other protected 
supermultiplets. On the other hand a $\quar$-BPS multiplet 
$\B^{{1\over 4},{1\over 4}}_{[q,p,q](0,0)}$ with $q\ge 2$ can only be part of a
unique long multiplet $\A^{p+2q-2}_{[q-2,p,q-2](0,0)}$. At present the knowledge
of what protected multiplets may appear in $\N=4$ SYM is rather limited,
for a discussion of $\quar$-BPS operators in field theory see \Hoksemi. 
More generally it seems probable that in any interacting
theory multiplets containing higher spin currents, such as
$\C^{1,1}_{[0,0,0](j,j)}$, for any $j$, illustrated in \Nall, should be absent
(this was implicitly assumed in \OPENP). This rule of course applies
in the simplest case $j=0$ for the Konishi multiplet, the Konishi anomaly
affects not just the usual spin 1 vector current but also currents in the same
multiplet with spin 2,3,4. In general protected operators belonging
to semi-short multiplets are essentially multi-trace operators
formed from products of single trace $\half$-BPS chiral primary operators \HH,
which are at least protected against perturbative corrections
to leading order in $N$ \NonP.

We may also note that the conditions imposed on the action of the $Q$
and $\bQ$ supercharges, such as \Ttwoj, \Ttwo\ and \Ttwobj\ for $\N=2$ 
and their extensions to $\N=4$, on the superconformal primary states for 
semi-short multiplets have a direct correspondence in terms of the action 
of the $D$ and $\bar D$ spinor derivatives on superfields \ESok, where they
are referred to as current like protected operators. In the simplest
case the $\N=1$ Konishi scalar field $\O$ satisfies $D^2 \O = {\bar D}{}^2
\O = 0$ in free theory, analogous to the semi-short conditions
for $j=\bj=0$, but there is the well known anomaly to ${\rm O}(g^2)$.

A remaining question is whether there are perhaps any additional shortening
conditions other than those discussed here. The semi-shortening
conditions which arise at unitarity bounds are clearly related to null
states so that the cases considered masy be expected to be complete.
 As a partial test of completeness for $\N=2$ we have considered
the decomposition of a $\N=4$ $\half$-BPS supermultiplet in terms of $\N=2$
supermultiplets when we obtain
\eqn\decompNN{\eqalign{
\B^{{1\over 2},{1\over 2}}_{[0,p,0](0,0)} \simeq {}&
(p+1)\, {\hat \B}_{{1\over 2}p} 
\oplus \E_{p(0,0)}\oplus {\bar \E}_{{-p}(0,0)} \cr
&{} \oplus (p-1)\, {\hat \C}_{{1\over 2}p{-1}(0,0)} 
\oplus p \big ( \D_{{1\over 2}(p-1)(0,0)} \oplus 
{\bar \D}_{{1\over 2}(p-1)(0,0)} \big ) \cr
&{}\oplus \bigoplus_{k=1}^{p-2} (k+1) \big ( \B_{{1\over 2}k,p-k(0,0)} \oplus
{\bar \B}_{{1\over 2}k,k-p(0,0)} \big ) \cr
&{}\oplus \bigoplus_{k=0}^{p-3} (k+1) \big ( \C_{{1\over 2}k,p-k-2(0,0)} \oplus
{\bar \C}_{{1\over 2}k,k-p+2(0,0)} \big ) \cr
&{}\oplus \bigoplus_{k=0}^{p-4} \bigoplus_{l=0}^{p-k-4} (k+1) \,
\A^p_{{1\over 2}k,p-k-4-2l(0,0)} \, .  \cr}
}
All possible supermultiplets discussed in section 4 arise.
We may note that, as expected, for $p=2$ the decomposition contains the
$\N=2$ energy momentum tensor multiplet ${\hat \C}_{0(0,0)}$.

\vfill\eject
\noindent
{\bf Acknowledgements}

HO would like to thank Massimo Bianchi, Gleb Arutyunov and Emery Sokatchev
for useful conversations at various stages during this work, in particular to
Massimo Bianchi for emphasising the importance of understanding
possible decompositions of long supermultiplets. FAD would like to
thank Michael Nirschl for helpful conversations and Hendryk Pfeiffer for
computer assistance in the initial stages of this work. He would also like 
to thank  Trinity College, Cambridge and the EPSRC for financial support.
\vfil\eject

\appendix{A}{Two Point Functions}

To construct the conformal two point functions from the results of section 2
we first note that from the definition \defst\ and taking the field
$\O_I(t,r n)$, for $n_\ha$ a unit vector, we may obtain
\eqn\DO{
| \Delta \r_I = \O_I (i,0) | 0 \r \, 2^\Delta \, , \qquad
{}_{\bar I} \l \Delta | = 2^\Delta \l 0 | \O_{\bar I} (-i,0) \, ,
}
which may be derived in the same fashion as other results obtained later. 
We then consider $\exp(-i\beta n_\ha \E^+{}_{\!\!\ha}- i\alpha \E^+{}_{\!\!d} ) 
|\Delta \r_I$.  To obtain an explicit expression we first consider
\eqn\EA{
e^{-i \alpha \E^+{}_{\!\!d}} \O_I (i,0) | 0 \r 
= \O_I (i\tau,0) | 0 \r f(\tau) \, .
}
{}From this we may derive
\eqn\diffa{\eqalign{
{\d \over \d \alpha} \big ( \O_I (i\tau,0) | 0 \r f(\tau) \big )
= {}& - i [ \E^+{}_{\!\!d}, \O_I (i\tau,0) ]  | 0 \r \, f(\tau) \cr
= {}& - \half \big ( (1+\tau)^2 \pr_\tau + 2 \Delta (1+\tau) \big ) 
\O_I (i\tau,0) | 0 \r \, f(\tau) \, , \cr}
}
using \tO. By solving the differential equation we easily obtain
\eqn\resa{
\tau+1 = {2\over 1+\alpha} \, , \qquad f(\tau) = 
\big ( \half(\tau+1)\big)^{2\Delta} \, .
}
Similarly for
\eqn\EB{
e^{-i \beta n_\ha \E^+{}_{\!\!\ha}}  \O_I (i\tau_0,0) | 0 \r
= \O_J (i\tau, rn) | 0 \r  \, g^J{}_{\! I} (\tau,r) \, ,
}
and the following equations are now obtained
\eqn\dB{\eqalign{
{\d \over \d \beta}( r - i \tau ) = {}&- 
\half \big ( r - i (\tau+1)\big )^2 \, , \cr
{\d \over \d \beta} g^J{}_{\! I} (\tau,r) = {}& - \Delta r \,
g^J{}_{\! I} (\tau,r) + (\tau+1) (s_{0\ha} n_\ha)^J{}_{\! K}
g^K{}_{\! I} (\tau,r) \, . \cr}
}
Hence we get
\eqnn\resAB
$$\eqalignno{
\exp(-i\beta n_\ha \E^+{}_{\!\!\ha}{- i}\alpha \E^+{}_{\!\!d} ) |\Delta \r_I
{}& = \O_J (i(\tau{-1}),rn) | 0 \r \big ((\exp(\tan^{-1} \! 
{\ts{r\over \tau}} \, 2s_{0\ha} n_\ha )\big ){}^J{}_{\! I} 
\big (\half(\tau^2+r^2)\big )^\Delta \, , \cr
r = {}& {2\beta \over \beta^2 + (1+\alpha)^2} \, , \qquad
\tau = {2(1+\alpha) \over \beta^2 + (1+\alpha)^2}  \, . & \resAB \cr}
$$
Since ${}_{\bar I} \l \Delta | 
\exp(-i\beta n_\ha \E^+{}_{\!\!\ha}{- i}\alpha \E^+{}_{\!\!d} ) |\Delta \r_I
= N_{\bar I I}$, from \HE\ and \norm, we then determine the two point function 
in the form
\eqn\twop{\eqalign{
\l 0 | \O_{\bar I} (-i,0) \, \O_I (i(\tau{-1}),rn) | 0 \r
= {} &\l 0 | \O_{\bar I} (0) \, \O_I (x) | 0 \r
=  \I_{\bar I I} (x) \, {1 \over (x^2)^\Delta } \, , \cr 
& x^a=(i\tau,rn) \, , \quad x^2= \tau^2 + r^2 \, , \cr}
}
for
\eqn\defI{
\I(x) = N \exp({- \tan^{-1}} \!{\ts{r\over \tau}} \, 2s_{0\ha} n_\ha )
= \exp({\tan^{-1}} \!{\ts{r\over \tau}} \, 2{\bar s}_{0\ha} n_\ha ) \, N \, ,
}
using \Ns\ and \defss.

As an illustration for a four dimensional spinor field, $\O_I= {\bar \psi}_\dal$,
${\bar \O}_{\bar I} = \psi_\alpha$, then $s_{ab}=\half i \bsi_{[a}\si_{b]}$
and we may take $N=\si_0$ so that \defI\ gives
\eqn\Ispin{
\I(x) = - i {1\over (x^2)^{1\over 2}} \, \si_a x^a \, .
}
Alternatively for a vector field $V_a$ we have $(s_{ab})^c{}_{\! d} =
i(\eta_{ad} \de^c{}_{\! b} - \eta_{bd} \de^c{}_{\! a} ) $ then 
using $N_{cd}=\de_{cd}$ from \defI\ we have
\eqn\Ivec{
\I_{cd}(x) = \eta_{cd} - 2 \, {x_c x_d \over x^2} \, ,
} 
with $x_c$ determined from \twop. Of course
\Ivec\ is just the expected result given by the inversion tensor.

\appendix{B}{Racah Speiser Algorithm}

We describe here in outline the Racah Speiser algorithm \RSpeis\ which provides
a succinct and simple way of decomposing tensor products\foot{For
a more pictorial description for the case of $SU(3)$ see \Swart\ and
in a mathematical context chapter VI, 24.4, exercise 9 in \Hump.}. A 
representation $R_{\underline \Lambda}$ with Dynkin labels 
${\underline \Lambda}= [\lambda_1, \dots, \lambda_r]$ is characterised by a 
highest weight vector $\Lambda$,
\eqn\HW{
\Lambda = \sum_{i=1}^r \lambda_i w_i \in W \quad \Leftrightarrow \quad
\lambda_i \in {\Bbb Z}, \ \lambda_i \ge 0 \, ,
}
where $w_i$ are the fundamental weights which are dual to the simple roots 
$\alpha_i$, $2 w_i \cdot \alpha_j / \alpha_j{}^{\! 2} = \de_{ij}$. 
For a tensor product $R_\uLambda \otimes R_{\uLambda'}$ we consider the set of 
weights $V_{\uLambda'} =
\{\lambda\}$ for all states in the representation space for $R_{\uLambda'}$
($\lambda = \sum_{i} \lambda_{i} w_i$ with $ \lambda_{i}$ positive or
negative integers and, allowing for multiplicity, there are $d_{\Lambda'}$
weights in $V_{\uLambda'}$ for $d_{\Lambda'}$ the 
dimension of $R_{\uLambda'}$). 
Then the Racah Speiser algorithm can be paraphrased by
expressing the tensor product as
\eqn\RRR{
R_\uLambda \otimes R_{\uLambda'} \simeq \sum_{\lambda \in V_{\uLambda'}} 
R_{\uLambda + {\underline \lambda}} = \sum_n R_{\uLambda_n} \, ,
\quad \Lambda_n \in W \, ,
}
where we require
\eqn\Rsi{
R_{{\underline \lambda}} = \hbox{sign}(\si) R_{{\underline \lambda}^\si} \, , \qquad
\lambda^\si = \si ( \lambda + \rho ) - \rho \, ,
}
for $\si$ an element of $\W$, the Weyl group, the symmetry group of the 
root space which permutes the different possible bases of simple roots,
and the Weyl vector $\rho$ is defined by
\eqn\drho{
\rho = \half \sum_{\alpha \in \Phi^+} \alpha = \sum_{i=1}^r w_i \, ,
}
for $\Phi^+$ the set of positive roots. 
$\W$ is generated by the reflections corresponding to the simple roots
\eqn\sW{
\sigma_i \lambda = \lambda - {2 \lambda \cdot \alpha_i \over \alpha_i{}^{\! 2}} \, 
\alpha_i \, ,
}
and $\hbox{sign}(\sigma)= (-1)^{N_\si}$
if $\sigma$ is formed from the product of ${N_\si}$ elementary reflections \sW.

For $\lambda = \sum_i \lambda_i w_i$ then $\lambda \in \pr W$ if $\lambda_i=0$
for some $i$, in which case $\si_i \lambda = \lambda$. For any weight vector 
$\lambda$ there is a $\si \in \W$ such that $\si \lambda \in W$.
If $\si \lambda \notin \pr W$ then $\si$ is unique, if $\si \lambda \in \pr W$ 
then there is a reflection $\si_\lambda$, with $ \hbox{sign}(\si_\lambda) = -$,
such that $\si_\lambda \lambda = \lambda$. Consequently if there is a 
$\si$ so that $\si(\lambda + \rho) \in W\setminus \pr W $ then in \Rsi, 
for all such $\lambda$, $\lambda^\si \in W$ is unique. Otherwise there is
a reflection $\si$, with $\hbox{sign}(\sigma)= -$,
such that $\lambda^\si = \lambda$ and in this case from \Rsi
\eqn\Rz{
R_{{\underline \lambda}} = - R_{{\underline \lambda}} = 0 \, .
}

Hence in \RRR, for any weights $\lambda \in  V_{\uLambda'}$,  we may let
$R_{\uLambda + {\underline \lambda}}  \to  \pm R_{\uLambda_n}$ 
for some unique $\Lambda_n \in W$ or alternatively
$R_{\uLambda + {\underline \lambda}} \to 0$. If $\Lambda + \lambda \in W $
then we may take, for some $n$, $\Lambda_n = \Lambda + \lambda$. 
Unless this is true for all  $\lambda \in V_{\uLambda'}$ there is a cancellation 
of representations in the sum over $\{\Lambda_n\}$ on the
right hand side of \RRR\ due to $\hbox{sign}(\si) = - $ for some cases.

It is easy to check that the algorithm is compatible with dimensions. Let
\eqn\Wdim{
d_{\underline \lambda} = \prod_{\alpha \in \Phi^+} {(\lambda+\rho)\cdot \alpha
\over \rho \cdot \alpha} \, .
}
For any $\Lambda\in W$ $d_\uLambda$ is just the Weyl formula the 
dimension of the representation $R_\uLambda$ with Dynkin labels $\uLambda$. 
For an elementary reflection as in \sW,
\eqn\rd{
\si_i (\Phi^+ \setminus \alpha_i) = \Phi^+ \setminus \alpha_i \, , \quad 
\si_i \alpha_i = - \alpha_i \quad \Rightarrow \quad 
d_{{\underline \lambda}^{\si_i}} = - d_{\underline \lambda} \, ,
}
since the scalar product is invariant under reflections.  In general then
$ d_{{\underline \lambda}^{\si}} = \hbox{sign}(\si) d_{\underline \lambda}$
and it is straightforward to see that (when \Rz\ holds $d_{\underline \lambda} =0$), 
\eqn\dimRR{
\sum_{\lambda \in V_{\uLambda'}} d_{\uLambda + {\underline \lambda}} 
= \sum_n d_{\uLambda_n} \, .
}

For illustration for the simplest case of $SU(2)$ where, following
convention, the representations are $R_j$, $j=0,\half,1,\dots$,  the
associated weights $V_j = \{-j, -j{+1}, \dots , j \}$, and the Weyl group
$\W = \Bbb Z_2$, where for $m\in V_j$, $\si m = -m$. In this case \RRR\ gives
\eqn\RRtwo{
R_j \otimes R_{j'} \simeq \cases{ \sum_{m \in V_{j'}} R_{j+m} \, ,  &
$j'\le j$ , \cr
\sum_{m \in V_{j'}} R_{j+m} = \sum_{m \in V_{j}} R_{j'+m} , &
$j'\ge j$  , \cr}
}
using from \Rsi,
\eqn\Rtwo{
R_{-n} = - R_{n-1} \, , \qquad R_{-{1\over2}} = 0 \, .
}
Of course this is just the standard result for tensor products in $SU(2)$.

Instead of \sW\ and \drho, defining for ${\underline \lambda} =
[\dots , \lambda_j, \dots ]$ for $\lambda = \sum_j \lambda_j w_j$, we may write
\eqn\sw{
\si_i {\underline \lambda} = [ \dots , \lambda_j - \lambda_i K_{ij}, \dots ] \, ,
\qquad {\underline \rho} = [ 1,1,\dots , 1] \, ,
}
for $[K_{ij}]$ the Cartan matrix. For the case of $SU(4)$ then with the Cartan
matrix \Car\ and with $\Rsi$ the elementary reflections give
\eqnn\sifour
$$\eqalignno{
[\lambda_1,\lambda_2,\lambda_3]^{\si_1} = {}& 
[-\lambda_1-2, \lambda_1+\lambda_2+1,\lambda_3] \, , \quad
[\lambda_1,\lambda_2,\lambda_3]^{\si_3}  =  
[\lambda_1, \lambda_2+\lambda_3+1 , -\lambda_3-2 ] \, ,\cr
[\lambda_1,\lambda_2,\lambda_3]^{\si_2}  = {}&
[\lambda_1+\lambda_2+1, -\lambda_2-2 , \lambda_2+\lambda_3+1] \, . & \sifour \cr}
$$
For this case $\W=\S_4$, the permutation group on four objects 
(in this case we have the relations $\si_i{}^{\!2}=e$, $\si_1\si_3 = \si_3\si_1,
\ \si_1 \si_2 \si_1 = \si_2 \si_1 \si_2, \ \si_3 \si_2 \si_3 = 
\si_2 \si_3 \si_2$). Note that
\eqn\reflect{\eqalign{
[\lambda_1,\lambda_2,\lambda_3]^{\si_1\si_2\si_1} = {}&
[-\lambda_2-2,-\lambda_1-2,\lambda_1+\lambda_2+\lambda_3+2] \, , \cr
[\lambda_1,\lambda_2,\lambda_3]^{\si_3\si_2\si_3} = {}&
[\lambda_1+\lambda_2+\lambda_3+2,-\lambda_3-2,-\lambda_2-2] \, , \cr
[\lambda_1,\lambda_2,\lambda_3]^{\si_1\si_2\si_3\si_2\si_1} ={}&
[-\lambda_2-\lambda_3-3, \lambda_2, -\lambda_1-\lambda_2-3 ] \, . \cr}
}
{}From \sifour\ and \reflect\ it is clear that we have 
${\underline\lambda}^{\si} = {\underline\lambda}$ with $\hbox{sign}\si
= -$ if one of the following conditions apply,
\eqn\condit{
\lambda_1 = -1 \, , \quad \lambda_2 = -1 \, , \quad\lambda_3 = -1 \, , \quad
\lambda_1+\lambda_2 = -2 \,  , \quad \lambda_2+\lambda_3 = -2 \, , \quad
\lambda_1+\lambda_2 +\lambda_3 = - 3 \, .
}
These correspond exactly to the zeros of the dimension formula \dimfour,
as expected since these cases correspond to zero contributions in the tensor
product expansion.

\vfil\eject

\appendix{C}{Tables of Representations}

\vbox{\tabskip=0pt \offinterlineskip
\hrule
\halign{&\vrule# &\strut \ \hfil#\  \cr
height2pt&\omit&&\omit&\cr
& $\ell+2q+p+2$\hfil && $\ell+2q+p+3$\hfil  &\cr
height2pt&\omit&&\omit&\cr
\noalign{\hrule}
height4pt&\omit&&\omit&\cr
& $ [q,p,q]_\ell~~~$ &
&  $\matrix{\ss [q-1,p,q-1]_{\ell+1},[q-1,p+2,q-1]_{\ell+1}, 
4[q,p,q]_{\ell+1},[q+1,p-2,q+1]_{\ell+1},[q+1,p,q+1]_{\ell+1}\cr
\ss [q-1,p,q-1]_{\ell-1},[q-1,p+2,q-1]_{\ell-1},
2[q,p,q]_{\ell-1},[q+1,p-2,q+1]_{\ell-1} }$  &\cr
height4pt&\omit&&\omit&\cr}
\hrule}

\vbox{\tabskip=0pt \offinterlineskip
\hrule
\halign{&\vrule# &\strut \ \hfil#\  \cr
height2pt&\omit&\cr
& $\ell+2q+p+4$\hfil &\cr
height2pt&\omit&\cr
\noalign{\hrule}
height4pt&\omit&\cr
&  $\matrix{\ss [q-2,p+2,q-2]_{\ell+2},2[q-1,p,q-1]_{\ell+2},
2[q-1,p+2,q-1]_{\ell+2},[q,p-2,q]_{\ell+2}\cr
\ss 6[q,p,q]_{\ell+2},[q,p+2,q]_{\ell+2},2[q+1,p-2,q+1]_{\ell+2},
2[q+1,p,q+1]_{\ell+2},[q+2,p-2,q+2]_{\ell+2}\cr
\ss [q-2,p,q-2]_{\ell},4[q-2,p+2,q-2]_{\ell},[q-2,p+4,q-2]_{\ell},
10[q-1,p,q-1]_\ell,8[q-1,p+2,q-1]_\ell, 4[q,p-2,q]_\ell\cr
\ss 15[q,p,q]_\ell,[q,p+2,q]_\ell,
8[q+1,p-2,q+1]_\ell, 2[q+1,p,q+1]_\ell,[q+2,p-4,q+2]_\ell,[q+2,p-2,q+2]_\ell\cr
\ss [q-2,p+2,q-2]_{\ell-2},2[q-1,p,q-1]_{\ell-2},[q,p-2,q]_{\ell-2},
[q,p,q]_{\ell-2}} $ &\cr
height4pt&\omit&\cr}
\hrule}

\vbox{\tabskip=0pt \offinterlineskip
\hrule
\halign{&\vrule# &\strut \ \hfil#\  \cr
height2pt&\omit&\cr
&$\ell+2q+p+5$\hfil &\cr
height2pt&\omit&\cr
\noalign{\hrule}
height4pt&\omit&\cr
&$\matrix{\ss [q-1,p,q-1]_{\ell+3},[q-1,p+2,q-1]_{\ell+3},4[q,p,q]_{\ell+3},
[q+1,p-2,q+1]_{\ell+3},[q+1,p,q+1]_{\ell+3}\cr
\ss [q-3,p+2,q-3]_{\ell+1},[q-3,p+4,q-3]_{\ell+1},2[q-2,p,q-2]_{\ell+1},
10[q-2,p+2,q-2]_{\ell+1},2[q-2,p+4,q-2]_{\ell+1},[q-1,p-2,q-1]_{\ell+1}\cr
\ss 20[q-1,p,q-1]_{\ell+1},15[q-1,p+2,q-1]_{\ell+1},
[q-1,p+4,q-1]_{\ell+1},
10[q,p-2,q]_{\ell+1},26[q,p,q]_{\ell+1},4[q,p+2,q]_{\ell+1}\cr
\ss [q+1,p-4,q+1]_{\ell+1},15[q+1,p-2,q+1]_{\ell+1},
6[q+1,p,q+1]_{\ell+1},2[q+2,p-4,q+2]_{\ell+1},4[q+2,p-2,q+2]_{\ell+1},
[q+3,p-4,q+3]_{\ell+1}\cr
\ss [q-3,p+2,q-3]_{\ell-1},[q-3,p+4,q-3]_{\ell-1},2[q-2,p,q-2]_{\ell-1},
8[q-2,p+2,q-2]_{\ell-1}\cr
\ss [q-1,p-2,q-1]_{\ell-1},15[q-1,p,q-1]_{\ell-1},4[q-1,p+2,q-1]_{\ell-1}, 
8[q,p-2,q]_{\ell-1}, 10[q,p,q]_{\ell-1}\cr
\ss [q+1,p-4,q+1]_{\ell-1},
4[q+1,p-2,q+1]_{\ell-1},[q+1,p,q+1]_{\ell-1},[q-1,p,q-1]_{\ell-3}\cr}$&
\cr height4pt&\omit&\cr}
\hrule}

\vbox{\tabskip=0pt \offinterlineskip
\hrule
\halign{&\vrule# &\strut \ \hfil#\  \cr
height2pt&\omit&\cr
& $\ell+2q+p+6$ \hfil &\cr
height2pt&\omit&\cr
\noalign{\hrule}
height4pt&\omit&\cr
& $\matrix{\ss [q,p,q]_{\ell+4},[q-2,p,q-2]_{\ell+2},4[q-2,p+2,q-2]_{\ell+2},
[q-2,p+4,q-2]_{\ell+2},10[q-1,p,q-1]_{\ell+2},8[q-1,p+2,q-1]_{\ell+2}\cr
\ss 4[q,p-2,q]_{\ell+2},15[q,p,q]_{\ell+2},[q,p+2,q]_{\ell+2},
8[q+1,p-2,q+1]_{\ell+2},2[q+1,p,q+1]_{\ell+2},
[q+2,p-4,q+2]_{\ell+2},[q+2,p-2,q+2]_{\ell+2}\cr
\ss [q-4,p+4,q-4]_{\ell},4[q-3,p+2,q-3]_{\ell},2[q-3,p+4,q-3]_{\ell},
\ss 6[q-2,p,q-2]_{\ell},15[q-2,p+2,q-2]_{\ell},[q-2,p+4,q-2]_{\ell}\cr
\ss 4[q-1,p-2,q-1]_{\ell},26[q-1,p,q-1]_{\ell},10[q-1,p+2,q-1]_{\ell},
[q,p-4,q]_\ell,15[q,p-2,q]_\ell,20[q,p,q]_\ell,[q,p+2,q]_\ell\cr
\ss 2[q+1,p-4,q+1]_{\ell},10[q+1,p-2,q+1]_{\ell},2[q+1,p,q+1]_{\ell},
[q+2,p-4,q+2]_{\ell},[q+2,p-2,q+2]_{\ell}\cr
\ss [q-2,p,q-2]_{\ell-2},[q-2,p+2,q-2]_{\ell-2},4[q-1,p,q-1]_{\ell-2},
[q,p-2,q]_{\ell-2},[q,p,q]_{\ell-2}}$ & \cr
height4pt&\omit&\cr}
\hrule}

\vbox{\tabskip=0pt \offinterlineskip
\hrule
\halign{&\vrule# &\strut \ \hfil#\  \cr
height2pt&\omit&\cr
& $\ell+2q+p+7$\hfil &\cr
height2pt&\omit&\cr
\noalign{\hrule}
height4pt&\omit&\cr
& $\matrix{\ss [q-1,p,q-1]_{\ell+3},[q-1,p+2,q-1]_{\ell+3},2[q,p,q]_{\ell+3},
[q+1,p-2,q+1]_{\ell+3}\cr
\ss [q-3,p+2,q-3]_{\ell+1},[q-3,p+4,q-3]_{\ell+1},2[q-2,p,q-2]_{\ell+1},
8[q-2,p+2,q-2]_{\ell+1},[q-1,p-2,q-1]_{\ell+1},15[q-1,p,q-1]_{\ell+1}\cr
\ss 4[q-1,p+2,q-1]_{\ell+1},8[q,p-2,q]_{\ell+1},10[q,p,q]_{\ell+1},
[q+1,p-4,q+1]_{\ell+1},4[q+1,p-2,q+1]_{\ell+1},[q+1,p,q+1]_{\ell+1}\cr
\ss [q-3,p+2,q-3]_{\ell-1},2[q-2,p,q-2]_{\ell-1},2[q-2,p+2,q-2]_{\ell-1},
[q-1,p-2,q-1]_{\ell-1}\cr
\ss 6[q-1,p,q-1]_{\ell-1},[q-1,p+2,q-1]_{\ell-1},2[q,p-2,q]_{\ell-1},
2[q,p,q]_{\ell-1},[q+1,p-2,q+1]_{\ell-1}}$&\cr  
height4pt&\omit&\cr}
\hrule}

\vbox{\tabskip=0pt \offinterlineskip
\hrule
\halign{&\vrule# &\strut \ \hfil#\  \cr
height2pt&\omit&&\omit&\cr
& $\ell+2q+p+8$\hfil && $\ell+2q+p+9$ \hfil &\cr
\noalign{\hrule}
height4pt&\omit&&\omit&\cr
& $\matrix{\ss[q-2,p+2,q-2]_{\ell+2},2[q-1,p,q-1]_{\ell+2},
[q,p-2,q]_{\ell+2},[q,p,q]_{\ell+2}\cr
\ss [q-2,p,q-2]_{\ell},[q-2,p+2,q-2]_{\ell},
4[q-1,p,q-1]_{\ell},[q,p-2,q]_{\ell},[q,p,q]_{\ell} }$  &
& $ [q-1,p,q-1]_{\ell+1} $  &\cr
height4pt&\omit&&\omit&\cr}
\hrule}
\noindent
Table 4. Diagonal representations in
$\C^{{1\over 4},{1\over 4}}_{[q,p,q]({1\over 2}\ell,{1\over 2}\ell)}$.

\vskip -12pt
\vbox{\tabskip=0pt \offinterlineskip
\hrule
\halign{&\vrule# &\strut \ \hfil#\  \cr
height2pt&\omit&&\omit&&\omit&\cr
& $p+2$\hfil && $p+3$\hfil  && $p+4$\hfil &\cr
height2pt&\omit&&\omit&&\omit&\cr
\noalign{\hrule}
height4pt&\omit&&\omit&&\omit&\cr
& $~[1,p,1]_0$ &&  $\matrix{\ss[0,p,0]_1,[0,p+2,0]_1\cr
\ss 2[1,p,1]_1,[2,p-2,2]_1}$  &
&  $\matrix{\ss 2[0,p,0]_2,[1,p-2,1]_2,[1,p,1]_2\cr
\ss [0,p,0]_0,[1,p-2,1]_0, 2[1,p,1]_0,
2[2,p-2,2]_0, [3,p-4,3]_0} $ &\cr
height4pt&\omit&&\omit&&\omit&\cr}
\hrule}

\vbox{\tabskip=0pt \offinterlineskip
\hrule
\halign{&\vrule# &\strut \ \hfil#\  \cr
height2pt&\omit&&\omit&&\omit&\cr
& $p+5$ \hfil
&& $p+6$\hfil && $p+7$\hfil &\cr
height2pt&\omit&&\omit&&\omit&\cr
\noalign{\hrule}
height4pt&\omit&&\omit&&\omit&\cr
& $\matrix{\ss[0,p,0]_3,[0,p-2,0]_1,
2[0,p,0]_1,4[1,p-2,1]_1\cr
\ss [1,p,1]_1,[2,p-4,2]_1,[2,p-2,2]_1}$ &
& $\matrix{
\ss [1,p-2,1]_2,2[0,p-2,0]_0\cr
\ss [1,p-4,1]_0,[1,p-2,1]_0}$&& $~[0,p-2,0]_1$ & \cr
height4pt&\omit&&\omit&&\omit&\cr}
\hrule}

\noindent
Table 5. Diagonal representations in $\B^{{1\over 4},{1\over 4}}_{[1,p,1](0,0)}$.

\vskip 4pt
\vbox{\tabskip=0pt \offinterlineskip
\hrule
\halign{&\vrule# &\strut \ \hfil#\  \cr
height2pt&\omit&&\omit&\cr
& $\ell+p+4$\hfil && $\ell+p+5$\hfil  &\cr
height2pt&\omit&&\omit&\cr
\noalign{\hrule}
height4pt&\omit&&\omit&\cr
& $~[1,p,1]_\ell$ &
&  $\matrix{\ss [0,p,0]_{\ell+1},[0,p+2,0]_{\ell+1}, 
4[1,p,1]_{\ell+1},[2,p-2,2]_{\ell+1},[2,p,2]_{\ell+1}\cr
\ss [0,p,0]_{\ell-1},[0,p+2,0]_{\ell-1},
2[1,p,1]_{\ell-1},[2,p-2,2]_{\ell-1} }$  &\cr
height4pt&\omit&&\omit&\cr}
\hrule}

\vbox{\tabskip=0pt \offinterlineskip
\hrule
\halign{&\vrule# &\strut \ \hfil#\  \cr
height2pt&\omit&\cr
& $\ell+p+6$\hfil &\cr
height2pt&\omit&\cr
\noalign{\hrule}
height4pt&\omit&\cr
&  $\matrix{\ss 2[0,p,0]_{\ell+2},
2[0,p+2,0]_{\ell+2},[1,p-2,1]_{\ell+2},6[1,p,1]_{\ell+2},
[1,p+2,1]_{\ell+2},2[2,p-2,2]_{\ell+2},
2[2,p,2]_{\ell+2},[3,p-2,3]_{\ell+2}\cr
\ss 6[0,p,0]_\ell,4[0,p+2,0]_\ell, 4[1,p-2,1]_\ell,13[1,p,1]_\ell,
[1,p+2,1]_\ell,
8[2,p-2,2]_\ell, 2[2,p,2]_\ell,[3,p-4,3]_\ell,[3,p-2,3]_\ell\cr
\ss 2[0,p,0]_{\ell-2},[1,p-2,1]_{\ell-2},[1,p,1]_{\ell-2}, } $ &\cr
height4pt&\omit&\cr}
\hrule}

\vbox{\tabskip=0pt \offinterlineskip
\hrule
\halign{&\vrule# &\strut \ \hfil#\  \cr
height2pt&\omit&\cr
&$\ell+p+7$\hfil &\cr
height2pt&\omit&\cr
\noalign{\hrule}
height4pt&\omit&\cr
&$\matrix{\ss [0,p,0]_{\ell+3},[0,p+2,0]_{\ell+3},4[1,p,1]_{\ell+3},
[2,p-2,2]_{\ell+3},[2,p,2]_{\ell+3}\cr
\ss [0,p-2,0]_{\ell+1},11[0,p,0]_{\ell+1},8[0,p+2,0]_{\ell+1},
[0,p+4,0]_{\ell+1},
10[1,p-2,1]_{\ell+1},22[1,p,1]_{\ell+1},4[1,p+2,1]_{\ell+1}\cr
\ss [2,p-4,2]_{\ell+1},15[2,p-2,2]_{\ell+1},
6[2,p,2]_{\ell+1},2[3,p-4,3]_{\ell+1},4[3,p-2,3]_{\ell+1},
[4,p-4,4]_{\ell+1}\cr
\ss [0,p-2,0]_{\ell-1},8[0,p,0]_{\ell-1},[0,p+2,0]_{\ell-1}, 
8[1,p-2,1]_{\ell-1}, 8[1,p,1]_{\ell-1},
[2,p-4,2]_{\ell-1},
4[2,p-2,2]_{\ell-1},[2,p,2]_{\ell-1},[0,p,0]_{\ell-3}\cr}$&
\cr height4pt&\omit&\cr}
\hrule}

\vbox{\tabskip=0pt \offinterlineskip
\hrule
\halign{&\vrule# &\strut \ \hfil#\  \cr
height2pt&\omit&\cr
& $\ell+p+8$ \hfil &\cr
height2pt&\omit&\cr
\noalign{\hrule}
height4pt&\omit&\cr
& $\matrix{\ss [1,p,1]_{\ell+4},6[0,p,0]_{\ell+2},4[0,p+2,0]_{\ell+2},
4[1,p-2,1]_{\ell+2},13[1,p,1]_{\ell+2},[1,p+2,1]_{\ell+2}\cr
\ss 8[2,p-2,2]_{\ell+2},2[2,p,2]_{\ell+2},
[3,p-4,3]_{\ell+2},[3,p-2,3]_{\ell+2}\cr
\ss 4[0,p-2,0]_{\ell},12[0,p,0]_{\ell},4[0,p+2,0]_{\ell},
[1,p-4,1]_\ell,15[1,p-2,1]_\ell,15[1,p,1]_\ell,[1,p+2,1]_\ell\cr
\ss 2[2,p-4,2]_{\ell},10[2,p-2,2]_{\ell},2[2,p,2]_{\ell},
[3,p-4,3]_{\ell},[3,p-2,3]_{\ell},
2[0,p,0]_{\ell-2},[1,p-2,1]_{\ell-2},[1,p,1]_{\ell-2}}$ & \cr
height4pt&\omit&\cr}
\hrule}

\vbox{\tabskip=0pt \offinterlineskip
\hrule
\halign{&\vrule# &\strut \ \hfil#\  \cr
height2pt&\omit&\cr
& $\ell+p+9$\hfil &\cr
height2pt&\omit&\cr
\noalign{\hrule}
height4pt&\omit&\cr
& $\matrix{\ss [0,p,0]_{\ell+3},[0,p+2,0]_{\ell+3},2[1,p,1]_{\ell+3},
[2,p-2,2]_{\ell+3}\cr
\ss [0,p-2,0]_{\ell+1},8[0,p,0]_{\ell+1},[0,p+2,0]_{\ell+1},8[1,p-2,1]_{\ell+1},
8[1,p,1]_{\ell+1},
[2,p-4,2]_{\ell+1},4[2,p-2,2]_{\ell+1},[2,p,2]_{\ell+1}\cr
\ss [0,p-2,0]_{\ell-1},3[0,p,0]_{\ell-1},[0,p+2,0]_{\ell-1},2[1,p-2,1]_{\ell-1},
2[1,p,1]_{\ell-1},[2,p-2,2]_{\ell-1}}$&\cr  
height4pt&\omit&\cr}
\hrule}

\vbox{\tabskip=0pt \offinterlineskip
\hrule
\halign{&\vrule# &\strut \ \hfil#\  \cr
height2pt&\omit&&\omit&\cr
& $\ell+p+10$\hfil && $\ell+p+11$ \hfil &\cr
\noalign{\hrule}
height4pt&\omit&&\omit&\cr
& $\matrix{\ss 2[0,p,0]_{\ell+2},
[1,p-2,1]_{\ell+2},[1,p,1]_{\ell+2}\cr
\ss 2[0,p,0]_{\ell},[1,p-2,1]_{\ell},[1,p,1]_{\ell} }$  &
& $~[0,p,0]_{\ell+1} $  &\cr
height4pt&\omit&&\omit&\cr}
\hrule}
\noindent
Table 6. Diagonal representations in
$\C^{{1\over 4},{1\over 4}}_{[1,p,1]({1\over 2}\ell,{1\over 2}\ell)}$.

\vskip 4pt
\vbox{\tabskip=0pt \offinterlineskip
\hrule
\halign{&\vrule# &\strut \ \hfil#\  \cr
height2pt&\omit&&\omit&&\omit&&\omit&\cr
& $\ell+4$\hfil && $\ell+5$\hfil  && $\ell+6$\hfil && $\ell+7$ \hfil &\cr
height2pt&\omit&&\omit&&\omit&&\omit&\cr
\noalign{\hrule}
height4pt&\omit&&\omit&&\omit&&\omit&\cr
& $~[0,2,0]_\ell$ &&  $\matrix{\ss 2[0,2,0]_{\ell+1},[1,0,1]_{\ell+1}\cr
\ss [1,2,1]_{\ell+1},[1,0,1]_{\ell-1}}$  &
&  $\matrix{\ss [0,0,0]_{\ell+2},3[0,2,0]_{\ell+2},[0,4,0]_{\ell+2}\cr
\ss 2[1,0,1]_{\ell+2},2[1,2,1]_{\ell+2},[2,0,2]_{\ell+2}\cr
\ss [0,0,0]_\ell,[0,2,0]_\ell, 3[1,0,1]_\ell,[2,0,2]_\ell\cr
\ss [0,0,0]_{\ell-2}} $ &
& $\matrix{\ss 2[0,2,0]_{\ell+3},[1,0,1]_{\ell+3},[1,2,1]_{\ell+3}\cr
\ss [0,0,0]_{\ell+1},2[0,2,0]_{\ell+1},4[1,0,1]_{\ell+1}\cr
\ss [1,2,1]_{\ell+1},[2,0,2]_{\ell+1}\cr
\ss [0,0,0]_{\ell-1},[1,0,1]_{\ell-1} }$ & \cr
height4pt&\omit&&\omit&&\omit&&\omit&\cr}
\hrule}
\vbox{\tabskip=0pt \offinterlineskip
\hrule
\halign{&\vrule# &\strut \ \hfil#\  \cr
height2pt&\omit&&\omit&&\omit&\cr
& $\ell+8$\hfil && $\ell+9$\hfil && $\ell+10$ \hfil &\cr
height2pt&\omit&&\omit&&\omit&\cr
\noalign{\hrule}
height4pt&\omit&&\omit&&\omit&\cr
& $\matrix{\ss [0,2,0]_{\ell+4},[0,0,0]_{\ell+2},[0,2,0]_{\ell+2},
3[1,0,1]_{\ell+2},[2,0,2]_{\ell+2}\cr
\ss [0,0,0]_\ell,[0,2,0]_\ell,[1,0,1]_\ell}$&
& $\matrix{\ss[1,0,1]_{\ell+3}\cr
\ss [0,0,0]_{\ell+1},[1,0,1]_{\ell+1}}$  && $~[0,0,0]_{\ell+2}$  &\cr
height4pt&\omit&&\omit&&\omit&\cr}
\hrule}
\noindent
Table 7. Diagonal representations in
$\C^{{1\over 2},{1\over 2}}_{[0,2,0]({1\over 2}\ell,{1\over 2}\ell)}$.

\vskip 4pt
\vbox{\tabskip=0pt \offinterlineskip
\hrule
\halign{&\vrule# &\strut \ \hfil#\  \cr
height2pt&\omit&&\omit&\cr
& $2q+p+2$\hfil && $2q+p+3$\hfil  &\cr
height2pt&\omit&&\omit&\cr
\noalign{\hrule}
height4pt&\omit&&\omit&\cr
& $~[q,p,q]_0$ &
&  $\matrix{\ss [q-1,p,q-1]_{1},[q-1,p+2,q-1]_{1},
4[q,p,q]_{1},[q+1,p-2,q+1]_{1},[q+1,p,q+1]_{1}\cr }$  &\cr
height4pt&\omit&&\omit&\cr}
\hrule}

\vbox{\tabskip=0pt \offinterlineskip
\hrule
\halign{&\vrule# &\strut \ \hfil#\  \cr
height2pt&\omit&\cr
& $2q+p+4$\hfil &\cr
height2pt&\omit&\cr
\noalign{\hrule}
height4pt&\omit&\cr
&  $\matrix{\ss [q-2,p+2,q-2]_{2},2[q-1,p,q-1]_{2},
2[q-1,p+2,q-1]_{2},[q,p-2,q]_{2}\cr
\ss 6[q,p,q]_{2},[q,p+2,q]_{2},2[q+1,p-2,q+1]_{2},
2[q+1,p,q+1]_{2},[q+2,p-2,q+2]_{2}\cr
\ss [q-2,p,q-2]_{0},[q-2,p+2,q-2]_{0},[q-2,p+4,q-2]_{0},
4[q-1,p,q-1]_0,4[q-1,p+2,q-1]_0, [q,p-2,q]_0\cr
\ss 8[q,p,q]_0,[q,p+2,q]_0,
4[q+1,p-2,q+1]_0, 2[q+1,p,q+1]_0,[q+2,p-4,q+2]_0,[q+2,p-2,q+2]_0} $ &\cr
height4pt&\omit&\cr}
\hrule}

\vbox{\tabskip=0pt \offinterlineskip
\hrule
\halign{&\vrule# &\strut \ \hfil#\  \cr
height2pt&\omit&\cr
&$2q+p+5$\hfil &\cr
height2pt&\omit&\cr
\noalign{\hrule}
height4pt&\omit&\cr
&$\matrix{\ss [q-1,p,q-1]_{3},[q-1,p+2,q-1]_{3},4[q,p,q]_{3},
[q+1,p-2,q+1]_{3},[q+1,p,q+1]_{3}\cr
\ss [q-3,p+2,q-3]_{1},[q-3,p+4,q-3]_{1},2[q-2,p,q-2]_{1},
8[q-2,p+2,q-2]_{1},2[q-2,p+4,q-2]_{1},[q-1,p-2,q-1]_{1}\cr
\ss 15[q-1,p,q-1]_{1},13[q-1,p+2,q-1]_{1},
[q-1,p+4,q-1]_{1},
8[q,p-2,q]_{1},22[q,p,q]_{1},4[q,p+2,q]_{1}\cr
\ss [q+1,p-4,q+1]_{1},13[q+1,p-2,q+1]_{1},
6[q+1,p,q+1]_{1},2[q+2,p-4,q+2]_{1},4[q+2,p-2,q+2]_{1},
[q+3,p-4,q+3]_{1}}$&
\cr height4pt&\omit&\cr}
\hrule}

\vbox{\tabskip=0pt \offinterlineskip
\hrule
\halign{&\vrule# &\strut \ \hfil#\  \cr
height2pt&\omit&\cr
& $2q+p+6$ \hfil &\cr
height2pt&\omit&\cr
\noalign{\hrule}
height4pt&\omit&\cr
& $\matrix{\ss [q,p,q]_{4},[q-2,p,q-2]_{2},4[q-2,p+2,q-2]_{2},
[q-2,p+4,q-2]_{2},10[q-1,p,q-1]_{2},8[q-1,p+2,q-1]_{2}\cr
\ss 4[q,p-2,q]_{2},15[q,p,q]_{2},[q,p+2,q]_{2},
8[q+1,p-2,q+1]_{2},2[q+1,p,q+1]_{2},
[q+2,p-4,q+2]_{2},[q+2,p-2,q+2]_{2}\cr
\ss [q-4,p+4,q-4]_{0},2[q-3,p+2,q-3]_{0},2[q-3,p+4,q-3]_{0},
\ss 3[q-2,p,q-2]_{0},8[q-2,p+2,q-2]_{0},[q-2,p+4,q-2]_{0}\cr
\ss 2[q-1,p-2,q-1]_{0},12[q-1,p,q-1]_{0},6[q-1,p+2,q-1]_{0},
[q,p-4,q]_0,8[q,p-2,q]_0,11[q,p,q]_0,[q,p+2,q]_0\cr
\ss 2[q+1,p-4,q+1]_{0},6[q+1,p-2,q+1]_{0},2[q+1,p,q+1]_{0},
[q+2,p-4,q+2]_{0},[q+2,p-2,q+2]_{0}}$ & \cr
height4pt&\omit&\cr}
\hrule}

\vbox{\tabskip=0pt \offinterlineskip
\hrule
\halign{&\vrule# &\strut \ \hfil#\  \cr
height2pt&\omit&\cr
& $2q+p+7$\hfil &\cr
height2pt&\omit&\cr
\noalign{\hrule}
height4pt&\omit&\cr
& $\matrix{\ss [q-1,p,q-1]_{3},[q-1,p+2,q-1]_{3},2[q,p,q]_{3},
[q+1,p-2,q+1]_{3}\cr
\ss [q-3,p+2,q-3]_{1},[q-3,p+4,q-3]_{1},2[q-2,p,q-2]_{1},
8[q-2,p+2,q-2]_{1},[q-1,p-2,q-1]_{1},15[q-1,p,q-1]_{1}\cr
\ss 4[q-1,p+2,q-1]_{1},8[q,p-2,q]_{1},10[q,p,q]_{1},
[q+1,p-4,q+1]_{\ell},4[q+1,p-2,q+1]_{1},[q+1,p,q+1]_{1}}$&\cr
height4pt&\omit&\cr}
\hrule}

\vbox{\tabskip=0pt \offinterlineskip
\hrule
\halign{&\vrule# &\strut \ \hfil#\  \cr
height2pt&\omit&&\omit&\cr
& $2q+p+8$\hfil && $2q+p+9$ \hfil &\cr
\noalign{\hrule}
height4pt&\omit&&\omit&\cr
& $\matrix{\ss[q-2,p+2,q-2]_{2},2[q-1,p,q-1]_{2},
[q,p-2,q]_{2},[q,p,q]_{2}\cr
\ss [q-2,p,q-2]_{0},[q-2,p+2,q-2]_{0},
4[q-1,p,q-1]_{0},[q,p-2,q]_{0},[q,p,q]_{0} }$  &
& $~[q-1,p,q-1]_{1} $  &\cr
height4pt&\omit&&\omit&\cr}
\hrule}

\noindent
Table 8. Diagonal representations in
$\C^{{1\over 4},{1\over 4}}_{[q,p,q](0,0)}$.

\vskip 4pt
\vbox{\tabskip=0pt \offinterlineskip
\hrule
\halign{&\vrule# &\strut \ \hfil#\  \cr
height2pt&\omit&&\omit&&\omit&\cr
& $p+2$\hfil && $p+3$\hfil  && $p+4$\hfil &\cr
height2pt&\omit&&\omit&&\omit&\cr
\noalign{\hrule}
height4pt&\omit&&\omit&&\omit&\cr
& $~[0,p,0]_0$ &&  $\matrix{\ss 2[0,p,0]_{1},[1,p-2,1]_{1}\cr
\ss [1,p,1]_{1}}$  &
&  $\matrix{\ss [0,p-2,0]_{2},3[0,p,0]_{2},[0,p+2,0]_{2},
2[1,p-2,1]_{2},2[1,p,1]_{2},[2,p-2,2]_{2}\cr
\ss [0,p,0]_0, 2[1,p-2,1]_0, [2,p-4,2]_0, [2,p-2,2]_0\cr} $ &\cr
height4pt&\omit&&\omit&&\omit&\cr}
\hrule}

\vbox{\tabskip=0pt \offinterlineskip
\hrule
\halign{&\vrule# &\strut \ \hfil#\  \cr
height2pt&\omit&&\omit&\cr
& $p+5$ \hfil
&& $p+6$\hfil &\cr
height2pt&\omit&&\omit&\cr
\noalign{\hrule}
height4pt&\omit&&\omit&\cr
& $\matrix{\ss 2[0,p,0]_{3},[1,p-2,1]_{3},[1,p,1]_{3}\cr
\ss 2[0,p-2,0]_{1},2[0,p,0]_{1},[1,p-4,1]_{1},
6[1,p-2,1]_{1}\cr
\ss [1,p,1]_{1},2[2,p-4,2]_{1}, 2[2,p-2,2]_{1},
[3,p-4,3]_{1}\cr }$ &
& $\matrix{\ss [0,p,0]_{4},[0,p-2,0]_{2},[0,p,0]_{2},
4[1,p-2,1]_{2},[2,p-4,2]_{2},[2,p-2,2]_{2}\cr
\ss [0,p-4,0]_0,3[0,p-2,0]_0,[0,p,0]_0\cr
\ss 2[1,p-4,1]_0,2[1,p-2,1]_0,[2,p-4,2]_0}$&\cr
height4pt&\omit&&\omit&\cr}
\hrule}
\vbox{\tabskip=0pt \offinterlineskip
\hrule
\halign{&\vrule# &\strut \ \hfil#\  \cr
height2pt&\omit&&\omit&\cr
& $p+7$\hfil && $p+8$ \hfil &\cr
\noalign{\hrule}
height4pt&\omit&&\omit&\cr
& $\matrix{\ss[1,p-2,1]_{3},2[0,p-2,0]_{1},
[1,p-4,1]_{1},[1,p-2,1]_{1}}$  && $~[0,p-2,0]_{2} $  &\cr
height4pt&\omit&&\omit&\cr}
\hrule}

\noindent
Table 9. Diagonal representations in
$\C^{{1\over 2},{1\over 2}}_{[0,p,0](0,0)}$.

\listrefs
\bye